\documentclass[aps,prc,amsfonts,preprintnumbers,superscriptaddress,showpacs,nofootinbib]{revtex4}

\usepackage{graphics}
\usepackage{amsmath}
\usepackage{amssymb}
\usepackage{psfrag}
\usepackage{epsfig}
\usepackage{epsf}
\usepackage{float}
\usepackage{slashed}
\allowdisplaybreaks

\newcommand{\fet}[1]{\mbox{\boldmath $#1$}}

\newcommand{\beq}{\begin{equation}}
\newcommand{\eeq}{\end{equation}}
\newcommand{\beqa}{\begin{eqnarray}}
\newcommand{\eeqa}{\end{eqnarray}}
\newcommand{\nn}{\nonumber \\ }

\DeclareMathOperator{\arccosh}{{\rm arccosh}}

\begin{document}

\title{Three-nucleon force in chiral EFT with
  explicit $\Delta$(1232) degrees of freedom: 
Longest-range contributions at fourth order}

\author{H.~Krebs}
\email[]{Email: hermann.krebs@rub.de}
\affiliation{Institut f\"ur Theoretische Physik II, Ruhr-Universit\"at Bochum,
  D-44780 Bochum, Germany}
\author{A.~M.~Gasparyan}
\email[]{Email: ashotg@tp2.rub.de}
\affiliation{Institut f\"ur Theoretische Physik II, Ruhr-Universit\"at Bochum,
  D-44780 Bochum, Germany}
\affiliation{Institute for Theoretical and Experimental Physics, B. Cheremushkinskaya 25, 117218 Moscow, Russia}
\author{E.~Epelbaum}
\email[]{Email: evgeny.epelbaum@rub.de}
\affiliation{Institut f\"ur Theoretische Physik II, Ruhr-Universit\"at Bochum,
  D-44780 Bochum, Germany}
\date{\today}

\begin{abstract}
We analyze the longest-range two-pion exchange contributions to the
three-nucleon force at leading-loop order in the framework of
heavy-baryon chiral effective field
theory with explicit $\Delta$(1232) degrees of freedom. All relevant
low-energy constants which appear in the calculation are determined
from pion-nucleon scattering. Comparing our results with the ones
obtained in the $\Delta$-less theory at N$^4$LO, we find  
effects of the $\Delta$ isobar for this particular 
topology to be rather well represented in terms of
resonance saturation of various low-energy constants
in the $\Delta$-less approach. 
\end{abstract}

\pacs{13.75.Cs,21.30.-x}

\maketitle

\vspace{-0.2cm}

\section{Introduction}
\def\theequation{\arabic{section}.\arabic{equation}}
\label{sec:intro}

Three-nucleon forces (3NF) and their impact on nuclear structure and
reactions became an important frontier in nuclear physics, see 
Refs.~\cite{Stephan:2010zz,Ciepal:2012zz,Viviani:2010mf,Witala:2013ioa,Navratil:2007we,Gazit:2008ma,Roth:2011ar,Hebeler:2010jx,Hagen:2012sh,Holt:2012fr,Tews:2012fj,Skibinski:2013uua,Ekstrom:2014iya,Golak:2014ksa,Hebeler:2015wxa,Binder:2015mbz,Ekstrom:2015rta,Lynn:2015jua,Witala:2016inj,Drischler:2016cpy,Ekstrom:2017koy,Drischler:2017wtt,Piarulli:2017dwd,Tews:2018kmu,Binder:2018pgl} 
for a selection of recent studies along these lines and
Refs.~\cite{KalantarNayestanaki:2011wz,Hammer:2012id}  for review
articles. Chiral effective field theory (EFT) provides a
model-independent and systematically improvable 
theoretical framework to describe nuclear forces
and low-energy nuclear structure and dynamics in harmony with the
symmetries of QCD \cite{Epelbaum:2008ga,Machleidt:2011zz}.  
Nucleon-nucleon (NN) scattering has been extensively studied in chiral EFT
in the past two decades following the pioneering work by Weinberg
\cite{Weinberg:1990rz} and Ordonez et al.~\cite{Ordonez:1993tn}. In
particular, NN potentials at
next-to-next-to-next-to-leading order (N$^3$LO) in the chiral
expansion are available since about one and a half decades \cite{Epelbaum:2004fk,Entem:2003ft} and served
as a basis for numerous \emph{ab initio} calculations of nuclear structure
and reactions. Recently, accurate and precise chiral EFT potentials up
to fifth order in the chiral expansion, i.e.~N$^4$LO, have been
developed
\cite{Entem:2014msa,Epelbaum:2014sza,Entem:2017gor,Reinert:2017usi}.
In particular, the semilocal N$^4$LO$^+$
potentials of Ref.~\cite{Reinert:2017usi} provide a 
description of the 2013 Granada
database of neutron-proton and proton-proton scattering data below
$E_{\rm lab} = 300$~MeV, which is comparable to or even better than that
based on the available high-precision phenomenological potentials.  

\begin{figure}[t]
\vskip 1 true cm
\includegraphics[width=0.9\textwidth,keepaspectratio,angle=0,clip]{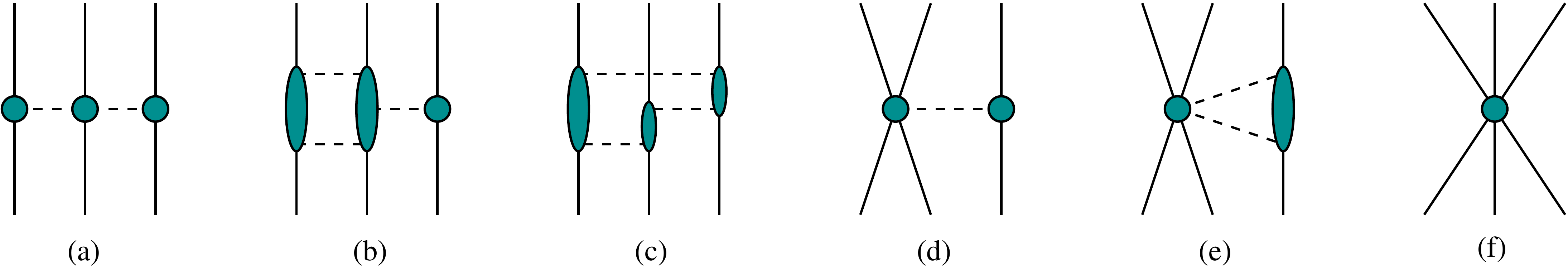}
    \caption{
          Various topologies contributing to the 3NF 
up-to-and-including N$^4$LO:  two-pion ($2\pi$) exchange (a), two-pion-one-pion ($2\pi$-$1\pi$)
         exchange (b), ring (c), one-pion-exchange-contact (d),
         two-pion-exchange-contact (e) and purely contact (f) diagrams.
Solid and dashed lines
         represent nucleons and pions, respectively. 
Shaded blobs represent the corresponding amplitudes. 
\label{fig0} 
 }
\end{figure}

The chiral expansion  of the 3NF at one-loop level, i.e.~up to
and including next-to-next-to-next-to-next-to-leading order (N$^4$LO) contributions, 
can be described in terms of six topologies depicted in
Fig.~\ref{fig0}. 
The first nonvanishing contributions emerge at
next-to-next-to-leading order (N$^2$LO) from tree-level
diagrams of type (a), (d) and (f) \cite{vanKolck:1994yi,Epelbaum:2002vt}. 
The resulting 3NF at N$^2$LO has been intensively explored in three-
and four-nucleon scattering calculations as well as in nuclear
structure calculations, see
\cite{Stephan:2010zz,Ciepal:2012zz,Viviani:2010mf,Witala:2013ioa,Navratil:2007we,Gazit:2008ma,Roth:2011ar,Hebeler:2010jx,Hagen:2012sh,Holt:2012fr,Tews:2012fj,Ekstrom:2014iya,Ekstrom:2015rta,Lynn:2015jua,Witala:2016inj,Drischler:2016cpy,Ekstrom:2017koy,Drischler:2017wtt,Piarulli:2017dwd,Tews:2018kmu} for some recent examples and the
review articles \cite{KalantarNayestanaki:2011wz,Hammer:2012id} and references therein.  
The first corrections to the 3NF emerge at N$^3$LO from all possible one-loop
diagrams of type (a)-(e) in Fig.~\ref{fig0} constructed from 
the lowest-order vertices. The resulting parameter-free
expressions have been worked out in Refs.~\cite{Bernard:2007sp,Bernard:2011zr}, see also
Ref.~\cite{Ishikawa:2007zz}. 
An interesting feature of the N$^3$LO 3NF contributions is their
rather rich isospin-spin-momentum structure emerging primarily from
the ring topology (c) in Fig.~\ref{fig0}. This is in  contrast with the quite
restricted operator structure of the N$^2$LO 3NF. Numerical implementation of the
N$^3$LO 3NF corrections requires their partial wave
decomposition \cite{Golak:2009ri,Hebeler:2015wxa} and a consistent
implementation of the regulator. This work is currently in progress, see
Refs.~\cite{Skibinski:2011vi,Witala:2013ioa,Skibinski:2013uua,Golak:2014ksa} for some preliminary
results. We further emphasize that four-nucleon forces also start to
contribute at N$^3$LO and have been worked out in
Refs.~\cite{Epelbaum:2006eu,Epelbaum:2007us}. Pioneering applications of the chiral four-nucleon
forces to the $\alpha$-particle binding energy \cite{Rozpedzik:2006yi,Nogga4N}
and neutron matter \cite{Tews:2012fj,Kaiser:2015lsa,Drischler:2017wtt}
indicate that their effects in these systems are fairly small. 

While the impact of the first corrections to the chiral 3NF
on few- and many-nucleon observables is yet to be investigated, one
may ask whether the chiral expansion of the 3NF at  
subleading order, i.e.~at N$^3$LO, provides a reasonable approximation
to the converged result. To clarify this issue we have
worked out the next-to-next-to-next-to-next-to-leading order (N$^4$LO) 
contributions to the long-range \cite{Krebs:2012yv} and intermediate-range
\cite{Krebs:2013kha} 3NF corresponding to diagrams (a) and
(b, c) in Fig.~\ref{fig0}, respectively. The corresponding potentials
at large distance emerge as parameter-free predictions as they are
completely determined by the chiral symmetry of QCD and experimental
information on pion-nucleon scattering needed 
to fix the relevant low-energy constants (LECs).  More precisely, for the two-pion-exchange
topology, the N$^4$LO 3NF contributions depend on some 
of the LECs $c_i$, $\bar d_i$ and $\bar e_i$ from
the order-$Q^2$, $Q^3$ and $Q^4$ effective pion-nucleon Lagrangians,
which have been extracted from the available $\pi N$ partial wave
analyses. The resulting longest-range 3NF was shown to converge
reasonably fast  \cite{Krebs:2012yv}. The situation appears to be very different for the
two-pion-one-pion ($2\pi$-$1\pi$) exchange and ring 3NF topologies corresponding to
diagrams (b) and (c) in Fig.~\ref{fig0}: the formally leading
contributions emerging at N$^3$LO turn out to be rather
small in magnitude while  the first corrections at  N$^4$LO are 
considerably larger \cite{Krebs:2013kha}. The origin of
such an unnatural convergence pattern can be understood if one assumes
the intermediate  $\Delta$(1232) excitation  as a dominant 3NF
mechanism, which is well in line with various phenomenological studies
\cite{Fujita:1957zz,Pieper:2001ap,Deltuva:2003fn}.  In the standard
formulation of chiral EFT based on pions and nucleons as the only
explicit degrees of freedom and used, in particular, in Refs.~\cite{Krebs:2012yv,Krebs:2013kha},
all effects of the $\Delta$ (and heavier resonances as well as heavy
mesons) are taken into account implicitly through (some of the) LECs 
starting from the subleading effective Lagrangian, i.e. $c_i, \, \bar
d_i, \, \bar e_i, \, \ldots$.  In particular, the values of the
LECs $c_{3,4}$, which contribute to the two-pion exchange 3NF at
N$^2$LO, are known to receive large contributions from the
$\Delta$. Thus, for this longest-range 3NF topology, effects of the
$\Delta$ are already, to a large extent, accounted for at the lowest
order (N$^2$LO). The first corrections at N$^3$LO emerge from
one-loop diagrams constructed from the leading-order  pion-nucleon
vertices, which are not affected by the $\Delta$, and the corresponding
potentials appear to be fairly small in magnitude. This explains the
observed good convergence pattern of the chiral expansion for the
two-pion exchange 3NF. On the other hand, for the intermediate-range
topologies, the expansion starts at N$^3$LO while the first effects of
the $\Delta$ appear at N$^4$LO  and lead to large
corrections. Moreover, since the  N$^4$LO contributions to the
$2\pi$-$1\pi$ and ring 3NFs are proportional to $c_i$, only effects
due to single-delta excitations are implicitly taken into account at
that order. This raises the question of whether the double- and
triple-delta excitations, which in the standard $\Delta$-less formulation
of chiral EFT are taken into account at even higher orders, 
might lead to sizable 3NF contributions. While this question could, at
least in principle, be clarified by extending the calculations to even
higher orders in the chiral expansion, this would require calculation
of two--loop diagrams and also dealing with a large number of new
LECs which makes this strategy hardly feasible. Instead, we follow a
different approach and use chiral EFT with
explicit $\Delta$ degrees of freedom, which offers a more
efficient way to resum the contributions due to intermediate $\Delta$-excitations.
To be specific, we employ a formulation in which the
delta-nucleon mass splitting is treated on the same footing as the
pion mass, which is known as the small-scale expansion (SSE) \cite{Hemmert:1997ye}.  
Following the pioneering calculations in Refs.~\cite{Ordonez:1995rz,Kaiser:1998wa},
we have already  worked out the contributions of the $\Delta$ to the two-
and three-nucleon forces up to N$^2$LO in the SSE
\cite{Krebs:2007rh,Epelbaum:2007sq} and also looked at
isospin-breaking corrections to the NN potential
\cite{Epelbaum:2008td}. These calculations confirmed a
better convergence of the $\Delta$-full EFT formulation compared to its
standard, $\Delta$-less version. Interestingly, for the 3NF, the only
nonvanishing $\Delta$ contribution up to N$^2$LO is the two-pion
exchange diagram with an 
intermediate $\Delta$-resonance, commonly called the Fujita-Miyazawa
force. This term is shifted in the $\Delta$-full theory to
next-to-leading order (NLO).

In this paper we, for the first time, extend the SSE for the
nuclear forces to N$^3$LO and concentrate on the longest-range
contribution to the 3NF corresponding to diagram (a) in
Fig.~\ref{fig0}. 
This topology is particularly challenging due to (i) the need to carry
out a non-trivial renormalization program as will be explained later 
and (ii) the need to re-consider pion-nucleon scattering in order to
determine the relevant LECs, see Ref.~\cite{Krebs:2012yv} where this program was
carried out in the standard, $\Delta$-less version of chiral EFT. We will
also discuss in detail renormalization within the $\Delta$-full framework
and work out the $\Delta$ contributions to the relevant low-energy
constants in the effective Lagrangian. Although we do not expect to
see large benefits from the explicit treatment of the $\Delta$ for the
$2\pi$-exchange 3NF,  where the standard chiral expansion already
shows a good convergence \cite{Bernard:2007sp,Krebs:2012yv}, this calculation is a necessary
prerequisite for analyzing the $\Delta$ contributions to the more problematic
intermediate-range diagrams. This work is in progress and will be
reported in a separate publication. 

Our paper is organized as follows. In section \ref{sec:lagr}, we describe the
framework and specify all terms in the effective Lagrangian that are needed in the calculation.
Renormalization of the lowest-order effective Lagrangian to leading loop order is
carried out in section \ref{sec:ren}. In section \ref{sec:piN} we provide analytic
expressions for the contribution of the $\Delta$ to the relevant LECs $c_i$, $\bar d_i$
and $\bar e_i$ and determine the numerical values of these LECs from
pion-nucleon scattering. $\Delta$ contributions to the $2\pi$-exchange
3NF at N$^3$LO are worked out in section \ref{sec:TPE}. In
particular, we provide here parameter-free expressions both in
momentum and coordinate spaces. A comparison of our findings with the
ones of Refs.~\cite{Bernard:2007sp,Krebs:2012yv}  is given in section
\ref{sec:results}. Finally, the main results of our work are briefly summarized
in section \ref{sec:summary}. 
The Appendices contain the unitary transformations of the nuclear Hamiltonian and
the delta-contributions up to N$^3$LO to the $\pi$N invariant amplitudes.

\section{The framework}
\def\theequation{\arabic{section}.\arabic{equation}}
\label{sec:lagr}

In the following,  we briefly describe the formalism we employ in our
analysis, namely  the
heavy-baryon formulation of chiral
EFT with explicit $\Delta(1232)$ degrees of
freedom~\cite{Hemmert:1997ye}. In this framework, the soft scales are
given by small external momenta $Q$, pion mass $M_\pi$ and the
delta-nucleon mass splitting $\Delta:=m_\Delta- m_N$. The resulting
expansion in powers of the small parameter $\epsilon$ defined as  
\beqa
\epsilon \in \left\{ \frac{Q}{\Lambda_\chi}, \,
\frac{M_\pi}{\Lambda_\chi}, \,
\frac{\Delta }{\Lambda_\chi} \right\}
\eeqa
with $\Lambda_\chi \sim 1$ GeV denoting the chiral  
symmetry breaking scale, is
known in the literature as the SSE.  

We begin with specifying the effective chiral Lagrangian for pions,
nucleons and the $\Delta$. 
It is well known that 
the free spin-$3/2$ Lagrangian is non-unique and can be written in the
form
\beq
{\cal L}_{\Delta}^{{\rm free}}=-\bar{\psi}_\alpha^i O_A^{\alpha
  \mu}\left[(i\slashed{\partial} -
  m_\Delta) g_{\mu
    \nu}-\frac{1}{4}\gamma_\mu \gamma_\lambda (i\slashed{\partial} -
 m_\Delta)\gamma^\lambda \gamma_\nu
\right]\xi_{3/2}^{i j}
O_A^{\nu \beta}\psi_\beta^j, 
\eeq
where the tensor
\beq
O_A^{\mu \nu} = g^{\mu \nu} + \frac{1}{2} \, A \,\gamma^\mu \gamma^\nu
\eeq
parametrizes non-uniqueness in the description of a spin-$3/2$
theory in terms of a parameter $A$, which can be chosen arbitrarily
subject to the
restriction $A\neq -1/2$. Further, the quantity $\xi_{3/2}^{i j}$
is the  isospin-$3/2$ projection operator given by 
\beq
\xi_{3/2}^{i j}=\delta^{i j} - \frac{1}{3}\tau^i \tau^j \,,
\eeq
where $\tau_i$ denote the isospin Pauli matrices. 
Physical observables do not
depend on the choice of the parameter $A$ since the entire dependence
on $A$ can be absorbed into a field redefinition of the delta
field. In practical calculations, the choice of $A$ is a matter of
convenience. In the \emph{covariant} approach, one usually chooses
$A=-1$, see 
e.g.~\cite{Bernard:2003xf,Bernard:2005fy,Ledwig:2011cx,Bernard:2012hb,Siemens:2014pma}, since in this case
the free Lagrangian takes the particularly simple form
\beq
{\cal L}_{\Delta}^{{\rm free}}=\bar{\psi}_i^\mu
\,(i \gamma_{\mu\nu\alpha}\,\partial^\alpha  - m_\Delta \gamma_{\mu\nu}) \xi_{3/2}^{ij} \, 
\psi_j^\nu,
\eeq
with
\beq
\gamma_{\mu \nu \alpha}=\frac{1}{4}\{[\gamma_\mu, \gamma_\nu],\gamma_\alpha\},\quad
\gamma_{\mu \nu}=\frac{1}{2}[\gamma_\mu, \gamma_\nu].  
\eeq
This form of the Lagrangian leads to a fairly compact and convenient expression for
the free propagator of the delta field 
\beq
S^{\mu\nu} = \frac{p\!\!/ + m_\Delta}{p^2-m^2_\Delta}\left( -g^{\mu\nu}
+\frac{1}{3} \gamma^\mu\gamma^\nu + \frac{1}{3m_\Delta}\left(\gamma^\mu p^\nu
- \gamma^\nu p^\mu\right) +   \frac{2}{3m_\Delta^2} p^\mu p^\nu \right)~.
\eeq 
For every interaction in the Lagrangian, one generally has 
a freedom to introduce an off-shell parameter. As a consequence, 
interaction terms depend, in addition to the point-transformation
parameter $A$, also on the off-shell parameters  $z_i$  via the tensor 
\beq
\tilde O_{z_i}^{\mu \lambda} O_{A \lambda}^{\,\,\,\,\,\,\,\,\,\nu}=g^{\mu
  \nu} + \left[z_i + \frac{1}{2} (1+4
  z_i)A\right]\gamma^{\mu}\gamma^{\nu} \,.
\eeq
All terms proportional to the off-shell parameters are
redundant \cite{Tang:1996sq,Pascalutsa:2000kd,Krebs:2009bf} meaning
that their contributions to observables can be absorbed into
a redefinition of the corresponding
low-energy constants (LECs). A particular choice of the off-shell
parameters in the calculations is, therefore, a matter of convention. 
For example, in the covariant calculation of Ref.~\cite{Bernard:2012hb} we have set 
$A^{{\rm Relativistic}}=-1$ and all $z_i^{{\rm Relativistic}}=0$. In
the present analysis,  we employ the heavy-baryon
$1/m$-expansion worked out by Hemmert et al.~\cite{Hemmert:1997ye},  where the choice 
 $A^{{\rm HB}}=0$ without specifying a particular value for the off-shell parameter
$z_0^{{\rm HB}}$ of the
leading-order pion-nucleon-delta coupling has been made. 
In order to be consistent with the convention used in the  
covariant calculation of Ref.~\cite{Bernard:2012hb},  we have to set
\beq
z_0^{{\rm HB}} + \frac{1}{2}\left(1+4 z_0^{{\rm HB}}\right)A^{{\rm
    HB}} = z_0^{{\rm Relativistic}} + \frac{1}{2}\left(1+4 z_0^{{\rm Relativistic}}\right)A^{{\rm
   Relativistic}} \quad\Longrightarrow\quad z_0^{{\rm HB}}=-\frac{1}{2}.
\eeq 
This choice will be used throughout this work.

The effective heavy-baryon Lagrangians which contribute to the nuclear forces up to 
N$^3$LO  are given by 
\beqa
\label{lagr_all}
{\cal L}_{{\rm SSE}}={\cal L}_{\pi\pi}^{(2)} +{\cal L}_{\pi\pi}^{(4)}+ {\cal L}_{\pi N}^{(1)} + {\cal L}_{\pi N}^{(2)} +
{\cal L}_{\pi N}^{(3)} +{\cal L}_{\pi N \Delta}^{(1)} + {\cal L}_{\pi
  N \Delta}^{(2)} +
{\cal L}_{\pi N \Delta}^{(3)} + {\cal L}_{\pi \Delta \Delta}^{(1)}  +
{\cal L}_{\pi \Delta \Delta}^{(2)} + \delta {\cal L}_{ \pi N}^{(2)},
\eeqa
where the subscripts refer to the small-scale dimension. 
Notice that the last term denotes the contribution to the pion-nucleon effective
Lagrangian induced by the non-propagating spin-1/2 components of the
Rarita-Schwinger field  for the delta. 
The relevant
terms in the pion Lagrangians have the form \cite{Gasser:1983yg}
\beqa
{\cal L}_{\pi\pi}^{(2)} &=& \frac{1}{2}\left(\partial_\mu \mathring{\fet
\pi} \cdot \partial^\mu \mathring{\fet \pi} - M^2 \mathring{\fet \pi} \cdot \mathring{\fet \pi} \right) +\frac{M^2}{8 F^2}(8 \alpha -1)(\mathring{\fet
\pi}\cdot \mathring{\fet \pi})^2 + \frac{1}{2
  F^2}(1-4\alpha)(\mathring{\fet\pi}\cdot\partial_\mu \mathring{\fet\pi})
(\mathring{\fet\pi}\cdot\partial^\mu \mathring{\fet\pi}) \nn
&-&\frac{\alpha}{F^2}\mathring{\fet\pi}\cdot \mathring{\fet\pi}\,\partial_\mu \mathring{\fet\pi}\cdot\partial^\mu \mathring{\fet\pi}
,\\
{\cal L}_{\pi\pi}^{(4)} &=&-\frac{l_3}{F^2}M^4 \mathring{\fet\pi}\cdot \mathring{\fet\pi} +
\frac{l_4}{F^2}M^2\left(\partial_\mu \mathring{\fet\pi}\cdot\partial^\mu \mathring{\fet\pi}
-M^2 \mathring{\fet\pi}\cdot \mathring{\fet\pi}\right) ,
\eeqa
where $\mathring{\fet
\pi}$, $M$ and $F$ refer to the pion fields in the chiral limit, the pion mass to leading
order in quark masses and  the pion decay coupling in the chiral
limit, while $l_i$ are further LECs.  Here and in what follows, 
$\mathring{X}$ indicates that the quantity $X$ is taken in the chiral
limit. Further, the parameter $\alpha$ reflects the freedom in the
choice of a particular parametrization for the pion field. All
physical quantities are, of course,  independent of this parameter.   
We do not give here explicitly ${\cal L}_{\pi N}^{(1)}, {\cal L}_{\pi N}^{(2)}, {\cal L}_{\pi
  N}^{(3)}$ as all relevant terms are listed in Ref.~\cite{Krebs:2012yv} where, in
order to be consistent with our notation, the LECs $c_i$ and $d_i$
should be replaced by $\mathring{c}_i$ and $\mathring{d}_i$.  
The remaining Lagrangians in Eq.~(\ref{lagr_all}) are given by \cite{Hemmert:1997ye}
\beqa
{\cal L}_{\pi N \Delta}^{(1)}&=&-\frac{\mathring{h}_A}{F} \mathring{N}_v^\dagger \mathring{T}_\mu^{i}\partial^\mu \mathring{\pi}^i 
+ h.c., \\
{\cal L}_{\pi N \Delta}^{(2)}&=&  \frac{i}{F}(b_3 + b_6)
\mathring{N}_v^\dagger \mathring{T}_\mu^{i}\partial^\mu \partial\cdot v\,\mathring{\pi}^i  + i
\frac{\mathring{h}_A}{F m} \mathring{N}_v^\dagger\partial^\mu \mathring{T}_\mu^i \partial\cdot v\,\pi^i
+ h.c.,\nn
{\cal L}_{\pi N \Delta}^{(3)}&=&\frac{2}{F}(2
\,h_7-h_8-2\,h_9-2\,h_{10})M^2 \mathring{N}_v^\dagger \mathring{T}_\mu^i \partial^\mu \mathring{\pi}^i -
\frac{1}{F}(h_{12} + h_{13}) \mathring{N}_v^\dagger \mathring{T}_\mu^i (\partial\cdot
v)^2\partial^\mu \mathring{\pi}^i + h.c., \nn
{\cal L}_{\pi\Delta \Delta}^{(1)}&=&-\mathring{T}_\mu^{i \dagger}
\left(i\,\partial\cdot v - \mathring{\Delta}-\frac{\mathring{g}_1}{F}\fet\tau\cdot
  (\partial_\alpha \mathring{\fet \pi}) S^\alpha \right) \mathring{T}_\nu^j g^{\mu\nu}\delta_{i
    j},\nn
{\cal L}_{\pi \Delta \Delta}^{(2)}&=& - 4 M^2 c_1^\Delta 
\mathring{T}_\mu^{i\dagger} \mathring{T}_\nu^j g^{\mu\nu}\delta_{i j}
+ \frac{1}{2 m} \mathring{T}_\mu^{i\dagger}\left(\partial^2-(\partial\cdot v)^2\right)\mathring{T}_\nu^j g^{\mu\nu}\delta_{i
    j},\nn
\delta {\cal L}_{\pi N}^{(2)}&=&\frac{\mathring{h}_A^2}{18 F^2 m} \mathring{N}_v^\dagger \left(
i \fet\tau\cdot(\partial_\mu \mathring{\fet\pi}\times\partial_\nu \mathring{\fet\pi}) -
2\,\partial_\mu \mathring{\fet\pi} \cdot \partial_\nu \mathring{\fet\pi}\right)\nn
&\times&\left(4\left(1+8\,
z_0^{{\rm HB}} + 12\, (z_0^{{\rm HB}})^2\right)\,S^\mu
S^\nu + \left(5-8\, z_0^{{\rm HB}} - 4\, (z_0^{{\rm HB}})^2\right)\, v^\mu v^\nu\right) \mathring{N}_v,
\eeqa
where $N_v$ and $T_\mu^i$ denote the large components of the nucleon and
delta field, respectively, $v$ is the four-velocity and $m$ is the
nucleon mass in the chiral limit.  For the
sake of compactness, we do not show the velocity index explicitly in the
case of the delta fields $T_\mu^i$. The quantity  $h_A$ denotes the $\pi N
\Delta$ axial coupling, $b_i$, $h_i$ and $c_i^\Delta$ are further LECs
and the covariant spin
operator  is defined via
\beq
S_\mu = \frac{1}{2} i \gamma_5 \sigma_{\mu \nu} v^\nu \,, \quad \quad
\sigma_{\mu \nu} = \frac{i}{2} [ \gamma_\mu , \; \gamma_\nu]\,. 
\eeq

Last but not least, we emphasize that we adopt in the present work the convention for the pion-nucleon LECs which
maintains an explicit decoupling of the delta. To be specific, the
results for a given amplitude  or  nuclear potential $\cal M$ have the
form  
\beq
{\cal M}  = {\cal M}_{\slashed\Delta} + {\cal M}_{\Delta},
\eeq
where  ${\cal M}_{\Delta}$ denotes the contribution associated with
the delta degrees of freedom while $ {\cal M}_{\slashed\Delta} $ is
the purely nucleonic part. As guaranteed by the decoupling theorem
\cite{Appelquist:1974tg}, all effects of the delta isobar at low
energy can be accounted for in an implicit way, i.e.~through its
contributions to the effective pion-nucleon Lagrangian. Expanding the
delta contribution  ${\cal M}_{\Delta}$ around $\Delta \to \infty$,
one generally finds terms with both positive and negative powers of
the $\Delta$.   While the latter ones can be identified with the
$\Delta$-resonance saturation of the pion-nucleon LECs, see section
\ref{sec:piN} for more details, terms with
positive powers of the delta-nucleon mass splitting can, as a matter
of convention, be eliminated by an appropriate redefinition of  the
pion-nucleon LECs. This is the convention we adopt in our analysis. 
It guarantees, that no positive powers of the $\Delta$ appear in
the finite expressions for all physical quantities and the
$\Delta(1232)$ contributions 
decouple (vanish) in the large-$\Delta$ limit. 
Stated differently, this convention ensures that
our results actually correspond to a partial
resummation of the $\Delta$-resonance contributions to the pion-nucleon
LECs within the $\Delta$-less formulation. 

\section{Renormalization of the effective Lagrangian to leading
  loop order}
\label{sec:ren}

We now discuss in detail renormalization of the lowest-order effective chiral Lagrangian
at the one-loop level which is achieved by expressing all quantities in terms 
of renormalized parameters rather than their chiral limit values. 
We do not consider here renormalization in the pionic sector as it is 
extensively discussed in the literature and concentrate entirely on
the nucleon and delta sectors. We begin with introducing the
renormalized fields and coupling constants via the relations 
\beqa
\label{zfactors}
\mathring{N}_v&=& \sqrt{Z_N} N_v, \quad
\mathring{T}_\mu^i\,=\,\sqrt{Z_\Delta}T_\mu^i, \quad
\mathring{\pi}^i\,=\,\sqrt{Z_\pi}\pi^i,\nn
M&=&M_\pi + \delta M, \quad m\,=\,m_N + \delta m, \quad
\mathring{\Delta}\,=\,\Delta+\delta\Delta, \nn
Z_\pi &=& 1+\delta Z_\pi, \quad Z_N\,=\,1+\delta Z_N, \quad Z_\Delta \,=\, 1+ \delta Z_\Delta, 
\nn
 F&=&F_\pi + \delta F,
\quad \mathring{g}_A\,=\,g_A + \delta g_A, \quad \mathring{h}_A\,=\,h_A +
\delta h_A, \quad \mathring{g}_1\,=\,g_1+\delta g_1.
\eeqa
and determine the shifts $\delta X$ with $X\in\{M, F, m, \Delta,  Z_N,
Z_\Delta, g_A, h_A, g_1\}$  order by order in the small scale
expansion. Notice that in this formulation, the heavy baryon expansion
corresponds to a $1/m_N$-expansion, where $m_N$ is now the physical nucleon
mass and ${\it not}$ the nucleon mass in the chiral limit, see \cite{Epelbaum:2002gb}
for more details. We further emphasize that $\mathring{T}_\mu^i$ does
not correspond to an interpolating field of an asymptotic state so that its renormalization prescription is
conventional. Even if $Z_\Delta$ is a complex number, the replacement
$\mathring{T}_\mu^i\,=\,\sqrt{Z_\Delta}T_\mu^i$ in the Lagrangian does
not lead to a violation of unitarity in the kinematical region we are
interested in simply because there are no ${\it external}$ delta lines. 
Indeed, the complex renormalization factor $Z_\Delta$, which shows up in
the delta propagator, is compensated by vertices to which this delta
propagator is attached so the amplitude does not depend on
$Z_\Delta$. This argument does not rely on whether the renormalization factor
is a real or complex number. The main motivation for us to make the replacement
$\mathring{T}_\mu^i\,=\,\sqrt{Z_\Delta}T_\mu^i$ is to ensure that 
we can treat delta fields in the same manner as stable
particles. This procedure is a matter of convention and does not affect
the final result for the amplitudes. In the following, we discuss in
detail renormalization of the various quantities in
Eq.~(\ref{zfactors}). 

\begin{itemize}
\item \emph{Nucleon mass and field renormalization} \\
To study nucleon-mass
  and field-renormalization up to order $\epsilon^3$ one needs to
  calculate the self-energy diagrams shown in 
  Fig.~\ref{fig:nucleon_self_energy}.
\begin{figure}[t]
\includegraphics[width=10.0cm,keepaspectratio,angle=0,clip]{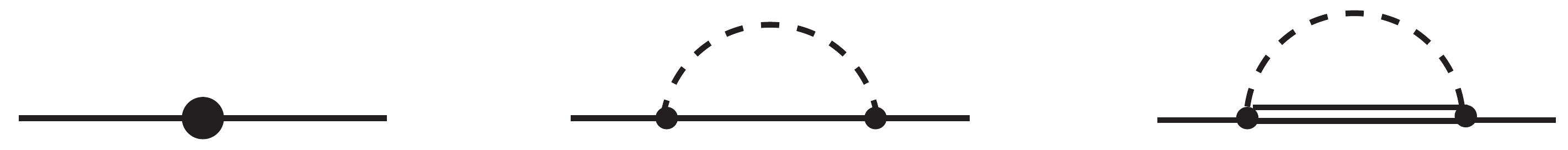}
\caption{Feynman diagrams which contribute to the nucleon self-energy
  up to order $\epsilon^3$.   Only nonvanishing diagrams are shown.
  Solid, dashed and double lines represent nucleons, pions and the
  delta, respectively.  Solid dots (filled circles) denote
  leading-order (subleading and higher-order) vertices from the
  effective Lagrangian.    
 }\label{fig:nucleon_self_energy} 
\end{figure}
The full nucleon propagator in the rest-frame of the nucleon can be
parametrized via
\beqa
D_N(p\cdot v)&=&\frac{1}{p\cdot v -\Sigma_N(p\cdot v)+i \epsilon},
\eeqa
where $\Sigma_N(p\cdot v)$ denotes the nucleon self-energy. 
In the vicinity of $p\cdot v=0$,  the propagator of the renormalized physical nucleon
fields has a simpler
form 
\beqa
D_N(p\cdot v)&=&\frac{1}{p\cdot v + i \epsilon} + {\cal O}((p\cdot
v)^0).
\eeqa
Making the Taylor expansion 
\beqa
p\cdot v - \Sigma_N(p\cdot v)=-\Sigma_N(0)+(1-\Sigma_N^\prime(0))
p\cdot v + {\cal O}((p\cdot v)^2)\,,
\eeqa
we obtain renormalization conditions for the  nucleon mass and the 
$Z$-factor $Z_N$:
\beqa
\Sigma_N(0)=0 \quad {\rm and}\quad \Sigma_N^\prime(0)=0.\label{nucleon_mass_renorm}
\eeqa
The contribution of the first diagram in
Fig.~\ref{fig:nucleon_self_energy}  to the nucleon self-energy is
given by
\beqa
\Sigma_N^{{\rm tree}}(p\cdot v)&=&\delta m-4 c_1 M_\pi^2 - \delta
Z_N p\cdot v.
\eeqa
The contribution of the nucleonic one-loop diagram, see second graph
in  Fig.~\ref{fig:nucleon_self_energy}, to the self-energy at the
order we are working is given by 
\beqa
\Sigma_N^{{\rm loop}, \pi N}(p\cdot v)&=&\frac{3 g_A^2}{4 F_\pi^2}
p\cdot v I(d:0)+\frac{3 g_A^2}{4 F_\pi^2}(M_\pi^2 - (p\cdot v)^2) I(d:
0; (p,0))\,,
\eeqa
while the delta-loop contribution emerging from the last diagram in
Fig.~\ref{fig:nucleon_self_energy}  is given by
\beqa
\Sigma_N^{{\rm loop}, \pi \Delta}(p\cdot v)&=&-\frac{2 (d-2)
  h_A^2}{(d-1)F_\pi^2}(\Delta - p\cdot v) I(d:0) \nn
&+& \frac{2(d-2)
  h_A^2}{(d-1) F_\pi^2}(M_\pi^2-\Delta^2+2\Delta p\cdot v - (p\cdot v)^2)I(d:0;(p,\Delta))\,. 
\eeqa
Here, scalar master integrals in $d$ dimensions are defined according to:
\beqa
I(d:p_1,\dots,p_n)&=&\frac{1}{i}\mu^{4-d}\int \frac{d^d
  l}{(2\pi)^d}\frac{1}{(l+p_1)^2-M_\pi^2+i
  \epsilon}\dots\frac{1}{(l+p_n)^2-M_\pi^2+i \epsilon}, \\
I(d:p_1,\dots,p_n;(p,\delta))&=&\mu^{4-d}\frac{1}{i}\int \frac{d^d
  l}{(2\pi)^d}\frac{1}{(l+p_1)^2-M_\pi^2+i
  \epsilon}\dots\frac{1}{(l+p_n)^2-M_\pi^2+i \epsilon}\frac{1}{(l+p)\cdot
  v-\delta+i \epsilon}.\nonumber
\eeqa
Using the renormalization conditions in Eq.~(\ref{nucleon_mass_renorm}) we
obtain the following expressions at order $\epsilon^3$ in four dimensions:
\beqa
\delta m &=& 4 c_1 M_\pi^2 + \frac{3 g_A^2 M_\pi^3}{32 \pi F_\pi^2} +
\frac{h_A^2 \Delta}{36 \pi^2 F_\pi^2}(2\Delta^2 - 3 M_\pi^2)+\frac{8
  h_A^2 \Delta}{3 F_\pi^2}(2\Delta^2-3 M_\pi^2)\lambda_\pi + \frac{4
  h_A^2}{3 F_\pi^2}(\Delta^2 - M_\pi^2)\bar{J}_0(-\Delta), \nn
\delta Z_N&=&-\frac{3 g_A^2 M_\pi^2}{32 \pi^2 F_\pi^2}+ \frac{h_A^2}{4 \pi^2
  F_\pi^2}(2\Delta^2 - M_\pi^2) + \frac{1}{2 F_\pi^2}\left(16
h_A^2(2\Delta^2 - M_\pi^2) -9 g_A^2 M_\pi^2\right)\lambda_\pi +
\frac{4 h_A^2 \Delta}{F_\pi^2}\bar{J}_0(-\Delta)\,,
\eeqa
where the quantities $\lambda_\pi$ and $\bar J_0$ are defined in
appendix \ref{appendix:piN_amplitude}.
\item 
\emph{Renormalization of the nucleon axial coupling} \\
To renormalize the axial-vector coupling constant $\mathring{g}_A$, we consider the axial-vector form
factor of the nucleon as shown in Fig.~\ref{fig:g_A}. 
\begin{figure}[t]
\includegraphics[width=5.0cm,keepaspectratio,angle=0,clip]{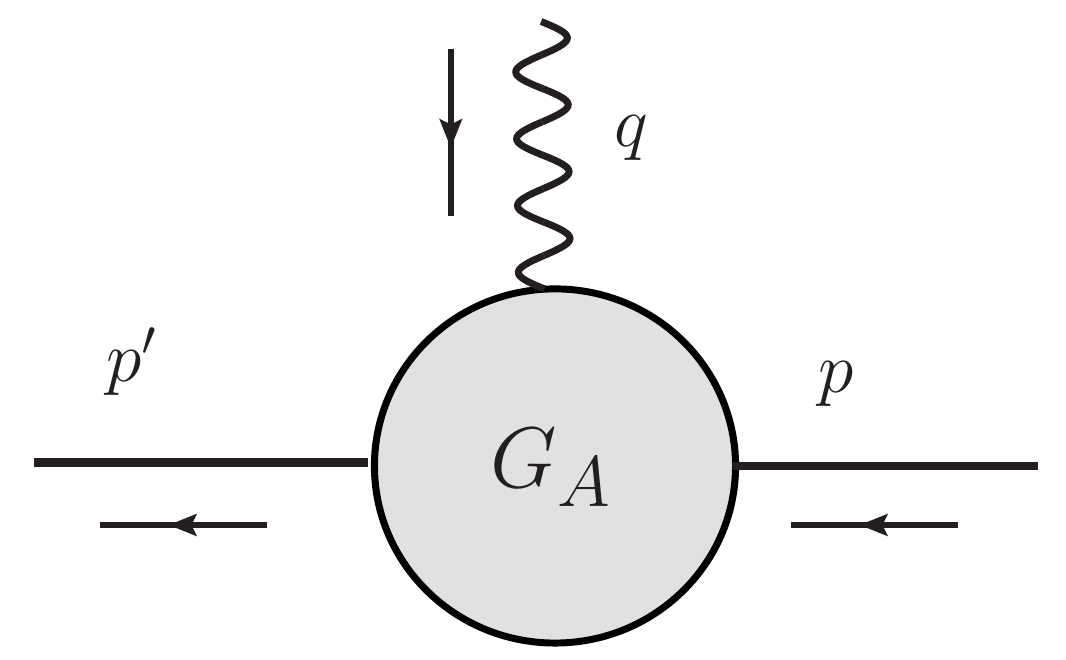}
\hskip 1.5 true cm
\includegraphics[width=10.0cm,keepaspectratio,angle=0,clip]{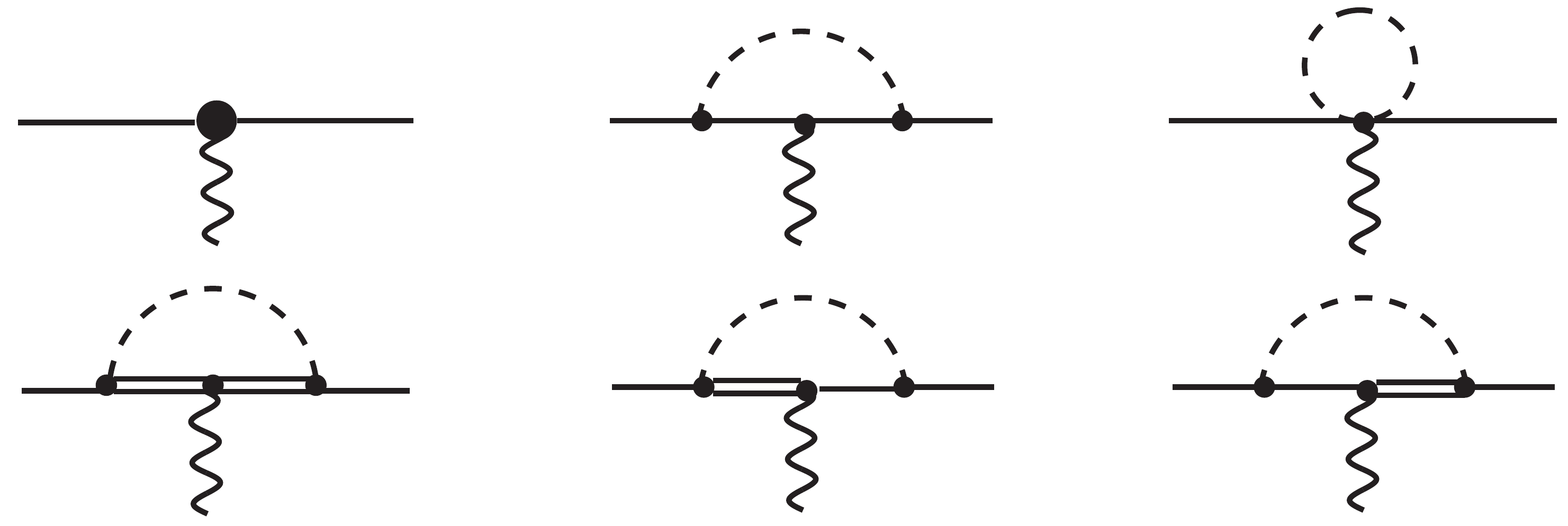}
\caption{Left panel: generic one-particle irreducible contribution to
  the axial-vector nucleon form factor.
Right panel: Nonvanishing Feynman diagrams which contribute to $G_A(0)$ up
  to order $\epsilon^3$.   Wavy lines represent external axial
  sources.  For the remaining notation see Fig.~\ref{fig:nucleon_self_energy}.          
 }\label{fig:g_A} 
\end{figure}
In the Breit frame ($q_0$=0), the matrix
element can be parametrized via~\cite{Bernard:1998gv}
\beqa
{\cal M}(p^\prime, p,
q)&=&-\tau^j\vec{\epsilon}_{A}^{\,j}\cdot\vec{\sigma} \frac{E}{2 m_N}G_A(q^2)+\dots,\label{GA_definition}
\eeqa
where the ellipses refer to terms which are of no relevance for
renormalization of $\mathring{g}_A$. 
In the above expression, $E$  denotes the energy of the incoming
nucleon (which in the Breit-frame is also equal to the energy of
the outgoing nucleon) while $\vec{\epsilon}_A^{\,j}$ is the polarization
vector of the $j$-th component of an isotriplet external axial
field.
The  physical value of the nucleon
axial coupling  $g_A$ is defined as 
\beqa
g_A=G_A(0).\label{gA_renorm_condition}
\eeqa
Up to order $\epsilon^3$, the contributions to the axial form factor
$G_A(0)$ emerge from the tree-level diagrams
\beqa
G_A^{{\rm tree}} (0)&=&g_A+\delta g_A+g_A \delta Z_N + 4 d_{16} M_\pi^2\,,
\eeqa
one-loop diagrams without delta excitations 
\beqa
G_A^{{\rm loop}, \pi N} (0)&=&-\frac{g_A}{4 F_\pi^2}(g_A^2(d-3)-4) I(d: 0)
\eeqa
and one-loop diagrams with intermediate delta excitations,  see
the second raw of the right panel of Fig.~\ref{fig:g_A},
\beqa
G_A^{{\rm loop}, \pi N \Delta}  (0)&=&-\frac{2 (d-2) h_A^2}{9(d-1)^2
  F_\pi^2}(24 g_A + 5 (d^2-2d-3) g_1) I(d:0) - \frac{16 (d-2) g_A
  h_A^2 M_\pi^2}{3 (d-1)^2 F_\pi^2 \Delta} I(d:0;(0,0))\nn
&+&\frac{2 (d-2) h_A^2}{9 (d-1)^2 F_\pi^2 \Delta}(24 g_A (M_\pi^2 -
\Delta^2)-5(d^2 - 2 d - 3) g_1 \Delta^2)I(d:0;(0,\Delta)).
\eeqa
Using the renormalization condition in Eq.~(\ref{gA_renorm_condition})
we obtain in four dimensions
\beqa
\delta g_A&=&-\delta Z_N g_A - 4 d_{16} M_\pi^2 - \frac{M_\pi^2}{7776
  \pi^2 F_\pi^2}(243 g_A^3 - 576 g_A h_A^2 + 1240 h_A^2
g_1)-\frac{h_A^2 \Delta^2}{486 \pi^2 F_\pi^2}(24 g_A - 155 g_1)\nn
 &-&
\frac{4 g_A h_A^2 M_\pi^3}{27 \pi F_\pi^2 \Delta}+\left(\frac{16
    h_A^2\Delta^2}{81 F_\pi^2}(24 g_A+25 g_1) - \frac{M_\pi^2}{162
    F_\pi^2}(81 g_A^3 + 36 g_A (32 h_A^2 - 9) + 400 h_A^2
  g_1)\right)\lambda_\pi \nn
&+&\left(\frac{4 h_A^2 \Delta}{81 F_\pi^2}(24
  g_A + 25 g_1) - \frac{32 g_A h_A^2 M_\pi^2}{27 F_\pi^2 \Delta}\right)\bar{J}_0(-\Delta).
\eeqa
\item 
\emph{Delta mass and field renormalization}\\ 
To study the delta mass and field renormalization one needs to
calculate the corresponding nonvanishing self-energy diagrams shown in
Fig.~\ref{fig:delta_self_energy}.
\begin{figure}[t]
\includegraphics[width=10.0cm,keepaspectratio,angle=0,clip]{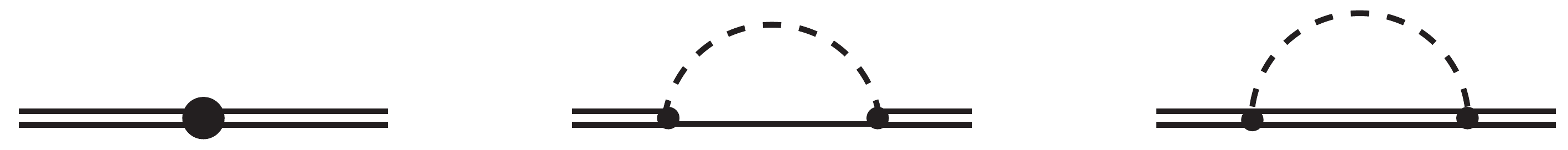}
\caption{Nonvanishing diagrams which contribute to delta self-energy
  up to order $\epsilon^3$.        
For the remaining notation see Fig.~\ref{fig:nucleon_self_energy}.          
 }\label{fig:delta_self_energy} 
\end{figure}
In general, the  self-energy of the $\Delta$-resonance in the rest-frame
can be parametrized via
\beqa
\Sigma_\Delta(p\cdot v)_{\mu\nu}^{i
  j}&=&P_{\mu\nu}^{3/2}\xi_{I=3/2}^{i j}\Sigma_\Delta(p\cdot v),
\eeqa
where the spin- and isospin-$3/2$ projector operators are defined by
\beqa
P_{\mu\nu}^{3/2}&=&g_{\mu\nu}-v_\mu v_\nu + \frac{4}{3} S_\mu
S_\nu\quad {\rm and}\quad \xi_{I=3/2}^{i j}\,=\,\delta^{i j} -
\frac{1}{3} \tau^{i} \tau^{j},
\eeqa
respectively.
The contribution of the tree-level diagram,  see the first graph in
Fig~\ref{fig:delta_self_energy}, to the  delta self-energy is
given by
\beqa
\Sigma_\Delta^{{\rm tree}}(p\cdot v)&=&-4 c_1^{\Delta} M^2.
\eeqa
The contributions of the two one-loop diagrams with the $\pi N$ and $\pi \Delta$ cuts
in $d$ space-time dimensions have the form 
\beqa
\Sigma_\Delta^{{\rm loop},\pi N}(p\cdot
v)&=&\frac{h_A^2}{(d-1)F_\pi^2}\left[p\cdot v\,I(d:0)+(M_\pi^2-(p\cdot
  v)^2)I(d:0;(p,0))\right]\,,\nn
\Sigma_\Delta^{{\rm loop},\pi \Delta}(p\cdot
v)&=&\frac{5 (d^2-2 d -3) g_1^2}{12 (d-1)^2 F_\pi^2}\left[(p\cdot v-\Delta
  ) I(d:0)+(M_\pi^2-(p\cdot v - \Delta)^2) I(d:0;(p,\Delta))\right]\,.
\eeqa
The full $\Delta$-propagator in the rest-frame of the $\Delta$-resonance
can be written as 
\beqa
D_\Delta(p\cdot v)_{\mu \nu}^{i j} = -D_\Delta(p\cdot
v)P_{\mu\nu}^{3/2}\xi_{I=3/2}^{i j}, \quad \mbox{with} \quad
D_\Delta(p\cdot v)=\frac{i}{p\cdot v
  -\mathring{\Delta}-\Sigma_\Delta(p\cdot v)}.
\eeqa 
In the vicinity of the pole, the full delta-propagator has a simpler
structure, namely
\beqa
D_\Delta(p\cdot v )\simeq \frac{i}{p\cdot v -\Delta+i \Gamma_\Delta / 2}.
\eeqa
Here, $\Delta$ and $\Gamma_\Delta$ denote the (pole-position)
mass and width of the delta
resonance, respectively. Expanding the full propagator around the
pole one extracts the mass, width and the complex $Z_\Delta$-factor:
\beqa
p\cdot v -\mathring{\Delta}-\Sigma_\Delta(p\cdot
v)&=&\Delta-i\frac{\Gamma_\Delta}{2}-\mathring{\Delta}-\Sigma_\Delta\left(\Delta-i\frac{\Gamma_\Delta}{2}\right)
+ \left(1 -
\Sigma_\Delta^{\prime}\left(\Delta-i\frac{\Gamma_\Delta}{2}\right)
\right)(p\cdot v - \Delta + i \frac{\Gamma_\Delta}{2}) \nn
&+& {\cal O}\left(\left(p\cdot v - \Delta + i \frac{\Gamma_\Delta}{2}\right)^2\right).
\eeqa
Renormalization of the delta mass and width is determined from the condition
\beqa
\Delta-i\frac{\Gamma_\Delta}{2}-\mathring{\Delta}-\Sigma_\Delta\left(\Delta-i\frac{\Gamma_\Delta}{2}\right)&=&0\,.
\eeqa
From the real part of this condition we deduce the delta-mass renormalization as
\beqa
\Delta - \mathring{\Delta} - {\rm Re}\,
\Sigma_\Delta\left(\Delta-i\frac{\Gamma_\Delta}{2}\right) &=&
0\label{delta_mass_ren} \,,
\eeqa
while the imaginary part of this condition yields the following
result for the width:
\beqa
\label{rendeltawidth}
\frac{\Gamma_\Delta}{2}+{\rm Im}\, \Sigma_\Delta\left(\Delta-i\frac{\Gamma_\Delta}{2}\right) &=& 0.\label{delta_width_determ}
\eeqa
The complex-valued $Z_\Delta$-factor is determined by the relation
\beqa
\label{rendeltaZ}
\Sigma_\Delta^{\prime}\left(\Delta-i\frac{\Gamma_\Delta}{2}\right)&=&0.\label{delta_z_factor_determ}
\eeqa
At the one-loop level we can replace the relations (\ref{delta_mass_ren}),  (\ref{rendeltawidth})
and  (\ref{rendeltaZ}) by  
\beqa
\Delta - \mathring{\Delta} - \, {\rm Re}\,
\Sigma_\Delta(\Delta)  &=& 0\,, \quad 
\frac{\Gamma_\Delta}{2}+ \, {\rm Im}\,
\Sigma_\Delta(\Delta)\;=\;
0\quad{\rm and}\quad \, \Sigma_\Delta^{\prime}(\Delta)\,=\, 0,\label{BW_cond_2}
\eeqa
One immediately sees that the above relations coincide with the Breit-Wigner
conditions. The pole conditions in Eqs.~(\ref{delta_mass_ren}),
(\ref{delta_width_determ}) and
(\ref{delta_z_factor_determ}) and the Breit-Wigner conditions start to
differ from each other at the two-loop level which is beyond the order
we are working at.  
From the conditions in Eq.~(\ref{BW_cond_2}) we finally obtain
\beqa
\delta\Delta &=& 4 c_1^{\Delta} M_\pi^2 + \frac{2 h_A^2 \Delta}{3
  F_\pi^2}\left(3 M_\pi^2 -
  2\Delta^2\right)\lambda_\pi 
+\frac{25 g_1^2 M_\pi^3}{864 \pi F_\pi^2} + \frac{h_A^2 \Delta (2 \Delta^2 -
  3 M_\pi^2)}{72\pi^2 F_\pi^2}
- \frac{h_A^2}{3 F_\pi^2}(M_\pi^2-\Delta^2) {\rm Re}
\bar J_0(\Delta),\nn
\delta Z_\Delta &=& -
\frac{1}{18 F_\pi^2}\left( 36 h_A^2 (M_\pi^2 - 2\Delta^2)+25 g_1^2 M_\pi^2\right)\lambda_\pi- \frac{65 g_1^2 M_\pi^2}{864
\pi^2 F_\pi^2}- \frac{h_A^2 \Delta}{F_\pi^2} \bar{J}_0(\Delta).
\eeqa
\item 
\emph{Renormalization of  the $\pi N \Delta $ axial coupling} \\
To renormalize the LEC $\mathring{h}_A$ we consider the axial-vector nucleon-delta transition form
factor, see Fig.~\ref{fig:h_A}, in the rest-frame of the delta:
\begin{figure}[t]
\includegraphics[width=5.0cm,keepaspectratio,angle=0,clip]{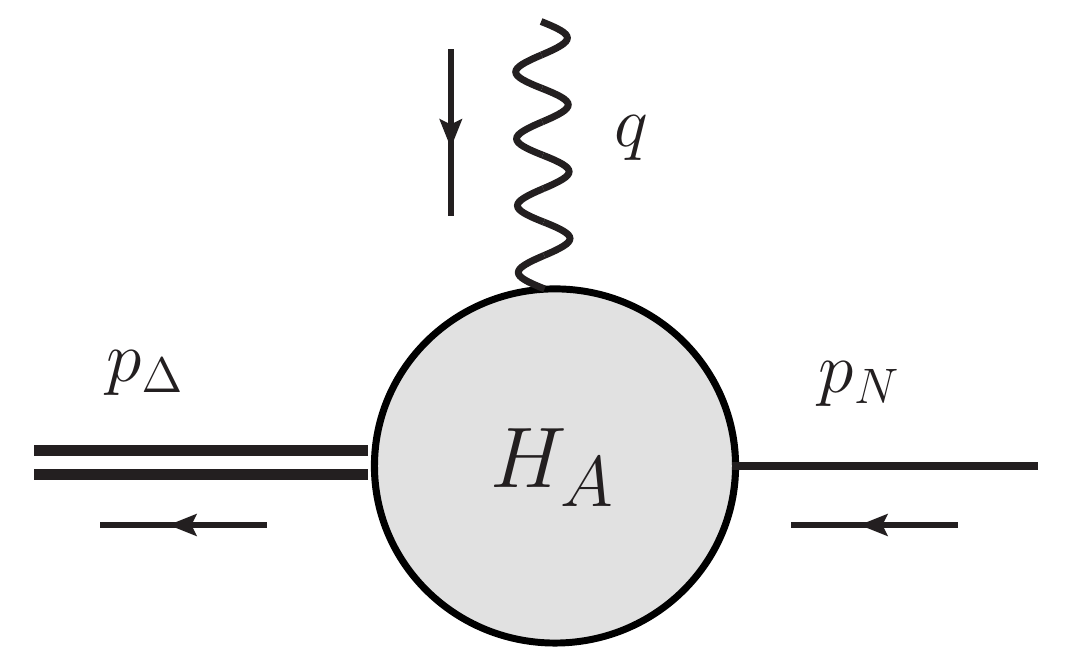}
\hskip 1.5 true cm
\includegraphics[width=10.0cm,keepaspectratio,angle=0,clip]{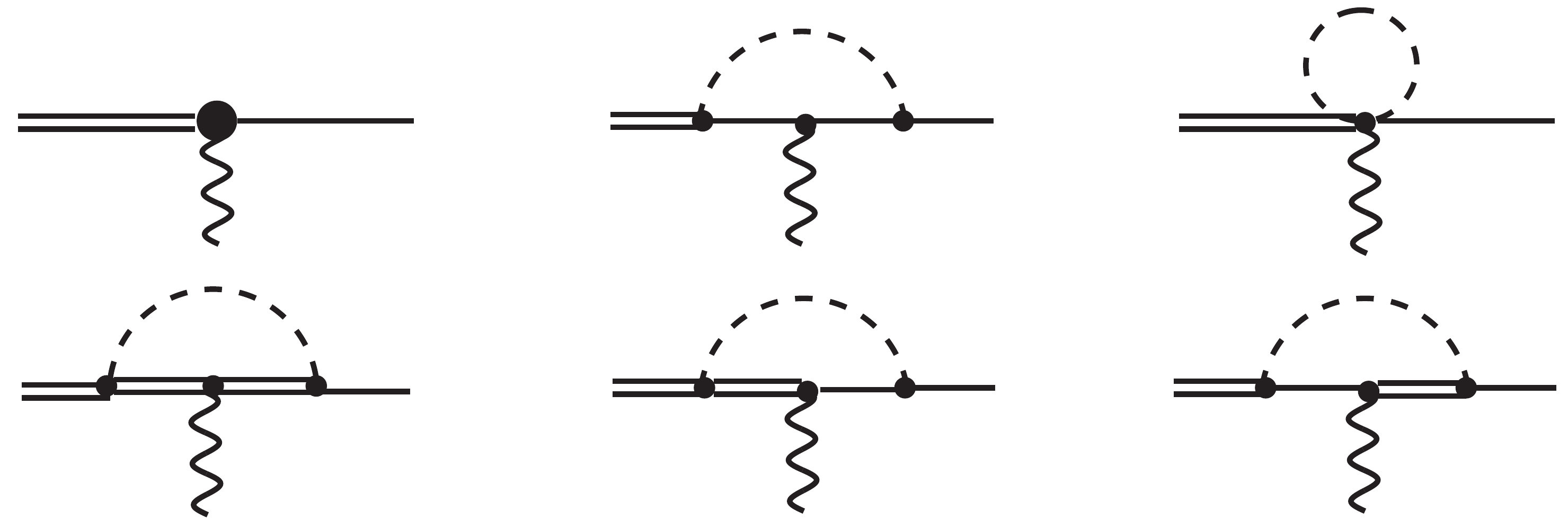}
\caption{Left panel: generic one-particle irreducible contribution to
  the axial-vector nucleon-delta
  transition form factor.
Right panel: Nonvanishing Feynman diagrams which contribute to
 $H_A(\Delta,\Delta^2,0)$ up to order $\epsilon^3$.        
Wavy lines represent external axial
  sources.  For the remaining notation see Fig.~\ref{fig:nucleon_self_energy}.          
 }\label{fig:h_A} 
\end{figure}
\beqa
{\cal M}(p_\Delta, p_N, q)_\mu^i&=&P_{\mu\nu}^{3/2}\xi_{I=3/2}^{i
  j}\epsilon_A(q)_j^\nu \,H_A(p_\Delta\cdot v, q^2, p\cdot v)+\dots,
\eeqa
where the ellipses refer to other terms which are not
relevant for renormalization of the $\pi N \Delta $ axial coupling constant. 
We analytically continue the form factor $H_A$ and choose the renormalization point to be 
\beqa
h_A={\rm Re}\,H_A\left (\Delta - i \frac{\Gamma_\Delta}{2},\;
  \left(\Delta - i \frac{\Gamma_\Delta}{2}\right)^2, \;0 \right)\,,
\eeqa
which,  in the one-loop approximation, becomes
\beqa
 h_A={\rm Re}\,H_A \left(\Delta, \; \Delta^2, \; 0
 \right).\label{hA_renorm_condition} \,.
\eeqa
Up to order $\epsilon^3$, the quantity $H_A(\Delta, \Delta^2,
0)$ receives contributions from the tree-level diagram (first
graph in the right panel of Fig.~\ref{fig:h_A})  
\beqa
H_A^{{\rm tree}}  (\Delta, \Delta^2,0)&=&h_A + \delta h_A -(b_2+b_7)\Delta +
\frac{h_A}{2}(\delta Z_N + \delta Z_\Delta) + 2 (h_8+2(h_9 + h_{10}))M_\pi^2,
\eeqa
one-loop diagrams without delta excitations (the remaining two
diagrams in the upper raw of  the right panel of Fig.~\ref{fig:h_A})   
\beqa
H_A^{{\rm loop}, \pi
  N} (\Delta, \Delta^2,0)&=&\frac{h_A(d-1-g_A^2)}{(d-1) F_\pi^2} I(d: 0) + \frac{g_A^2 h_A
    M_\pi^2}{(d-1) F_\pi^2 \Delta} I(d:0;(0,0)) \nn
&-& \frac{g_A^2 h_A
    (M_\pi^2 
- \Delta^2)}{(d-1) F_\pi^2 \Delta} I(d: 0; (\Delta, 0))
\eeqa
and one-loop graphs with pions, nucleons and delta degrees of
freedom (diagrams in the second raw of  the right panel of Fig.~\ref{fig:h_A}) 
\beqa
H_A^{{\rm loop}, \pi
  N \Delta} (\Delta, \Delta^2,0)&=&-\frac{(d-3) h_A}{36 (d-1)^3 F_\pi^2}(12 (d-1) h_A^2 +
5(d+1) g_1(3(d-1)^2g_A + 4 g_1))I(d:0) \nn
&-& \frac{5(d^2-2d-3)h_A g_1^2 M_\pi^2}{9(d-1)^3 F_\pi^2\Delta}I(d:0;(0,0)) - \frac{(d-3)h_A^3
  (M_\pi^2-\Delta^2)}{6(d-1)^2 F_\pi^2 \Delta} I(d: 0; (\Delta,0))
\nn
&+&\frac{(d-3)h_A (M_\pi^2 - \Delta^2)}{18(d-1)^3 F_\pi^2
  \Delta}(10(d+1)g_1^2+3(d-1)h_A^2)I(d: 0;(0,\Delta)).
\eeqa
Substituting these expressions into the renormalization condition
given in Eq.~(\ref{hA_renorm_condition}), we obtain the following
order-$\epsilon^3$ expression for $\delta h_A$ in four dimensions: 
\beqa
\delta h_A&=&-\frac{h_A}{2}\left( \delta Z_N+{\rm Re}\,\delta
  Z_\Delta\right)+\Delta \left(b_2 + b_7\right)-2\left( h_8 + 2 (h_9 +
  h_{10})\right) M_\pi^2+\left(3 h_A^2+5 g_1^2 -27
  g_A^2\right)\frac{h_A \Delta^2}{972\pi^2 F_\pi^2}\nn
&-&\frac{h_A M_\pi^2}{2592 \pi^2 F_\pi^2}\left(12 h_A^2-108 g_A^2 + 20
  g_1^2 + 195 g_A g_1\right) + \left(81 g_A^2 - 25
  g_1^2\right)\frac{h_A M_\pi^3}{1944 \pi F_\pi^2 \Delta}\nn
&+&\left[h_A(81 g_A^2 + 9 h_A^2 + 25 g_1^2)\frac{4 \Delta^2}{243 F_\pi^2}
-h_A(100 g_1^2 + 225 g_A g_1 + 36 (h_A^2 + 9 (g_A^2 -
1)))\frac{M_\pi^2}{162 F_\pi^2}\right]\lambda_\pi\nn
&-&\left(9 h_A^2 + 50 g_1^2\right)\frac{h_A (M_\pi^2-\Delta^2)}{486
  F_\pi^2 \Delta} \bar{J}_0(-\Delta) + \left(h_A^2 + 18 g_A^2
 \right)\frac{ h_A (M_\pi^2-\Delta^2)}{54 F_\pi^2 \Delta} {\rm
  Re}\, \bar{J}_0(\Delta).
\eeqa
\end{itemize}

\section{Determination of the LECs from $\pi N$ scattering}
\label{sec:piN}

Given that the LECs in the effective Lagrangian with and without
explicit delta degrees of freedom have a different meaning, we 
cannot use the values of the various LECs from our earlier work
\cite{Krebs:2012yv}  based
on the $\Delta$-less formulation and have to redo the analysis of the
pion-nucleon system utilizing the small-scale expansion.  
Specifically, we need to calculate  the $\pi N$ scattering amplitude up to order $\epsilon^3$.

Before discussing renormalization of the $\pi N$ amplitude in the
explicit decoupling scheme as explained in section \ref{sec:lagr}, we
first perform the following shifts in the LECs in order to get rid of redundant
terms:
\beqa
h_A&\rightarrow&h_A-\Delta(b_2+b_3+b_6 + b_7) + \Delta^2(h_{12}
+ h_{13}) + 4 M_\pi^2 h_7,\nn
\mathring{c}_2&\rightarrow&\mathring{c}_2 + \frac{4(d-2)}{3(d-1)}h_A (b_3+b_6)-\frac{2(d-2)}{3(d-1)}\Delta
(b_3+b_6)^2-\frac{4(d-2)}{3(d-1)} \Delta h_A (h_{12}+h_{13}),\nn
\mathring{c}_3&\rightarrow&\mathring{c}_3 - \frac{4(d-2)}{3(d-1)}h_A (b_3+b_6) + \frac{2(d-2)}{3(d-1)}\Delta
(b_3+b_6)^2+\frac{4(d-2)}{3(d-1)} \Delta h_A(h_{12}+h_{13}),\nn
\mathring{c}_4&\rightarrow&\mathring{c}_4 + \frac{4}{3(d-1)}h_A (b_3+b_6)-\frac{2}{3(d-1)}\Delta
(b_3+b_6)^2 - \frac{4}{3(d-1)} \Delta h_A (h_{12}+h_{13}),\nn
\mathring{d}_1 + \mathring{d}_2&\rightarrow&\mathring{d}_1 + \mathring{d}_2+\frac{d-2}{6(d-1)}(b_3+b_6)^2- \frac{d-2}{3(d-1)} h_A (h_{12} + h_{13}), \nn
\mathring{d}_3&\rightarrow&\mathring{d}_3-\frac{d-2}{6(d-1)}(b_3+b_6)^2+ \frac{d-2}{3(d-1)} h_A (h_{12} + h_{13}),\nn
\mathring{d}_{14}-\mathring{d}_{15}&\rightarrow&\mathring{d}_{14}-\mathring{d}_{15}-\frac{2}{3(d-1)}(b_3+b_6)^2+ \frac{4}{3(d-1)} h_A (h_{12} + h_{13}).\label{LECs_shift_eq}
\eeqa
Notice that these replacements are performed in the amplitude written
in $d$ dimensions. After this shift the amplitude does not depend on
the LECs $b_3 + b_6, b_2+b_7, h_{12} + h_{13}$ and $h_7$ anymore.

Let us now discuss renormalization of the pion-nucleon amplitude. All
divergencies which remain after expressing the amplitude in terms of
physical quantities as discussed in the previous section are absorbed
into redefinition of the LECs $c_i$ and $d_i$ entering the order-$Q^2$
and $Q^3$ effective pion-nucleon Lagrangians. While the LECs $c_i$ are
finite in the $\Delta$-less framework provided one uses  dimensional
regularization with the $\overline{MS}$-scheme, this does not hold true
anymore in the $\Delta$-full theory due to the appearance of
ultraviolet divergencies $\propto \Delta$ at $\epsilon^3$ and higher
powers of $\Delta$  at orders beyond $\epsilon^3$.  We parametrize the bare LECs $\mathring{c}_i$
and $\mathring{d}_i$  via
\beq
\mathring{c}_i=c_i+\Delta
\left(\frac{\beta_i^c}{F_\pi^2}\lambda_\pi+\frac{1}{(4\pi F_\pi)^2}c_i^\Delta\right),
\quad \quad \mathring{d}_i =
\frac{\beta_i^{d,N}+\beta_i^{d,\Delta}}{F_\pi^2}\lambda_\pi +
d_i+\frac{1}{(4\pi F_\pi)^2} d_i^\Delta, 
\eeq
where the various $\beta$-functions relevant for pion-nucleon
scattering are given by
\beqa
\beta_1^c&=&2\,h_A^2, \nn
\beta_2^c&=&-\frac{80}{2187}
h_A^2\left(9\,g_A - 5\, g_1\right)^2,\nn
\beta_3^c&=&\frac{16}{2187} h_A^2\left(729 + 5\left(9\,g_A - 5\,
    g_1\right)^2\right),\nn
\beta_4^c&=&-\frac{2}{2187} h_A^2\left(972 + 2349 \,g_A^2 + 1152 \,h_A^2 -
  2250\, g_A\, g_1 + 125 \,g_1^2\right),\nn
\beta_1^{d,N}&=&-\frac{1}{6}g_A^4,\nn
\beta_2^{d,N}&=&-\frac{1}{12}-\frac{5}{12}g_A^2, \nn
\beta_3^{d,N}&=&\frac{1}{2}+\frac{1}{6}g_A^4, \nn
\beta_5^{d,N}&=&\frac{1}{24}+\frac{5}{24}g_A^2,\nn
 \beta_{14}^{d,N}&=&\frac{1}{3}g_A^4,\nn
 \beta_{15}^{d,N}&=&\beta_{18}^{d,N}=0,\nn
\beta_1^{d,\Delta} + \beta_2^{d,\Delta}
+\beta_3^{d,\Delta}&=&\frac{10}{27}h_A^2, \nn
\beta_3^{d,\Delta}&=&\frac{h_A^2}{2187}(125 \,g_1^2+288 \,h_A^2 - 243\, g_A^2
- 450 \,g_A g_1),\nn 
\beta_5^{d, \Delta}&=&-\frac{5}{27}h_A^2,
\nn
\beta_{18}^{d,\Delta}&=&0,\nn
\beta_{14}^{d,\Delta} - \beta_{15}^{d,
  \Delta}&=&\frac{2\,h_A^2}{2187}\left(288\,h_A^2 - 243\,g_A^2 - 450
  \,g_A\,g_1 + 125\,g_1^2\right).
\eeqa
This particular form of the $\beta$-functions guarantees that the
amplitude remains finite in the $d \to 4$ limit. We use here the
notation, in which the divergencies associated with loop
diagrams without delta excitations (with delta excitations) are
cancelled by terms $\propto \beta_i^{d, N}$ ($\propto \beta_i^{c, \Delta}$
and $\propto \beta_i^{d, \Delta}$).  Furthermore, in order to maintain
the explicit decoupling scheme,
we have introduced additional finite dimensionless shifts $\bar{c}_i^\Delta$
and $\bar{d}_i^\Delta$.  The explicit decoupling scheme is defined by the
requirement that all observables calculated in the SSE include 
only nucleonic
contributions after taking the $\Delta\to \infty$ limit. In this limit
all contributions emerging from the intermediate delta excitations
have to vanish 
(in the explicit decoupling scheme) so that the delta isobar explicitly decouples from the
theory. In order to satisfy the explicit decoupling, the
values of the LECs $\bar{c}_i^\Delta$ and $\bar{d}_i^\Delta$ have
to be chosen as 
\beqa
c_1^\Delta&=&2 h_A^2\log\left(\frac{2 \Delta}{\mu}\right), \nn
c_2^\Delta&=&-\frac{2\, h_A^2}{6561}\left(6399 \,g_A^2 - 8910 \,g_A
 \, g_1 + 3575\, g_1^2\right)-\frac{80\, h_A^2}{2187}\left(9\, g_A -
 5\, g_1\right)^2\log\left(\frac{2 \Delta}{\mu}\right),\nn
c_3^\Delta&=&\frac{2\, h_A^2}{6561}\left(6399 \,g_A^2 - 8910 \,g_A
 \, g_1 + 3575\, g_1^2\right)+\frac{16\, h_A^2}{2187}\left(729+5\,\left(9\, g_A -
 5\, g_1\right)^2\right)\log\left(\frac{2 \Delta}{\mu}\right),\nn
c_4^\Delta&=&\frac{h_A^2}{6561}\left(4860-35559 \,g_A^2 + 1728\,
  h_A^2+28350\,g_A
 \, g_1 - 4775\, g_1^2\right)\nn
&-&\frac{2\, h_A^2}{2187}\left(972+2349
 \,g_A^2 + 1152\, h_A^2-2250\, g_A\,g_1 +
 125\, g_1^2\right)\log\left(\frac{2 \Delta}{\mu}\right),
\eeqa
and 
\beqa
d_1^\Delta+d_2^\Delta&=&-\frac{h_A^2}{6561}\left(-2106 +
  5103\, g_A^2 + 216\, h_A^2 - 3870\, g_A\, g_1 + 925\, g_1^2\right)
\nn
&+&
\frac{h_A^2}{2187}\left(810 + 243\, g_A^2-288\, h_A^2 + 450\, g_A\,
  g_1 - 125\, g_1^2\right) \log\left(\frac{2 \Delta}{\mu}\right),\nn
d_3^\Delta&=&\frac{h_A^2}{6561}\left(5103\, g_A^2 + 216\, h_A^2
  - 3870\, g_A \, g_1 + 925\, g_1^2\right)\nn
&+&\frac{h_A^2}{2187}\left(-243 \,g_A^2 + 288\,h_A^2 - 450 \, g_A\,
  g_1 + 125\, g_1^2\right) \log\left(\frac{2 \Delta}{\mu}\right),\nn
d_5^\Delta &=&-\frac{13\,h_A^2}{81} - \frac{5\,h_A^2}{27}
\log\left(\frac{2 \Delta}{\mu}\right),\nn
d_{14}^\Delta-d_{15}^\Delta&=&\frac{h_A^2}{6561}\left(5589\,
g_A^2 + 432\,h_A^2 - 9090\,g_A\, g_1 +
3425\,g_1^2\right)\nn
&+&\frac{2\,h_A^2}{2187}\left(-243\,g_A^2 + 288\,h_A^2
- 450\,g_A\,g_1 + 125\,g_1^2\right) \log\left(\frac{2 \Delta}{\mu}\right),
\eeqa
respectively. Clearly, the above expressions are unique modulo terms
that vanish in the $\Delta \to \infty$ limit.  On top of the
  explicit decoupling scheme, we put a constraint on negative powers of
  $\Delta$. Specifically, we require that the $1/\Delta$-expansion of the pion-nucleon amplitude
 is consistent with the resonance saturation. This means that the
 $1/\Delta$-expansion of the $\Delta$-full pion-nucleon amplitude should
 be equal to the $\Delta$-less amplitude with the LECs $c_i$ and $\bar{d}_i$
being replaced by Eqs.~(\ref{res_c}) and (\ref{res_d}), respectively. In order to
achieve this also for relativistic corrections, we have to perform
additional shifts of $c_i$ and $d_i$-LECs, namely
\beqa
c_2&\to & c_2+\frac{8 h_A^2}{9 m_N} ,\nn
\bar{d}_1+\bar{d}_2 &\to & \bar{d}_1+\bar{d}_2 + \frac{h_A^2}{18 m_N
  \Delta}, \nn 
\bar{d}_3 &\to & \bar{d}_3 - \frac{2 h_A^2}{9 m_N \Delta},\nn
\bar{d}_5 &\to &\bar{d}_5+\frac{h_A^2}{12 m_N \Delta}, \nonumber\\ 
\bar{d}_{14}-\bar{d}_{15} &\to & \bar{d}_{14}-\bar{d}_{15} -\frac{2
  h_A^2}{9 m_N \Delta}.
\eeqa

We now turn to pion-nucleon scattering. In the center-of-mass system (cms), the amplitude for the reaction 
$\pi^a (q_1) + N(p_1) \to \pi^b(q_2) + N (p_2)$ with $p_{1,2}$ and
$q_{1,2}$ being the corresponding four-momenta and $a,b$ referring to
the pion isospin quantum numbers, takes the form:
\beq
T_{\pi N}^{ba} = \frac{E + m}{ 2 m }   \bigg( \delta^{ba} \Big[ g^+
(\omega, t) + i \vec \sigma \cdot \vec q_2 \times \vec q_1 \, h^+
(\omega, t) \Big]
+ i \epsilon^{bac} \tau^c \Big[ g^-
(\omega, t) + i \vec \sigma \cdot \vec q_2 \times \vec q_1 \, h^-
(\omega, t) \Big] \bigg)\,.
\eeq 
Here, $\omega = q_1^0 = q_2^0$ is the pion cms energy, 
$E_1 = E_2 \equiv E = ( \vec q \, ^2 + m^2 )^{1/2}$ the nucleon energy and
$\vec q_1 \, ^2 = \vec q_2 \, ^2 \equiv \vec q \, ^2 = ((s - M_\pi^2  -
m^2)^2  - 4 m^2 M_\pi^2)/(4s)$. 
Further, $t = (q_1  - q_2)^2$  is the invariant momentum transfer
squared while $s$ denotes the total  cms energy squared. The
quantities $g^\pm (\omega ,t)$ ($h^\pm (\omega ,t)$) refer to the isoscalar and isovector
non-spin-flip (spin-flip) amplitudes and  can be calculated  
in chiral perturbation theory. The contributions to the amplitudes
which do not involve intermediate delta excitations up to order $Q^4$
(i.e.~subleading one-loop order) are given in Ref.~\cite{Krebs:2012yv}. 
In Appendix \ref{appendix:piN_amplitude}, we give explicitly the 
delta-isobar contributions up to order $\epsilon^3$.  
In a complete analogy to  the $\Delta$-less calculation reported in
Ref.~\cite{Krebs:2012yv}, the phase shifts are obtained  from the
partial-wave amplitudes 
in the isospin basis $f^{I}_{l\pm}(s)$ by means of the following unitarization prescription
 \begin{eqnarray}
&& \delta^{I}_{l\pm}(s)=\arctan\left(|\vec{q} \, |\,\Re \,f^{I}_{l\pm}(s)\right)\,.
\end{eqnarray}
Determination of the LECs is carried out using exactly the same
procedure as in our $\Delta$-less calculations
\cite{Krebs:2012yv}. While the $\pi N$
scattering amplitude is worked out here only to order
$\epsilon^3$, we decided to include also the order-$Q^4$ terms
obtained within the $\Delta$-less theory when fitting the phase shifts in
order to facilitate a direct comparison with the results of
Ref.~\cite{Krebs:2012yv}. This way we make sure that the differences
between the values of the LECs obtained in the two analyses are solely due to the explicit treatment of
the delta degrees of freedom.  The impact of the $Q^4$
terms on the 3NF will be discussed  in section \ref{sec:results}. 

For the pion-nucleon contributions to the scattering amplitude we
proceed in exactly the same way as in Ref.~\cite{Krebs:2012yv}. We remind the
reader that certain LECs $\bar{e}_i$ from $\mathcal{L}_{\pi N}^{(4)}$ enter the amplitude only in
linear combinations with the LECs $c_i$ and, therefore, cannot be
determined from $\pi N$ scattering data.    Following
Refs.~\cite{Fettes:2000xg} and \cite{Krebs:2012yv}, these
$\bar{e}_i$-contributions are  absorbed into redefinition of the $c_i$'s by setting 
\beq
\label{ei_conv}
e_{22}-4 e_{38}-\frac{l_3 c_1}{F_{\pi}^2} =0, \quad \quad
e_{20}+e_{35} =0 , \quad \quad
2 e_{19}- e_{22}- e_{36}+2\frac{l_3 c_1}{F_{\pi}^2} =0, \quad \quad
2 e_{21}- e_{37} =0\,,
\eeq
without loss of generality. The LEC $ d_{18}$ from
$\mathcal{L}_{\pi N}^{(4)}$ can be fixed by means of the Goldberber-Treiman discrepancy
\begin{eqnarray}
g_{\pi N N}=\frac{g_{A}\, m_N}{F_{\pi}}\left(1-\frac{2M_{\pi}^{2}\,d_{18}}{g_{A}}\right)\,,
\end{eqnarray}
where for $g_{\pi N N}$ we adopt the value from
Ref.~\cite{Timmermans:1990tz} of $g^2_{\pi N N}/(4\pi )\simeq 13.54$,
which also agrees with the determination in Ref.~\cite{Baru:2010xn} 
based on the Goldberger-Miyazawa-Oehme sum rule and utilizing the 
most accurate available data on the pion-nucleon scattering lengths. 
We set $\bar{d}_{18}=0$ and use the effective, larger value for $g_A$ of
\beq
\label{ga_conv}
g_A = \frac{F_{\pi}g_{\pi N N}}{m_N} \simeq 1.285 
\eeq
in all expressions. This is a legitimate procedure at the order we are
working. 
This leaves us with 13 independent (linear combinations of the) low
energy constants in the nucleonic contributions to the
scattering amplitude which have to be fixed from a fit to the data, 
namely $c_{1,2,3,4}$, $\bar d_1 + \bar d_2$, $\bar d_3$, $\bar d_5$,
$\bar d_{14} - \bar d_{15}$ and $\bar e_{14,15,16,17,18}$, see
Ref.~\cite{Krebs:2012yv}  for more details and explicit expressions. 
We also use the same values for the pion mass
and decay constant as in that reference, namely $M_\pi = 138.03$ MeV
and $F_\pi = 92.4$ MeV.  

The contributions to the amplitude associated with the delta
excitations and given in appendix  \ref{appendix:piN_amplitude} involve
further LECs, namely $h_A$ and $g_1$.  For the $\pi N \Delta$ axial
vector constant we adopt the value of  $h_A=1.34$ which is fixed from
the  width of the delta-resonance and also agrees well  with the
large-$N_c$ prediction \cite{Dashen:1993jt,Dashen:1993as}. Notice that similarly to the
convention adopted for $g_A$, the $\pi N
\Delta$  Goldberger-Treiman discrepancy is implicitly taken into
account by using the above value of the LEC $h_A$\footnote{The results for the 3NF are expected to be much
  less sensitive to the precise value of $h_A$ than to the value of
  $g_A$. This is because the changes in $h_A$ can, to some extent, 
  be compensated by the changes in the LECs from the subleading and
  higher-order effective Lagrangian.}.
Our fits to $\pi N$ data turn out to be fairly insensitive
to a particular value of $g_1$. For this reason we decided to fix it to its
large-$N_c$  value $g_1=9/5\,g_A\approx2.31$ \cite{Dashen:1993jt,Dashen:1993as,Siemens:2016jwj}. 
We, therefore, have to finally fix exactly the same combinations of the low-energy
constants as in the $\Delta$-less theory. 

\begin{figure}[tb]
\vskip 1 true cm
\includegraphics[width=15.0cm,keepaspectratio,angle=0,clip]{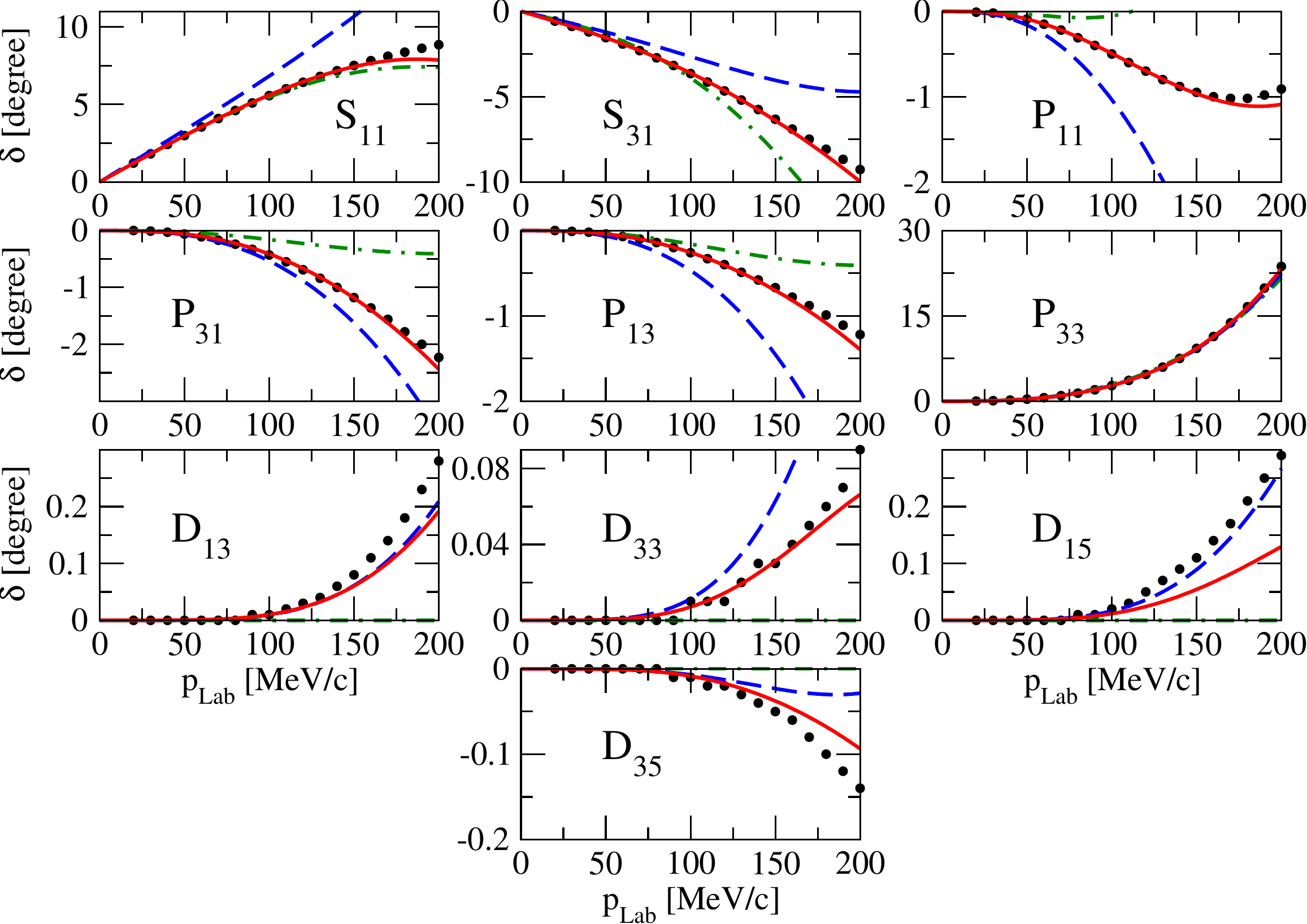}
\caption{Results of the fit for $\pi N$ $s$, $p$ and $d$-wave phase shifts using the GW partial wave analysis  of Ref.~\cite{Arndt:2006bf}.
The solid curves correspond to the $\epsilon^3+Q^4$ results, the dashed curves
to the order-$\epsilon^3$ results, and the dashed-dotted curves to the order-$\epsilon^2$ calculation.         
 }\label{fig:SAID} 
\end{figure}

\begin{figure}[tb]
\vskip 1.5 true cm
\includegraphics[width=15.0cm,keepaspectratio,angle=0,clip]{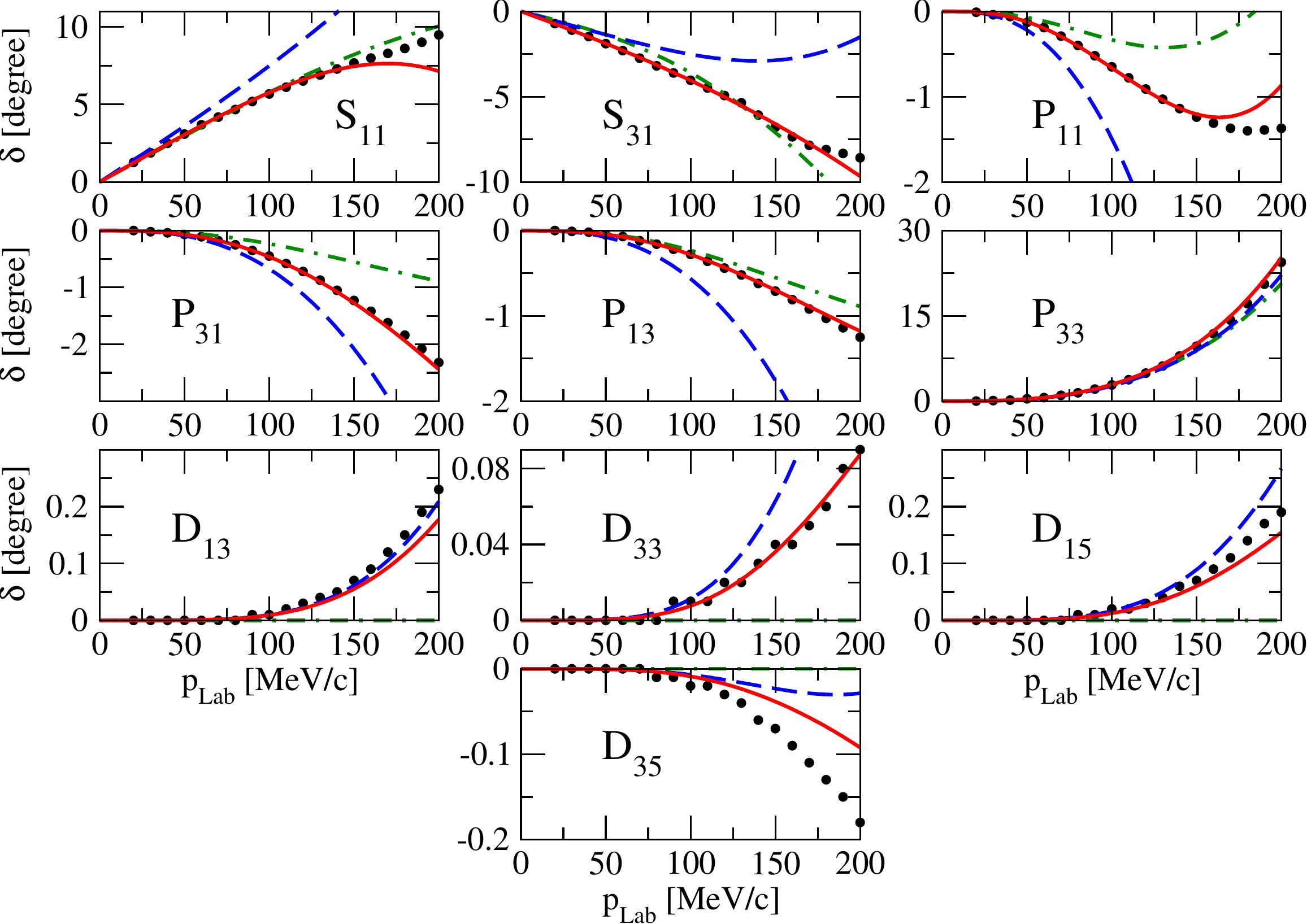}
\caption{Results of the fit for $\pi N$ $s$, $p$ and $d$-wave phase shifts using the KH partial wave analysis  of Ref.~\cite{Koch:1985bn}.
The solid curves correspond to the full $\epsilon^3+Q^4$ results, the dashed curves
to the order-$\epsilon^3$ results, and the dashed-dotted curves to the order-$\epsilon^2$ calculation.         
 }\label{fig:KA84}
\end{figure}

As in Ref.~\cite{Krebs:2012yv} we performed a combined fit for all $s$-, $p$-, and $d$-waves.
We remind the reader that it is crucial to include $d$-waves in the fit as they impose severe constraints
on some of the $e_i$ constants, especially on $\bar e_{14}$ and $\bar e_{17}$,
which also enter the  N$^4$LO expressions for the three-body force. 
The results of the fits using the partial wave analysis (PWA) by the
George-Washington University group (GW) \cite{Arndt:2006bf} and the
Karlsruhe-Helsinki group (KH) \cite{Koch:1985bn} 
are presented in Figs.~\ref{fig:SAID} and \ref{fig:KA84},
respectively. 
In these figures we show the full, i.e.~order-$\epsilon^3+Q^4$ results (solid curves) as well 
as the phase shifts calculated up to order $\epsilon^3$ without $Q^4$ terms (dashed curves) and $\epsilon^2$ (dashed-dotted curves) 
using the same parameters (from the order-$\epsilon^3+Q^4$ fit) in all curves. 
We fit the data points from threshold up to $p_{Lab}=150$ MeV/c, and obtain a description of the phase shifts
similar to the $\Delta$-less case. Naturally, the description of the $P_{33}$ partial wave (delta $s$-channel) is significantly improved.

Notice that more sophisticated studies of pion-nucleon scattering employing a covariant formulation of baryon
chiral effective field theory with and without explicit $\Delta$(1232)
degrees of freedom have
been carried out recently, see Refs.~\cite{Alarcon:2011kh,Siemens:2016hdi,Yao:2016vbz,Siemens:2016jwj,Siemens:2017opr}.  
Also, more reliable ways to extract the low-energy constants from the $\pi N$ reaction
and to estimate their uncertainties have been explored as compared to
the ones employed in our analysis.
Those include, in particular, analytic extrapolations of the
scattering amplitude into the subthreshold region using the
solutions to the Roy-Steiner equation and a 
direct determination of the LECs from the available $\pi N$ scattering
data in the physical region  instead of using partial-wave analyses, 
see Refs.~\cite{Hoferichter:2015hva,Siemens:2016jwj,Yao:2016vbz,Siemens:2016hdi,Wendt:2014lja,Carlsson:2015vda}.
Future studies of nuclear forces and few-nucleon systems should,
obviously, employ
the most reliable available values of the  $\pi N$ LECs such as e.g.~the ones
from
Refs.~\cite{Hoferichter:2015hva,Siemens:2016jwj,Siemens:2017opr}. In
this paper we, however, focus mainly on the 
$\Delta$ contributions to the 3NF. To facilitate a comparison between
the  $\Delta$-full and  $\Delta$-less calculation of
Refs.~\cite{Krebs:2012yv,Krebs:2013kha} and to allow for an unambiguous
interpretation of our results in terms of resonance saturation, we
follow here the same
procedure for the determination of various LECs as adopted  
in Ref.~\cite{Krebs:2012yv}.

We finally turn to the discussion of the extracted parameters. The obtained values of the 
low energy constants are collected in Table~\ref{table:parameters}.
\begin{table}[b]
\begin{tabular*}{\textwidth}{@{\extracolsep{\fill}}lrrrrrrrrrrrrr}
\hline 
\hline
\noalign{\smallskip} 
 & $c_{1}$ & $c_{2}$ & $c_{3}$ & $c_{4}$ & $\bar{d}_{1}+\bar{d}_{2}$ & $\bar{d}_{3}$ & $\bar{d}_{5}$ & $\bar{d}_{14}-\bar{d}_{15}$ & $\bar{e}_{14}$ & $\bar{e}_{15}$ & $\bar{e}_{16}$ & $\bar{e}_{17}$ & $\bar{e}_{18}$\\[3pt]
\hline \\
&&&&&&&&&&&&& \\ [-22pt]
fit to the GW PWA \cite{Arndt:2006bf}&$-1.32$&$0.39$&$-2.68$&$1.86$&$1.46$&$-1.01$&$-0.10$&$-2.16$&$0.06$&$-2.47$&$-0.05$&$-0.56$&$0.54$\\ [2pt]
statistical
  error&$0.45$&$1.34$&$0.16$&$0.07$&$0.17$&$0.31$&$0.19$&$0.33$&$0.03$&$0.07$&$0.80$&$0.38$&$4.66$\\ [2pt]
\hline  \\
&&&&&&&&&&&&& \\ [-22pt]
fit to the KH PWA
                                                                                                       \cite{Koch:1985bn}&$-0.85$&$0.45$&$-1.91$&$1.49$&$2.07$&$-2.45$&$0.66$&$-3.86$&$-0.12$&$-7.05$&$3.39$&$-0.38$&$2.85$
\\ [2pt]
statistical error&$0.50$&$1.47$&$0.18$&$0.10$&$0.19$&$0.33$&$0.20$&$0.36$&$0.03$&$0.08$&$0.90$&$0.48$&$5.12$\\ [2pt]
\hline \hline 
\end{tabular*}
\caption{Low-energy constants 
obtained from a fit to the empirical $s$, $p$- and $d$-wave pion-nucleon phase shifts using partial wave analysis  of Ref.~\cite{Arndt:2006bf}
and of  Ref.~\cite{Koch:1985bn}.
Values of the LECs are given in GeV$^{-1}$, GeV$^{-2}$ and  GeV$^{-3}$
for the $c_i$, $\bar{d}_i$ and $\bar{e}_i$, respectively.
}
\label{table:parameters}
\end{table}
We also looked at the statistical errors of the fitted parameters in order to see qualitatively which low-energy constants
(or their linear combinations) are well constrained by the data and which of them are poorly determined.
Similarly to the strategy utilized in Ref.~\cite{Fettes:1998ud} we
assigned the same relative error to each data point from the partial
wave analyses equal to $5\%$. 
This ansatz is somewhat arbitrary but seems reasonable  for an estimate of the relative uncertainties of different low-energy constants.
Notice further that the statistical errors are calculated in the linearized approximation, 
i.e.~the covariance matrix is taken to be the inverse of the Hessian
matrix of the 
$\chi^2$ function at its minimum. Such an approximation is sufficient
for a qualitative analysis that we are going to perform.
The resulting statistical uncertainties for all low energy constants
are listed in Table~\ref{table:parameters} and appear to be almost the
same for both the KH and GW analyses. Moreover, 
they change very little when the fit is performed in the $\Delta$-less case as in ref.~\cite{Krebs:2012yv}.
One can see that the low energy constants $c_{2}$, $\bar e_{15}$ and
$\bar e_{16}$ have the largest errors ($0.8-5.1$) in the corresponding
natural units 
(GeV$^{-1}$, GeV$^{-2}$ and  GeV$^{-3}$ for the $c_i$, $\bar{d}_i$ and
$\bar{e}_i$, respectively), indicating that these parameters are not well determined 
in the fit. On the other hand, $\bar e_{17}$, $\bar e_{14}$, which are
the only LECs $\bar e_i$ contributing to the 3NF at order $Q^5$,
and the LEC $c_{4}$ are
strongly constrained by the data.

Another important quantity is a correlation between pairs of
parameters. The largest in magnitude values
for the correlation coefficients are obtained for the pairs $c_1-c_2$ ($0.99$), $c_1-\bar e_{16}$ ($-0.98$), $c_2-\bar e_{16}$ ($-0.99$),
$\big((\bar{d}_{1}+\bar{d}_{2})-(\bar{d}_{14}-\bar{d}_{15})\big)$ ($-0.97$).
In order to get a more detailed information on the correlation among
various parameters we have computed eigenvalues of the  
covariance matrix. The square roots of their numerical values in natural units are
$5.41,0.59,0.45,0.35,0.29,0.12,0.05,0.03,0.03,0.02,0.02,0.01,0.01$ for the KH analysis and 
$4.92,0.51,0.37,0.32,0.27,0.11,0.04,0.03,0.02,0.02,0.01,0.01,0.01$
for the GW analysis. One can see that the first eigenvalue is at least two orders of magnitude larger than any of the other eigenvalues.
This indicates that fixing certain linear combination of parameters
results in very slow changes in the $\chi^2$ even if the individual
values of the LECs entering this linear combination change
significantly. 
This combination is the corresponding eigenvector and is approximately equal to 
$-0.1c_1-0.3c_2-0.1\bar e_{15}+0.9\bar e_{16}$ (the other constants enter with much smaller coefficients). The coefficients are given in natural units.
We indeed observe that these four parameters are strongly correlated
and one can obtain fits comparable with the best one
with those parameters being significantly shifted.
The appearance of such a strong correlation among the parameters
reflects the fact that one cannot fully resolve the energy
dependence of the amplitude with a good accuracy in the low-energy
regime. This interpretation is confirmed by performing a fit to higher
energy, namely $p_\mathrm{Lab}=200$~MeV/c,
see Table~\ref{table:parameters200MeV}. In this case, both the statistical errors and the correlations (including the ones among
$c_1$, $c_2$, $\bar e_{15}$, $\bar e_{16}$) do become significantly
smaller. Unfortunately, the purely perturbative approach cannot be
expected to be applicable at such energies
as the phase shifts become quite large.

\begin{table}[t]
\begin{tabular*}{\textwidth}{@{\extracolsep{\fill}}lrrrrrrrrrrrrr}
\hline 
\hline
\noalign{\smallskip} 
 & $c_{1}$ & $c_{2}$ & $c_{3}$ & $c_{4}$ & $\bar{d}_{1}+\bar{d}_{2}$ & $\bar{d}_{3}$ & $\bar{d}_{5}$ & $\bar{d}_{14}-\bar{d}_{15}$ & $\bar{e}_{14}$ & $\bar{e}_{15}$ & $\bar{e}_{16}$ & $\bar{e}_{17}$ & $\bar{e}_{18}$\\[3pt]
\hline \\
&&&&&&&&&&&&& \\ [-22pt]
fit to the GW PWA
  \cite{Arndt:2006bf}&$-1.31$&$0.11$&$-2.54$&$1.85$&$1.43$&$-0.90$&$-0.16$&$-2.09$&$0.07$&$-3.44$&$1.65$&$-0.46$&$0.47$
\\ [2pt]
statistical error&$0.19$&$0.48$&$0.08$&$0.04$&$0.14$&$0.19$&$0.11$&$0.28$&$0.02$&$0.04$&$0.33$&$0.17$&$1.48$\\ [2pt]
\hline  \\
&&&&&&&&&&&&& \\ [-22pt]
fit to the KH PWA \cite{Koch:1985bn}&$-1.35$&$-0.89$&$-2.19$&$1.63$&$2.08$&$-2.13$&$0.45$&$-3.69$&$-0.05$&$-6.59$&$7.22$&$-0.35$&$1.88$
\\ [2pt]
statistical error&$0.21$&$0.51$&$0.08$&$0.05$&$0.15$&$0.20$&$0.11$&$0.29$&$0.02$&$0.04$&$0.37$&$0.22$&$1.57$\\ [2pt]
\hline \hline
\end{tabular*}
\caption{Low-energy constants 
obtained from a fit to the empirical $s$, $p$- and $d$-wave pion-nucleon phase shifts up to $p_\mathrm{Lab}=200$~MeV/c using partial wave analysis  of Ref.~\cite{Arndt:2006bf}
and of  Ref.~\cite{Koch:1985bn}.
Values of the LECs are given in GeV$^{-1}$, GeV$^{-2}$ and  GeV$^{-3}$
for the $c_i$, $\bar{d}_i$ and $\bar{e}_i$, respectively.
}
\label{table:parameters200MeV}
\end{table}
Finally, it is interesting to compare the values of the LECs with the ones
obtained in the $\Delta$-less approach. As already pointed out before,
one expects to find more natural values of the LECs in the $\Delta$-full
theory. This is indeed the case as one can see from table 
\ref{table:parametersSaturation}, where such a comparison is carried out for the
KH fits. The situation for the GW fits is similar, see table I of
\cite{Krebs:2012yv}.  
\begin{table}[b]
\begin{tabular*}{\textwidth}{@{\extracolsep{\fill}}lrrrrrrrrrrrrr}
\hline 
\hline
\noalign{\smallskip} 
 & $c_{1}$ & $c_{2}$ & $c_{3}$ & $c_{4}$ & $\bar{d}_{1}+\bar{d}_{2}$ & $\bar{d}_{3}$ & $\bar{d}_{5}$ & $\bar{d}_{14}-\bar{d}_{15}$ & $\bar{e}_{14}$ & $\bar{e}_{15}$ & $\bar{e}_{16}$ & $\bar{e}_{17}$ & $\bar{e}_{18}$\\[3pt]
\hline \\
&&&&&&&&&&&&& \\ [-22pt]
$Q^4$, KH PWA
  \cite{Koch:1985bn}&$-0.75$&$3.49$&$-4.77$&$3.34$&$6.21$&$-6.83$&$0.78$&$-12.02$&$1.52$&$-10.41$&$6.08$&$-0.37$&$3.26$
\\ [2pt]
$\epsilon^3+Q^4$,  KH PWA \cite{Koch:1985bn}&$-0.85$&$0.45$&$-1.91$&$1.49$&$2.07$&$-2.45$&$0.66$&$-3.86$&$-0.12$&$-7.05$&$3.39$&$-0.38$&$2.85$\\ [2pt]
$\Delta$-contribution&$0$&$2.81$&$-2.81$&$1.40$&$2.39$&$-2.39$&$0$&$-4.77$&$1.87$&$-4.15$&$4.15$&$-0.17$&$1.32$\\ [2pt]
\hline \hline
\end{tabular*}
\caption{Low-energy constants 
obtained from a fit to the empirical $s$, $p$- and $d$-wave
pion-nucleon phase shifts using the partial wave analysis of
Ref.~\cite{Koch:1985bn} and the corresponding delta-resonance contributions given in
Eqs.~(\ref{res_c}), (\ref{res_d}) and (\ref{e_sat}). 
Values of the LECs are given in GeV$^{-1}$, GeV$^{-2}$ and  GeV$^{-3}$ for the $c_i$, $\bar{d}_i$ and $\bar{e}_i$, respectively.}
\label{table:parametersSaturation}
\end{table}
Comparing the second and the third raws of this table, one observes a
sizable reduction in magnitude for most of the LECs when the delta is
included as an explicit degree of freedom. This raises the question
of whether these differences can be understood analytically. In the
following, we address
this question by isolating the contributions of the delta to the
various LECs.  To this aim, we make a $1/\Delta$-expansion of the
delta-resonance contributions to the $\pi N$  amplitude and
match the expanded expressions to the amplitude obtained in the
$\Delta$-less theory up to order $Q^4$. We decompose the various
renormalized LECs
into $\Delta$-less ($\Delta\!\!\!\!/$) and delta
contributions ($\Delta$) via 
\beqa
c_i&=&c_i(\Delta\!\!\!\!/)+c_i(\Delta),
\quad d_i\,=\,d_i(\Delta\!\!\!\!/)+d_i(\Delta), \quad e_i=e_i(\Delta\!\!\!\!/)+e_i(\Delta).
\eeqa
Expanding the $\epsilon^1$-result up to order
$1/\Delta$ we recover the well-known results for
the $c_i$'s \cite{Bernard:1996gq} 
\beqa
\label{res_c}
c_1(\Delta)&=&0, \quad c_2(\Delta)\,=\,\frac{4\, h_A^2}{9\,\Delta},
\quad c_3(\Delta)\,=\,-\frac{4\,h_A^2}{9\,\Delta}, \quad c_4(\Delta)\,=\,\frac{2\,h_A^2}{9\,\Delta}.
\eeqa
From the $1/\Delta^2$ terms of the order-$\epsilon^1$ $\pi N$-amplitude, we obtain the
delta contributions to the LECs  $\bar{d}_i$ given by
\beqa
\label{res_d}
d_1(\Delta) +
d_2(\Delta)&=&\frac{h_A^2}{9\,\Delta^2},\quad
d_3(\Delta)\,=\,-\frac{h_A^2}{9 \,\Delta^2}, \quad
d_{14}(\Delta)- d_{15}(\Delta)\,=\,-\frac{2\, h_A^2}{9\,\Delta^2}.
\eeqa
In principle,  one could also
expect $1/\Delta$-contributions from the order-$\epsilon^2$ $\pi
N$-amplitude. 
However, all such terms turn out to contribute to renormalization of
$h_A$ and do not lead to  resonance saturation 
of $d_i$. One observes from table \ref{table:parameters} that
the delta contributions explain at least a half of the
size of the LECs $d_1+d_2, d_{3}$ and
$d_{14}-d_{15}$, which appear to be unnaturally large in
the $\Delta$-less approach, see also Ref.~\cite{Siemens:2016jwj} for
similar conclusions.  

To explore delta-resonance saturation of the LECs $e_i$
from ${\cal L}_{\pi N}^{(4)}$ which enter  the order-$Q^4$
pion-nucleon amplitude, we need to
analyze the following terms:
\begin{itemize}
\item $1/\Delta^3$-contributions from $\epsilon^1$-amplitude
\item $1/\Delta^2$-contributions from $\epsilon^2$-amplitude (these
  terms vanish after renormalization of $h_A$),
\item $1/\Delta$-contributions from $\epsilon^3$-amplitude.
\end{itemize}
The complete contribution of the delta to these LECs is
given by a sum of these terms and has the form:
\beqa
\label{e_sat}
\bar{e}_{14}(\Delta)&=&\frac{h_A^2}{864\,F_\pi^2\,\pi^2\,\Delta}\left(7+10\,\log\left(\frac{2\,\Delta}{M_\pi}\right)\right),\nn
\bar{e}_{15}(\Delta)&=&-\frac{h_A^2}{18\,\Delta^3}-\frac{h_A^2}{839808\,F_\pi^2\,\pi^2\,\Delta}\left(3969
\,g_A^2 - 4050\,g_A\,g_1 + 1225\,g_1^2\right),\nn
\bar{e}_{16}(\Delta)&=&\frac{h_A^2}{18\,\Delta^3} +
\frac{h_A^2}{839808\,F_\pi^2\pi^2\,\Delta}\left(3969\,g_A^2 - 4050\,g_A\,g_1 +
1225\,g_1^2\right),\nn
\bar{e}_{17}(\Delta)&=&-\frac{h_A^2}{1728\,F_\pi^2\,\pi^2\,\Delta}\left(1+2
  \log\left(\frac{2\,\Delta}{M_\pi}\right)\right),\nn
\bar{e}_{18}(\Delta)&=&\frac{h_A^2}{36\,\Delta^3} +
\frac{h_A^2}{839808\,F_\pi^2\,\pi^2\,\Delta}\left(2025\,g_A^2+3456\,h_A^2
  -450\,g_A\,g_1 +
  425\,g_1^2\right)\nn
&-&\frac{h_A^2\,g_A^2}{108\,F_\pi^2\,\pi^2\,\Delta}\log\left(\frac{2\,\Delta}{M_\pi}\right),\nn
\bar{e}_{19}(\Delta)-\frac{1}{2}\bar{e}_{36}(\Delta)-2\bar{e}_{38}(\Delta)&=&-\frac{h_A^2}{93312\,F_\pi^2\,\pi^2\,\Delta}
\left(351+1296\,g_A^2+400\,g_1^2\right)\nn
&+&\frac{h_A^2}{5184\,F_\pi^2\,\pi^2\,\Delta}\left(-33 +
  81\,g_A^2-50\,g_A\,g_1 + 25\,g_1^2\right)
\log\left(\frac{2\,\Delta}{M_\pi}\right),\nn
\bar{e}_{20}(\Delta)+\bar{e}_{35}(\Delta)&=&\frac{h_A^2}{5832\,F_\pi^2\,\pi^2\,\Delta}\left(81\,g_A^2+25\,
g_1^2\right)-\frac{h_A^2}{5184\,F_\pi^2\,\pi^2\,\Delta}\left(81\,g_A^2-50\,g_A\,g_1+25\,g_1^2\right)
\log\left(\frac{2\,\Delta}{M_\pi}\right),\nn
\bar{e}_{21}(\Delta)-\frac{1}{2}\bar{e}_{37}(\Delta)&=&-\frac{h_A^2}{62208\,F_\pi^2\,\pi^2\,\Delta}
\left(72-999\,g_A^2+384\,h_A^2+750\,g_A\,g_1-175\,g_1^2\right) \nn
&+&
\frac{h_A^2}{10368\,F_\pi^2\,\pi^2\,\Delta}\left(-24+135\,g_A^2+50\,g_A\,g_1-25\,g_1^2\right)
\log\left(\frac{2\,\Delta}{M_\pi}\right),\nn
\bar{e}_{22}(\Delta)-4\,\bar{e}_{38}(\Delta)&=&-\frac{h_A^2}{72\,F_\pi^2\,\pi^2\,\Delta}\left(1+\log\left(\frac{2\,\Delta}{M_\pi}\right)\right).
\eeqa 
The appearance of logarithms of the physical pion mass in the above expressions is
due to our choice of the renormalization scale $\mu = M_\pi$
in the definitions of  $c_i$, $\bar d_i$ and $\bar e_i$. 
The LECs $\bar{e}_{15}(\Delta), \bar{e}_{16}(\Delta)$ and
$\bar{e}_{18}(\Delta)$ receive $1/\Delta^3$-contributions from
the order-$\epsilon^1$  $\pi N$-amplitude. Numerically, these
terms dominate over the loop
contributions (as one would expect from naive dimensional analysis)
and explain a half of the size of the LECs $\bar{e}_{15},
\bar{e}_{16}$ and $\bar{e}_{18}$ which appear to be unnaturally large in the $\Delta$-less
theory, see Table~\ref{table:parametersSaturation}. 
It is comforting to see that the delta contributions to the LECs
$c_i$, $\bar d_i$ and $\bar e_i$ given in the above expressions, whose
numerical values are listed in table \ref{table:parametersSaturation},
are in a very good agreement with the differences between
the $\Delta$-less and $\Delta$-full fits. Clearly, one should not expect
this agreement to be perfect since the delta contributions to the
amplitude involve terms beyond the order-$Q^4$ $\Delta$-less
result. Our findings, however, indicate that these resummed
contributions are likely to be small and the most important terms are well
represented by the $\Delta$-resonance contributions to the LECs
$c_i$, $\bar d_i$ and $\bar e_i$.

Last but not least, we emphasize that  
the (linear combinations of the) LECs $\bar
e_{19,20,21,22,35,36,37,38}$ and $c_1$ absorbed into redefinition of
$c_i$'s, see Eq.~(\ref{ei_conv}), do receive contributions due to  the delta
resonance, see Eq.~(\ref{e_sat}). Assuming that these LECs are
saturated by the delta, we may estimate the shifts in the $c_i$
induced by absorbing these order-$Q^4$ contributions via
\beqa
c_1&\to& c_1 + 2\,M_\pi^2\left(\bar{e}_{22}-4\,\bar{e}_{38} +
  \frac{ l_3\, c_1}{F_\pi^2}\right),\nn
c_2&\to& c_2 - 8\,M_\pi^2\left(\bar{e}_{20}+\bar{e}_{35}\right),\nn
c_3&\to& c_3 - 4\,
M_\pi^2\left(2\,\bar{e}_{19}-\bar{e}_{22}-\bar{e}_{36}\right),\nn
c_4&\to& c_4 - 4\,M_\pi^2\left(2\,\bar{e}_{21}-\bar{e}_{37}\right). 
\eeqa
Since the delta-resonance contributions to the induced shifts of
$c_i$'s start with the loop corrections and do not have any
$1/\Delta^3$ contribution from the order-$\epsilon^1$ terms, the
induced shifts appear to be rather small:
\beqa
2\,M_\pi^2\left(\bar{e}_{22}(\Delta)-4\,\bar{e}_{38}(\Delta) +
  \frac{\bar l_3\, c_1(\Delta)}{F_\pi^2}\right)&=&-0.10\,{\rm GeV}^{-1},\nn
-
8\,M_\pi^2\left(\bar{e}_{20}(\Delta)+\bar{e}_{35}(\Delta)\right)&=&-0.14\,{\rm
  GeV}^{-1},\nn
- 4\,
M_\pi^2\left(2\,\bar{e}_{19}(\Delta)-\bar{e}_{22}(\Delta)-\bar{e}_{36}(\Delta)\right)&=&0.10\,{\rm
  GeV}^{-1},\nn
-
4\,M_\pi^2\left(2\,\bar{e}_{21}(\Delta)-\bar{e}_{37}(\Delta)\right)&=&-0.26\,{\rm
  GeV}^{-1}. 
\eeqa

\section{$\Delta$(1232) contributions to the two-pion exchange 3NF}
\def\theequation{\arabic{section}.\arabic{equation}}
\label{sec:TPE}

After these preparations, we are now in the position to discuss the
contributions to the two-pion exchange 3NF emerging from the
intermediate $\Delta$-excitations up to the leading one-loop order
(i.e.~N$^3$LO).

\begin{figure}[tb]
\vskip 1 true cm
\includegraphics[width=15.0cm,keepaspectratio,angle=0,clip]{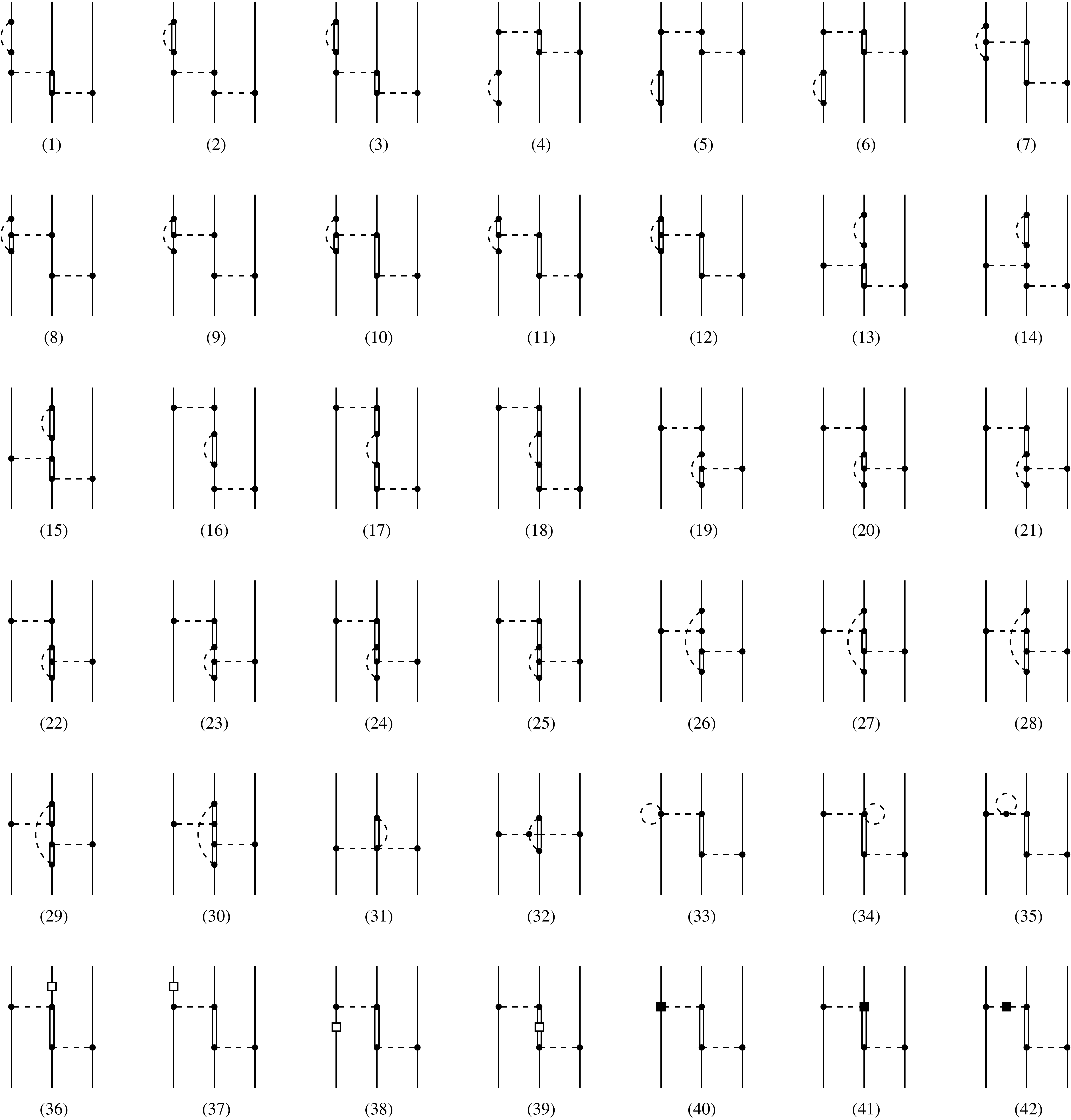}
\caption{Two-pion exchange 3N diagrams involving intermediate delta
  excitations at N$^3$LO. 
Solid dots and filled 
rectangles denote vertices from $\mathcal{L}_{\pi
   N}^{(1)}+ \mathcal{L}_{\pi
   N \Delta}^{(1)} + \mathcal{L}_{\pi
   \Delta \Delta}^{(1)} $ and $\mathcal{L}_{\pi
   N}^{(3)} + \mathcal{L}_{\pi
   N \Delta}^{(3)} $, respectively.  Open rectangles refer to
 $1/m$-vertices from $\mathcal{L}_{\pi
   N}^{(2)}+ \mathcal{L}_{\pi
   N \Delta}^{(2)} $. Diagrams which result from the interchange of
 the nucleon lines and/or application of the time reversal operation
 are not shown. Also not shown are diagrams which lead to vanishing
 contributions to the 3NF. 
 For the remaining notation see Fig.~\ref{fig:nucleon_self_energy}.   
 }\label{diagrams_3NF} 
\end{figure}

In the isospin and static limits, the general structure of the
two-pion exchange 3NF in momentum space 
has the following form (modulo terms of a shorter range, see
Ref.~\cite{Krebs:2012yv} for more details):
\beq
\label{2pi_general}
V_{2 \pi} = \frac{\vec \sigma_1 \cdot \vec q_1\,  \vec \sigma_3 \cdot
  \vec q_3}{[q_1^2 + M_\pi^2 ] \, [q_3^2 + M_\pi^2 ]}  \Big( \fet
  \tau_1 \cdot \fet \tau_3  \, {\cal A}(q_2) + \fet \tau_1 \times  \fet
\tau_3 \cdot \fet \tau_2  \,   \vec q_1 \times  \vec q_3   \cdot \vec
\sigma_2 \, {\cal B}(q_2) \Big)  \,,
\eeq
where $\vec \sigma_i$ ($\fet \tau_i $)  denote the Pauli
spin (isospin) matrices for the nucleon $i$ while $\vec q_{i}$ is the
momentum transfer, $\vec q_{i} = \vec p_i \,
' - \vec p_i$,  with $\vec p_i \, '$ and $\vec p_i$ being the final and initial momenta of the nucleon $i$. 
Here and in what follows, we use the notation: $q_i \equiv | \vec q_i
|$.  
Unless stated otherwise, the expressions for the 3NF are 
given for a particular choice of the nucleon labels. The complete result 
can then be obtained by taking into account all possible permutations of the
nucleons,
\beq
V_{\rm 3N}^{\rm full} = V_{\rm 3N} + \mbox{5 permutations}\,.
\eeq
The quantities ${\cal A} (q_2)$ and 
${\cal B} (q_2)$ in Eq.~(\ref{2pi_general}) are scalar
functions of the momentum transfer $q_2$ of the second nucleon whose explicit form is
derived within the chiral expansion. In the $\Delta$-less framework,
this expansion starts at N$^2$LO which corresponds  to the order
$Q^3$. The explicit expressions for    ${\cal A} (q_2)$ and 
${\cal B} (q_2)$ up to N$^4$LO, i.e. up to order $Q^5$, can be found
in Ref.~\cite{Krebs:2012yv}.  In the $\Delta$-full framework, the leading
contributions are shifted from N$^2$LO to NLO, i.e. to order
$\epsilon^2$.  These leading delta contributions have the form 
\beqa
{\cal A}_\Delta^{(2)}(q_2)&=&-\frac{g_A^2 h_A^2 }{18 \,\Delta\,
  F_\pi^4}\left(2 M_\pi^2+q_2^2\right),\nn
{\cal B}_\Delta^{(2)}(q_2)&=&\frac{g_A^2 h_A^2}{36 \,\Delta \, F_\pi^4}\,,
\eeqa
and are known to provide the dominant long-range mechanism of the  
3NF \cite{Fujita:1957zz}. 
There are no contributions of the delta to  ${\cal A} (q_2)$ and 
${\cal B} (q_2)$ at N$^2$LO \cite{Epelbaum:2007sq}, i.e. at order $\epsilon^3$, except for
the shift of the LEC $h_A$ as discussed in section \ref{sec:piN}, see Eq.~(\ref{LECs_shift_eq}).
At N$^3$LO ($\epsilon^4$) one has to take into account the
contributions emerging from the diagrams shown in Fig.~\ref{diagrams_3NF}. 
These graphs are analogous to the $\Delta$-less ones shown in Fig.~2 of
Ref.~\cite{Bernard:2007sp} but involve at least one intermediate delta
excitation.  Notice that in contrast to that work, we do not show in
Fig.~\ref{diagrams_3NF} certain diagrams which yield vanishing results
for the sake of compactness. This concerns, for example,
one-loop graphs leading to integrals involving an odd power of the
loop momentum to be integrated over.

The last three diagrams in Fig.~\ref{diagrams_3NF} contribute to
renormalization of the pion field and the lowest-order pion-nucleon and
pion-nucleon-delta vertices and also give rise to the corresponding
Goldberger-Treiman discrepancy relations.  These contributions are
automatically taken into account by expressing the 3NF in terms of
physical quantities and using the effective values of the LECs $g_A$ and $h_A$
which account for the Goldberger-Treiman discrepancy, see the
discussion in the previous section. We are, therefore, left with
one-loop diagrams constructed out of the lowest-order and tree-order
graphs which involve a single insertion of the $1/m_N$-vertices which
give rise to the leading relativistic corrections. We remind the
reader that the power counting scheme used to derive the nuclear
forces and currents in Refs.~\cite{Epelbaum:2004fk,Epelbaum:2006eu,Bernard:2007sp,Epelbaum:2007us,Bernard:2011zr,Krebs:2012yv,Krebs:2013kha,Epelbaum:2014efa,Epelbaum:2014sza,Reinert:2017usi} and \cite{Kolling:2009iq,Kolling:2011mt,Krebs:2016rqz} makes the assignment 
$Q/m_N \sim Q^2 / \Lambda_\chi^2 $ for the nucleon mass,
see Ref.~\cite{Weinberg:1991um} for more details. This implies that
$1/m_N$-corrections to the nuclear forces and currents are shifted to higher
orders compared to the corresponding static
contributions.\footnote{Notice that the same power
  counting is employed to determine the LECs from pion-nucleon scattering
  in Ref.~\cite{Krebs:2012yv} and in this work.}    
In particular, the leading relativistic corrections appear at the same
order with the leading one-loop diagrams. 

It is important to keep in mind that, in order to derive the genuine
3NF contributions, one needs to separate  the irreducible parts in the
corresponding amplitudes in order to avoid double counting when iterating the
potentials in the scattering equation. While this can be achieved in
different ways, see Ref.~\cite{Epelbaum:2005pn} for more details, we
employ  here the method of unitary transformation which was first applied in
the context of chiral EFT in Ref.~\cite{Epelbaum:1998ka}. A comprehensive discussion
of this approach can be found in Ref.~\cite{Epelbaum:2010nr}. 
The same  method  was used in
our earlier work on the derivation of the three-
\cite{Bernard:2007sp,Bernard:2011zr,Krebs:2012yv,Krebs:2013kha} and
four-nucleon forces \cite{Epelbaum:2006eu,Epelbaum:2007us} 
and electroweak nuclear current operators \cite{Kolling:2009iq,Kolling:2011mt,Krebs:2016rqz}. 
We remind the reader that in this approach one first applies the
canonical formalism to the effective chiral Lagrangian expressed in
terms of renormalized fields to derive the Hamilton
density in the pion-nucleon sector. In the second step, one decouples the purely nucleonic
subspace of the Fock space from the rest via a suitably chosen
unitary transformation. The determination of the unitary operator and
the resulting nuclear potentials is carried out perturbatively within
the EFT expansion. Clearly, there is always certain ambiguity in the choice of
the unitary operator. However, as was found in Refs.~\cite{Epelbaum:2006eu,Epelbaum:2007us}, most of the
choices of the unitary operator lead to nuclear potentials which
cannot be renormalized, i.e. the corresponding matrix elements 
involve ultraviolet-divergent integrals even after expressing all LECs in terms of their physical
values. While this is, of course, not a fundamental problem since 
nuclear potentials do not correspond to observable quantities, it is
desirable to have a well-defined and finite nuclear
Hamiltonian. Enforcing renormalizability at the level of the 
Hamiltonian strongly restricts the unitary ambiguity mentioned
above. In particular, the renormalizability requirement was found to lead to an
unambiguous result for the static parts of the three- \cite{Bernard:2007sp,Bernard:2011zr,Krebs:2012yv,Krebs:2013kha} and
four-nucleon potentials \cite{Epelbaum:2006eu,Epelbaum:2007us}, while the leading relativistic
corrections still depend on two arbitrary constants, which
parametrize the unitary ambiguity at this order in the chiral
expansion \cite{Bernard:2011zr}.  Explicit expressions for the
nuclear Hamiltonian in the operator form after fixing the unitary
ambiguity up to N$^4$LO in the $\Delta$-less approach can be found in
Refs.~\cite{Epelbaum:2007us,Bernard:2011zr,Krebs:2012yv}.  

To employ the method of unitary transformation within the small scale
expansion one can follow the lines of
Ref.~\cite{Epelbaum:2007us}. The crucial difference is that one now
needs to decouple not only pions but also the delta degrees of
freedom. As discussed in that work, it is convenient to start with the
minimal parametrization of the unitary operator using the ansatz
proposed by Okubo \cite{Okubo:1954zz}, see Eq.~(2.12) of
Ref.~\cite{Epelbaum:2007us}. Using this parametrization, the unitary
operator can be calculated via a perturbative solution of the decoupling
equation (2.13) of Ref.~\cite{Epelbaum:2007us} within the
small-scale expansion.  The resulting rather lengthy expressions for
the delta contributions to the nuclear force in the operator form  are not
listed in this work but can be made available as a Mathematica
notebook upon
request from the authors. Notice further that the resulting
nuclear Hamiltonian is defined unambiguously within this ansatz but is
not renormalizable as explained before. Following the lines of
Ref.~\cite{Epelbaum:2007us}, we exploit the unitary ambiguity to
ensure renormalizability of the nuclear potentials. This is achieved
by applying all possible additional unitary transformations acting on the
nucleonic subspace of the Fock space which can be constructed at a
given order in the SSE, see Ref.~\cite{Epelbaum:2007us} for more details. The
corresponding transformation angles are to be chosen in such a way that the
resulting Hamiltonian is finite. In the $\Delta$-less approach,
we had to introduce six such additional unitary 
transformations (plus two more transformations involving
$1/m_N$-corrections) 
whose generators are given in
Ref.~\cite{Epelbaum:2007us} (Ref.~\cite{Kolling:2011mt}). The
inclusion of the delta excitations in the intermediate states allows
for much more flexibility in the construction of the additional
unitary transformations. In particular, we were able to write 50
antihermitian generators $S_i^\Delta$ which
are listed in Eq.~(\ref{generators}). The corresponding unitary
transformations generate additional contributions to the nuclear
Hamiltonian which depend on 50 real parameters $\alpha_i^\Delta$.  
In order to derive nuclear potentials, one has to evaluate the
corresponding matrix elements of the nuclear Hamilton operator written
in second-quantized form. Calculating the 3NF contributions,
expressing them in terms of physical parameter and requiring
that there are no ultraviolet divergencies lead to constraints on
$\alpha_i^\Delta$, which are given in Eq.~(\ref{constraints}). In particular, we
find that 23 specific linear combinations of the $\alpha_i^\Delta$'s
have to vanish. While these constraints obviously do not allow for a
unique determination of these parameters, we find that they lead to an
unambiguous result for the 3NF, which does not depend on any 
of the undetermined linear combinations of $\alpha_i^\Delta$'s. 

We now turn to the results for the 3NF and consider first the static terms. We obtain the following contributions of the
delta-isobar to the functions ${\cal A}_\Delta (q_2)$ and ${\cal
  B}_\Delta (q_2)$ at leading one-loop order:
\beqa
{\cal A}_\Delta^{(4)}(q_2)&=&-\frac{g_A^2 h_A^2 }{139968 \pi^2 \Delta^3 F_\pi^6}\left(81 g_A^2 \left(40 \Delta^4+34 
M_\pi^4-\pi  \Delta  M_\pi^3-13 \Delta^2 M_\pi^2\right) \left(2 
M_\pi^2+q_2^2\right)\right.\nn
&-&\left. 450 g_A g_1 \left(8 \Delta^4+2 M_\pi^4-\pi  
\Delta  M_\pi^3-5 \Delta^2 M_\pi^2\right) \left(2 M_\pi^2+q_2^2\right)
+36 \Delta  \left(20 \pi  h_A^2 M_\pi^3 \left(2 M_\pi^2+q_2^2\right)\right.\right.\nn
&-&\left. \left. 27 \left(2 \Delta  M_\pi^2-\Delta^3\right) \left(M_\pi^2+2 
q_2^2\right)\right)+25 g_1^2 \left(40 \Delta^4+34 M_\pi^4-17 \pi  
\Delta  M_\pi^3-13 \Delta^2 M_\pi^2\right) \left(2 M_\pi^2+q_2^2
\right)\right)\nn
&+&\frac{g_A^2 h_A^2 }{144 \pi^2 F_\pi^6}\Delta\, D(q_2) \left(M_\pi^2+2 q_2^2\right) 
\left(-2 \Delta^2+2 M_\pi^2+q_2^2\right) -\frac{ g_A^2 h_A^2 }{144 \pi^2 F_\pi^6}\Delta \,L(q_2) 
\left(M_\pi^2+2 q_2^2\right)\nn
&+&\frac{g_A^2 
h_A^2 
}{139968 \pi^2 \Delta^3 F_\pi^6}H(0) \left(81 g_A^2 \left(40 \Delta^4+34 M_\pi^4-47 \Delta^2
    M_\pi^2
\right) \left(2 M_\pi^2+q_2^2\right)\right.\nn
&-&\left. 450 g_A g_1 \left(8 
\Delta^4+2 M_\pi^4-7 \Delta^2 M_\pi^2\right) \left(2
M_\pi^2+q_2^2\right)
+25 g_1^2 \left(40 \Delta^4+34 M_\pi^4-47 \Delta^2 M_\pi^2\right) 
\left(2 M_\pi^2+q_2^2\right)\right.\nn
&-&\left. 1944 \Delta^2 \
\left(M_\pi^2-\Delta^2\right) \left(M_\pi^2+2 q_2^2\right)\right) \nn
&+&\frac{  g_A^2 
h_A^2 }{34992 \pi^2 
F_\pi^6}\Delta\,\log \left(\frac{2 \Delta }{M_\pi}\right) \left(M_\pi^2 
\left(1620 g_A^2-1800 g_A g_1+500 g_1^2+729\right)\right.\nn
&+&\left.2 q_2^2 \left(405 
g_A^2-450 g_A g_1+125 g_1^2+729\right)\right), \nn
{\cal B}_\Delta^{(4)}(q_2)&=&\frac{g_A^2 h_A^2 }{279936 \pi^2 
\Delta^3 F_\pi^6}\left(81 
g_A^2 \left(58 \Delta^4+34 M_\pi^4-\pi  \Delta  M_\pi^3+50 \Delta^2 
M_\pi^2\right)\right.\nn
&-&\left. 450 g_A g_1 \left(10 \Delta^4+10 M_\pi^4-5 \pi  \Delta 
 M_\pi^3+2 \Delta^2 M_\pi^2\right)-144 h_A^2 \left(-16 \Delta^4+8 
M_\pi^4-9 \pi  \Delta  M_\pi^3+16 \Delta^2 M_\pi^2\right)\right. \nn
&+&\left. 250 
\Delta^4 g_1^2+850 g_1^2 M_\pi^4-425 \pi  \Delta  g_1^2 M_\pi^3+50 
\Delta^2 g_1^2 M_\pi^2-972 \Delta^2 M_\pi^2\right)\nn
&-&\frac{  g_A^2 h_A^2
}{576 \pi^2 F_\pi^6}\Delta\,  D(q_2) \left(-4 \Delta^2+4 M_\pi^2+q_2^2
\right) +\frac{  
g_A^2 h_A^2}{288 \pi^2 F_\pi^6}\Delta\, L(q_2)\nn
&-&\frac{g_A^2 h_A^2 }{139968 \pi^2 \Delta^3 F_\pi^6}H(0)\left(486 
\Delta^4+81 g_A^2 \left(29 \Delta^4+17 M_\pi^4+8 \Delta^2 M_\pi^2
\right)\right.\nn
&-&\left. 450 g_A g_1 \left(5 \Delta^4+5 M_\pi^4-4 \Delta^2
  M_\pi^2\right)
-576 h_A^2 \left(-2 \Delta^4+M_\pi^4+\Delta^2 M_\pi^2\right)+125 
\Delta^4 g_1^2+425 g_1^2 M_\pi^4\right. \nn
&-&\left. 400 \Delta^2 g_1^2 M_\pi^2-486 
\Delta^2 M_\pi^2\right)\nn
&-&\frac{  g_A^2 h_A^2 }{139968 \pi^2 F_\pi^6}\Delta\,\log \left(\frac{2 \Delta 
}{M_\pi}\right)\left(2349 g_A^2-2250 g_A 
g_1+1152 h_A^2+125 g_1^2+972\right),\label{AB_n3lo_delta}
\eeqa
where we use the notation for the various loop functions introduced in
Ref.~\cite{Kaiser:1998wa} which, in the case of dimensional regularization, reads: 
\beqa
L(q) &=& \frac{\sqrt{q^2 + 4 M_\pi^2}}{q} \ln \frac{\sqrt{q^2 + 4
    M_\pi^2}+q}{2 M_\pi} \,, \nn[5pt]
D(q) &=& \frac{1}{\Delta} \int_{2 M_\pi}^\infty \frac{d \mu}{q^2 +
  \mu^2} \arctan \frac{\sqrt{\mu^2 - 4 M_\pi^2}}{2 \Delta}\,, \nn[5pt]
H(q) &=& \frac{4 M_\pi^2 + 2 q^2 - 4 \Delta^2}{4 M_\pi^2
+ q^2 - 4 \Delta^2} \left[ L (q) - L \left(2 \sqrt{\Delta^2 - M_\pi^2}
\right) \right]\,.
\eeqa
The expressions for the loop functions resulting in the framework of spectral function regularization
introduced in Ref.~\cite{Epelbaum:2003gr} can be found in
\cite{Krebs:2007rh}. $1/\Delta$-expansion of the loop functions is
given by
\beqa
{D}(q)&=&\frac{-L(q)+\log \left(\frac{2 \Delta }{M_\pi}\right)+1}{2 
\Delta ^2} + \frac{-3 L(q) \left(4 M_\pi^2+q^2\right)+3 \left(6 
M_\pi^2+q^2\right) \log \left(\frac{2 \Delta }{M_\pi}\right)+3 
M_\pi^2+q^2}{72 \Delta ^4}\nn
&+&\frac{-10 L(q) \left(4 M_\pi^2+q^2
\right)^2-15 M_\pi^4+10 M_\pi^2 q^2+10 \left(30 M_\pi^4+10 M_\pi^2 
q^2+q^4\right) \log \left(\frac{2 \Delta }{M_\pi}\right)+2 q^4}{1600 
\Delta ^6}\nn
&+& {\cal O}(1/\Delta^8),\nn
H(0)&=&1 -\log \
\left(\frac{2 \Delta }{M_\pi}\right) +\frac{2 M_\pi^2-4 \
M_\pi^2 \log \left(\frac{2 \Delta }{M_\pi}\right)}{8 \Delta ^2} +\frac{7 M_\pi^4-12 M_\pi^4 \log \
\left(\frac{2 \Delta }{M_\pi}\right)}{32 \Delta ^4}\nn
&+& \frac{74 M_\pi^6-120 M_\pi^6 \log \left(\frac{2 \Delta \
}{M_\pi}\right)}{384 \Delta ^6} +  {\cal O}(1/\Delta^8).
\eeqa
These expressions indicate that the $\log \left(\frac{2 \Delta
  }{M_\pi}\right)$ terms in
${\cal A}_\Delta^{(4)} (q_2)$ and ${\cal
  B}_\Delta^{(4)} (q_2)$  are essential for vanishing of ${\cal A}_\Delta^{(4)} (q_2)$ and ${\cal
  B}_\Delta^{(4)} (q_2)$ in the $\Delta \to \infty$ limit as required
by the decoupling theorem. Notice further that the two-pion exchange diagrams shown in
Fig.~\ref{diagrams_3NF} also induce shorter-range
contributions which will be discussed in a separate publication.  

The static contributions discussed above depend only on the momentum
transfers $\vec q_i$ and are, therefore, local. It is thus 
natural to switch to the coordinate space representation of these 3NF
terms. The Fourier transform of a local potential is given by
\beqa
\label{temp5}
\tilde V_{3N} (\vec r_{12}, \, \vec r_{32} \, ) &=& \int \frac{d^3 q_1}{(2
  \pi)^3} \,  \frac{d^3 q_3}{(2 \pi)^3} \, e^{i \vec
  q_1 \cdot \vec r_{12}} \; e^{i \vec   q_3 \cdot \vec r_{32}} \;V_{3N}(\vec q_{1}, \, \vec q_{3} )\,, 
\eeqa
where $\vec r_{ij} \equiv \vec r_i - r_j $ is the distance
between the nucleons $i$ and $j$.
For the two-pion-exchange contribution, we obtain from Eq.~(\ref{2pi_general})
\beq
\tilde V_{\rm 2\pi} (\vec r_{12}, \, \vec r_{32} \,
)=-\vec\sigma_1\cdot\vec\nabla_{12}\;\vec\sigma_3\cdot\vec\nabla_{32}
\left(\fet\tau_1\cdot\fet\tau_3\;\tilde{\cal A}(\vec r_{12}, \vec r_{32})
-\fet \tau_1\times\fet\tau_3\cdot\fet\tau_2\; \vec
\nabla_{12}\times\vec\nabla_{32}\cdot\vec\sigma_2\; \tilde{\cal B}(\vec r_{12}, \vec r_{32})\right)\,.
\eeq
Here and in what follows, the differential operators $\vec \nabla_{ij}$ are
defined in terms of dimensionless variables $\vec x_{ij}=\vec r_{ij}
M_\pi$ while the functions
$\tilde {\cal A}$ and $\tilde {\cal B}$ are defined via
\beqa
\tilde{\cal A}(\vec r_{12}, \vec r_{32})& =&\int \frac{d^3
  q_1}{(2\pi)^3}\frac{d^3 q_3}{(2\pi)^3} \, e^{i \vec
  q_1 \cdot \vec r_{12}} \; e^{i \vec   q_3 \cdot \vec r_{32}}
\;\frac{1}{q_1^2+M_\pi^2}\frac{1}{q_3^2+M_\pi^2}\;{\cal A}(q_2),\nn
\tilde{\cal B}(\vec r_{12}, \vec r_{32})& =&\int \frac{d^3
  q_1}{(2\pi)^3}\frac{d^3 q_3}{(2\pi)^3} \, e^{i \vec
  q_1 \cdot \vec r_{12}} \; e^{i \vec   q_3 \cdot \vec r_{32}}
\;\frac{1}{q_1^2+M_\pi^2}\frac{1}{q_3^2+M_\pi^2}\;{\cal B}(q_2).
\eeqa
To perform the integrations, we employ the spectral-function
representation of the functions ${\cal A}$ and ${\cal B}$. The only
non-polynomial in $q_2$ terms in Eq.~(\ref{AB_n3lo_delta}) emerge from
the scalar loop functions $L(q_2)$ and $D(q_2)$. Their spectral-function representation is
given by
\beqa
L(q_2)&=&1+q_2^2\int_{2\,M_\pi}^\infty d\mu
\frac{\rho_L(\mu)}{q_2^2+\mu^2}, \quad
\rho_L(\mu)\,=\,\frac{\sqrt{\mu^2-4\,M_\pi^2}}{\mu^2},\nn
D(q_2)&=&\int_{2\,M_\pi}^\infty d\mu
\frac{\rho_D(\mu)}{q_2^2+\mu^2}, \quad \rho_D(\mu)\,=\,\frac{1}{\Delta}\arctan\left(\frac{\sqrt{\mu^2-4\,M_\pi^2}}{2\,\Delta}\right).
\eeqa
Therefore, the Fourier transform of the 3NF terms involving these functions can be written as 
\beqa
\int \frac{d^3 q_1}{(2
  \pi)^3} \,  \frac{d^3 q_3}{(2 \pi)^3} \, e^{i \vec
  q_1 \cdot \vec r_{12}} \; e^{i \vec   q_3 \cdot \vec r_{32}}
\;\frac{1}{q_1^2+M_\pi^2}\frac{1}{q_3^2+M_\pi^2}L(q_2)&=&
\frac{M_\pi^2}{(4\pi)^2}U_1(x_{12}) U_1(x_{32})\nn
&-&\frac{M_\pi}{(4\pi)^3}(\vec{\nabla}_{12}+\vec{\nabla}_{32})^2\int
d^3 x \,U_1(|\vec{x}_{12}+\vec{x}|) U_1(|\vec{x}_{32}+\vec{x}|)
V_1(x),\nn
\int \frac{d^3 q_1}{(2
  \pi)^3} \,  \frac{d^3 q_3}{(2 \pi)^3} \, e^{i \vec
  q_1 \cdot \vec r_{12}} \; e^{i \vec   q_3 \cdot \vec r_{32}}
\;\frac{1}{q_1^2+M_\pi^2}\frac{1}{q_3^2+M_\pi^2}D(q_2)&=&\frac{1}{(4\pi)^3}
\int
d^3 x \,U_1(|\vec{x}_{12}+\vec{x}|) U_1(|\vec{x}_{32}+\vec{x}|)
Q_1(x),
\eeqa
where the profile functions are given by 
\beqa
U_1(x)&=&\frac{4\pi}{M_\pi}\int \frac{d^3 q}{(2\pi)^3}\frac{e^{i\vec q\cdot
  \vec x/M_\pi} }{q^2+M_\pi^2}=\frac{e^{-x}}{x}, \nn [5pt]
V_1(x)&=&\frac{4\pi}{M_\pi}\int \frac{d^3 q}{(2 \pi )^3} \, e^{i
    \vec q \cdot \vec x/M_\pi}\, \int_{2M_\pi}^\infty
  d\mu\frac{1}{\mu^2+q^2}\frac{1}{\mu^2}\sqrt{\mu^2-4M_\pi^2}=\frac{1}{x}\int_{2}^\infty
  d\mu \frac{e^{-x\,\mu}}{\mu^2}\sqrt{\mu^2-4},\nn [5pt]
Q_1(x)&=&\int_{2}^\infty d \mu \frac{e^{-\mu x}}{x}\frac{M_\pi}{\Delta}\arctan\left(\frac{M_\pi}{\Delta}\frac{\sqrt{\mu^2-4}}{2}\right).
\eeqa
We are interested here only in the long-range terms and, therefore,
restrict ourselves to the case $x_{i j}\neq 0$. All terms
involving positive powers of  momenta
$\vec{q}_1$ and $\vec{q}_3$ can be expressed through gradients 
$-i M_\pi\vec\nabla_{12}$ and  $-i M_\pi \vec\nabla_{32}$ which can be
taken out of the integrals. 

For the NLO delta contributions to 3NF, we obtain the following
coordinate-space expressions: 
\beqa
\tilde{\cal A}^{(2)}_\Delta(\vec{r}_{12},\vec{r}_{32})&=&\frac{g_A^2 h_A^2}{288 \pi ^2 \Delta  F_\pi^4}M_\pi^6 
\left(\left(\vec{\nabla}_{12}+\vec{\nabla}_{32}\right)^2-2\right) 
U_1(x_{12}) U_1(x_{32}),\nn
\tilde{{\cal 
B}}^{(2)}_\Delta(\vec{r}_{12},\vec{r}_{32})&=&\frac{g_A^2 h_A^2 
}{576 \pi ^2 \Delta  F_\pi^4}M_\pi^6 U_1(x_{12}) U_1(x_{32})\,.
\eeqa
The N$^3$LO delta contributions have the form
\beqa
\tilde{{\cal 
A}}^{(4)}_\Delta(\vec{r}_{12},\vec{r}_{32})&=&\frac{g_A^2 h_A^2 
}{2239488 \pi ^5 \Delta ^3 
F_\pi^6}M_\pi^4 \left(243 \Delta ^4 \left(2 \left(\vec{\nabla}_{12}+\vec{
\nabla}_{32}\right)^2-1\right) \right.\nn
&\times&\left. \left(2 \Delta ^2+M_\pi^2 
\left(\left(\vec{\nabla}_{12}+\vec{\nabla}_{32}\right)^2-2\right)
\right) {\cal Q}_1(\vec{x}_{12},\vec{x}_{32})\right.\nn
&+&\left. \pi  M_\pi^2 
\left(-25 g_1^2 \left(\left(\vec{\nabla}_{12}+\vec{\nabla}_{32}
\right)^2-2\right) \left(40 \Delta ^4 (H(0)-1)+34 (H(0)-1) 
M_\pi^4\right.\right.\right.\nn
&+&\left.\left.\left. \Delta ^2 (13-47 H(0)) M_\pi^2+17 \pi  \Delta  M_\pi^3
\right)+1944 \Delta ^2 (H(0)-1) \left(2 \left(\vec{\nabla}_{12}+\vec{
\nabla}_{32}\right)^2-1\right) (M_\pi^2-\Delta^2 ) \right.\right. \nn
&+&\left. \left. 4 \Delta 
^4 \left(729 \left(1-2 \left(\vec{\nabla}_{12}+\vec{\nabla}_{32}
\right)^2\right)-250 g_1^2 \left(\left(\vec{\nabla}_{12}+\vec{
\nabla}_{32}\right)^2-2\right)\right) \log \left(\frac{2 \Delta 
}{M_\pi}\right)\right) U_1(x_{12}) U_1(x_{32})\right)\nn
&+&\frac{  g_A^2 h_A^2}{9216 
\pi ^5 F_\pi^6} \Delta M_\pi^6 \left(\vec{
\nabla}_{12}+\vec{\nabla}_{32}\right)^2 \left(1-2 \left(\vec{
\nabla}_{12}+\vec{\nabla}_{32}\right)^2\right)
{\cal V}_1(\vec{x}_{12},\vec{x}_{32})\nn
&-&\frac{g_A^4 h_A^2 }{27648 \pi ^4 \Delta ^3 
F_\pi^6}M_\pi^6 \left(\left(\vec{
\nabla}_{12}+\vec{\nabla}_{32}\right)^2-2\right) U_1(x_{12}) 
U_1(x_{32}) \left(40 \Delta ^4 (H(0)-1)+34 (H(0)-1)
  M_\pi^4\right.\nn
&+& \left. \Delta ^2 
(13-47 H(0)) M_\pi^2+\pi  \Delta  M_\pi^3+40 \Delta ^4 \log 
\left(\frac{2 \Delta }{M_\pi}\right)\right)\nn
&+&\frac{25 g_A^3 h_A^2 g_1 }{124416 \pi ^4 \Delta ^3 
F_\pi^6}M_\pi^6 \left(\left(\vec{
\nabla}_{12}+\vec{\nabla}_{32}\right)^2-2\right) U_1(x_{12}) 
U_1(x_{32}) \left(8 \Delta ^4 (H(0)-1)+2 (H(0)-1) M_\pi^4\right. \nn
&+&\left. \Delta ^2 
(5-7 H(0)) M_\pi^2+\pi  \Delta  M_\pi^3+8 \Delta ^4 \log 
\left(\frac{2 \Delta }{M_\pi}\right)\right)\nn
&+&\frac{5 g_A^2 h_A^4}{15552 \pi ^3 \Delta ^2 F_\pi^6} M_\pi^9 \left(\left(\vec{
\nabla}_{12}+\vec{\nabla}_{32}\right)^2-2\right) U_1(x_{12}) 
U_1(x_{32}),\nn
\tilde{{\cal \
B}}^{(4)}_\Delta(\vec{r}_{12},\vec{r}_{32})&=&-\frac{  g_A^2 
h_A^2 }{18432 \pi ^5 F_\pi^6}\Delta M_\pi^6 \left(\vec{\nabla}_{12}+\vec{\nabla}_{32}\right)^2 
{\cal V}_1(x_{12},x_{32})\nn
&+&\frac{  g_A^2 
h_A^2 }{36864 \pi ^5 F_\pi^6}\Delta M_\pi^4 \left(4 \Delta ^2+M_\pi^2 \left(\left(\vec{
\nabla}_{12}+\vec{\nabla}_{32}\right)^2-4\right)\right) {\cal 
Q}_1(x_{12},x_{32})\nn
&-&\frac{g_A^4 h_A^2 }{55296 \pi ^4 \Delta ^3 F_\pi^6}M_\pi^6 
U_1(x_{12}) U_1(x_{32}) \left(58 \Delta ^4 (H(0)-1)+34 (H(0)-1) 
M_\pi^4\right. \nn
&+&\left. 2 \Delta ^2 (8 H(0)-25) M_\pi^2+\pi  \Delta  M_\pi^3
\right)-\frac{29  g_A^4 h_A^2 
}{27648 \pi ^4 F_\pi^6}\Delta M_\pi^6 U_1(x_{12}) U_1(x_{32}) \log \left(\frac{2 \Delta }{M_\pi}
\right)\nn
&+&\frac{25 g_A^3 h_A^2 g_1 }{248832 \pi ^4 \Delta ^3 F_\pi^6}M_\pi^6 
U_1(x_{12}) U_1(x_{32}) \left(10 \Delta ^4 (H(0)-1)+10 (H(0)-1) 
M_\pi^4\right.\nn
&-&\left. 2 \Delta ^2 (4 H(0)+1) M_\pi^2+5 \pi  \Delta  M_\pi^3
\right)+\frac{125  g_A^3 h_A^2 
g_1 }{124416 \pi ^4 F_\pi^6}\Delta  M_\pi^6 U_1(x_{12}) U_1(x_{32}) \log \left(\frac{2 \Delta 
}{M_\pi}\right)\nn
&+&\frac{g_A^2 h_A^4 }{31104 \pi ^4 \Delta ^3 F_\pi^6}M_\pi^6 
U_1(x_{12}) U_1(x_{32}) \left(-16 \Delta ^4 (H(0)-1)+8 (H(0)-1) 
M_\pi^4\right.\nn
&+&\left. 8 \Delta ^2 (H(0)-2) M_\pi^2+9 \pi  \Delta  M_\pi^3
\right)-\frac{25 g_A^2 h_A^2 }{4478976 \pi ^4 \Delta ^3 F_\pi^6}g_1^2 
M_\pi^6 U_1(x_{12}) U_1(x_{32}) \left(10 \Delta ^4 (H(0)-1)\right. \nn
&+&\left. 34 
(H(0)-1) M_\pi^4-2 \Delta ^2 (16 H(0)+1) M_\pi^2+17 \pi  \Delta  
M_\pi^3\right)\nn
&+&\frac{g_A^2 
h_A^2 }{4608 \pi ^4 \Delta  F_\pi^6}M_\pi^6 (M_\pi^2-\Delta^2 ) (H(0)-1)  U_1(x_{12}) 
U_1(x_{32})-\frac{  g_A^2 h_A^4 }{1944 \pi ^4 F_\pi^6}\Delta
M_\pi^6 U_1(x_{12}) U_1(x_{32}) \log \left(\frac{2 \Delta }{M_\pi}
\right)\nn
&-&\frac{125  g_A^2 h_A^2 g_1^2 }{2239488 \pi ^4 F_\pi^6}\Delta
M_\pi^6 U_1(x_{12}) U_1(x_{32}) \log \left(\frac{2 \Delta }{M_\pi}
\right)-\frac{  g_A^2 h_A^2 }{2304 \pi ^4 F_\pi^6}\Delta M_\pi^6 
U_1(x_{12}) U_1(x_{32}) \log \left(\frac{2 \Delta }{M_\pi}
\right),
\eeqa
where the scalar integrals  ${\cal Q}_1(\vec{x}_{12},\vec{x}_{32})$ and ${\cal V}_1(\vec{x}_{12},\vec{x}_{32})$ 
are defined as 
\beqa
{\cal Q}_1(\vec{x}_{12},\vec{x}_{32})&=&\int
d^3 x \,U_1(|\vec{x}_{12}+\vec{x}|) U_1(|\vec{x}_{32}+\vec{x}|) 
Q_1(x),\nn
{\cal V}_1(\vec{x}_{12},\vec{x}_{32})&=& \int
d^3 x\, U_1(|\vec{x}_{12}+\vec{x}|) U_1(|\vec{x}_{32}+\vec{x}|)
V_1(x).
\eeqa

At N$^3$LO one also has to take into account relativistic
corrections. Nucleonic contributions are already discussed in Ref.~\cite{Krebs:2012yv}. Here we give
only the corresponding $\Delta$-contributions from the diagrams $(36-39)$ of
Fig.~\ref{diagrams_3NF} proportional to $1/m_N$:
\beqa
V_{2\pi, 1/m_N}&=&
\frac{{\vec{q}_1}\cdot{\vec\sigma_1}{\vec{q}_3}\cdot{
\vec\sigma_3}}{\left[q_1^2
+M_\pi^2\right]\left[q_3^2+M_\pi^2\right]}\frac{{g_A}^2{h_A}^2}{72\Delta^2F_\pi^4{m_N}}\Bigg\{{\fet\tau_1}\cdot{\fet
\tau_3}\Big[-4({\vec{q}_1}\cdot{\vec{q}_3})^2 \nn
&+&  i\left(2{\vec{k}_1}\cdot{\vec{
q}_1}-{\vec{k}_1}\cdot{\vec{q}_3}+{\vec{k}_3}\cdot{\vec{q}_1}-2{\vec{
k}_3}\cdot{\vec{q}_3}\right){\vec{q}_1}\cdot{\vec{q}_3}\times{\vec\sigma_2}
\Big] \nn
&-& i{\vec{q}_1}\cdot{\vec{q}_3}{\fet\tau_1}\cdot{\fet\tau_2}
\times{\fet
\tau_3}\left(2{\vec{k}_1}\cdot{\vec{q}_1}-{\vec{k}_1}\cdot{\vec{q}_3}+{\vec{
k}_3}\cdot{\vec{q}_1}-2{\vec{k}_3}\cdot{\vec{q}_3}-i{\vec{q}_1}\cdot{
\vec{q}_3}\times{\vec\sigma_2}
\right) \Bigg\}.\label{relativistic_corr}
\eeqa
At this stage several comments are in order. 
\begin{itemize}
\item There are no contributions from $1/m_N$-corrections to 
pion-nucleon and pion-nucleon-delta vertices since both of them are
proportional to zeroth components of momenta and for this reason
vanish in the kinematics relevant for nuclear forces. This argument
is, however, 
only applicable to irreducible topologies since the corresponding 3NF contributions
can be calculated using  Feynman rules.  
All diagrams with intermediate delta excitations involving $1/m_N$-corrections to 
pion-nucleon or pion-nucleon-delta vertices are indeed irreducible. The situation is
different in the case of nucleonic contributions, where we used the unitary transformation technique
to extract the corresponding irreducible pieces,  see~\cite{Bernard:2011zr} for
more details. 
\item Since the Fujita-Miyazawa 3NF corresponds to an irreducible diagram, we can not
construct additional unitary transformations which would affect relativistic
corrections to it. For this reason, the
expression in Eq.~(\ref{relativistic_corr}) is unambiguous. This is, again,
different for nucleonic contributions, where additional unitary
transformations can be employed and the corresponding
relativistic corrections do depend on arbitrary parameters $\bar{\beta}_8$
and $\bar{\beta}_9$~\cite{Bernard:2011zr}.
\item 
Finally, we emphasize that Eq.~(\ref{relativistic_corr}) is consistent with the resonance
saturation of the nucleonic N$^5$LO two-pion-exchange tree-level diagram with one of the vertices
taken from the order-$Q^3$ $\pi N$-Lagrangian proportional to $d_i$.
Indeed if we replace the $d_i$-constants of this diagram with their
resonance saturation values given in 
Eq.~(\ref{res_d}), we reproduce the result of Eq.~(\ref{relativistic_corr}).
\end{itemize}
In coordinate space, the corresponding relativistic corrections are given by
\beqa
\tilde V_{2\pi, 1/m_N}&=&\frac{{g_A}^2{h_A}^2M_\pi^7}{1152\pi^2\Delta^2F_\pi^4{m_N}}{\vec{
\nabla}_{12}}\cdot{\vec\sigma_1}{\vec{
\nabla}_{32}}\cdot{\vec\sigma_3}\left({\vec{\nabla}_{12}}\cdot{\vec{
\nabla}_{32}}{\fet\tau_1}\cdot{\fet\tau_2}\times{\fet
\tau_3}(-2{\vec{k}_1}\cdot{\vec{\nabla}_{12}}+{\vec{k}_1}\cdot{\vec{
\nabla}_{32}}-{\vec{k}_3}\cdot{\vec{
\nabla}_{12}}\right. \nn 
&+&\left. 2{\vec{k}_3}\cdot{\vec{\nabla}_{32}}+M_\pi{\vec{
\nabla}_{12}}\cdot{\vec{\nabla}_{32}}\times{\vec\sigma_2})+{\fet\tau_1}
\cdot{\fet\tau_3}(2{\vec{k}_1}\cdot{\vec{
\nabla}_{12}}-{\vec{k}_1}\cdot{\vec{
\nabla}_{32}}+{\vec{k}_3}\cdot{\vec{
\nabla}_{12}}-2{\vec{k}_3}\cdot{\vec{\nabla}_{32}}){\vec{
\nabla}_{12}}\cdot{\vec{\nabla}_{32}}\times{\vec\sigma_2}\right.\nn
&+&\left. 4M_\pi({\vec{
\nabla}_{12}}\cdot{\vec{\nabla}_{32}})^2{\fet\tau_1}\cdot{\fet\tau_3}
\right){U_1}({x_{12}}){U_1}({x_{32}}).
\eeqa

\section{Discussion}
\label{sec:results}
\begin{figure}[tb]
\vskip 1 true cm
\includegraphics[width=0.95\textwidth,keepaspectratio,angle=0,clip]{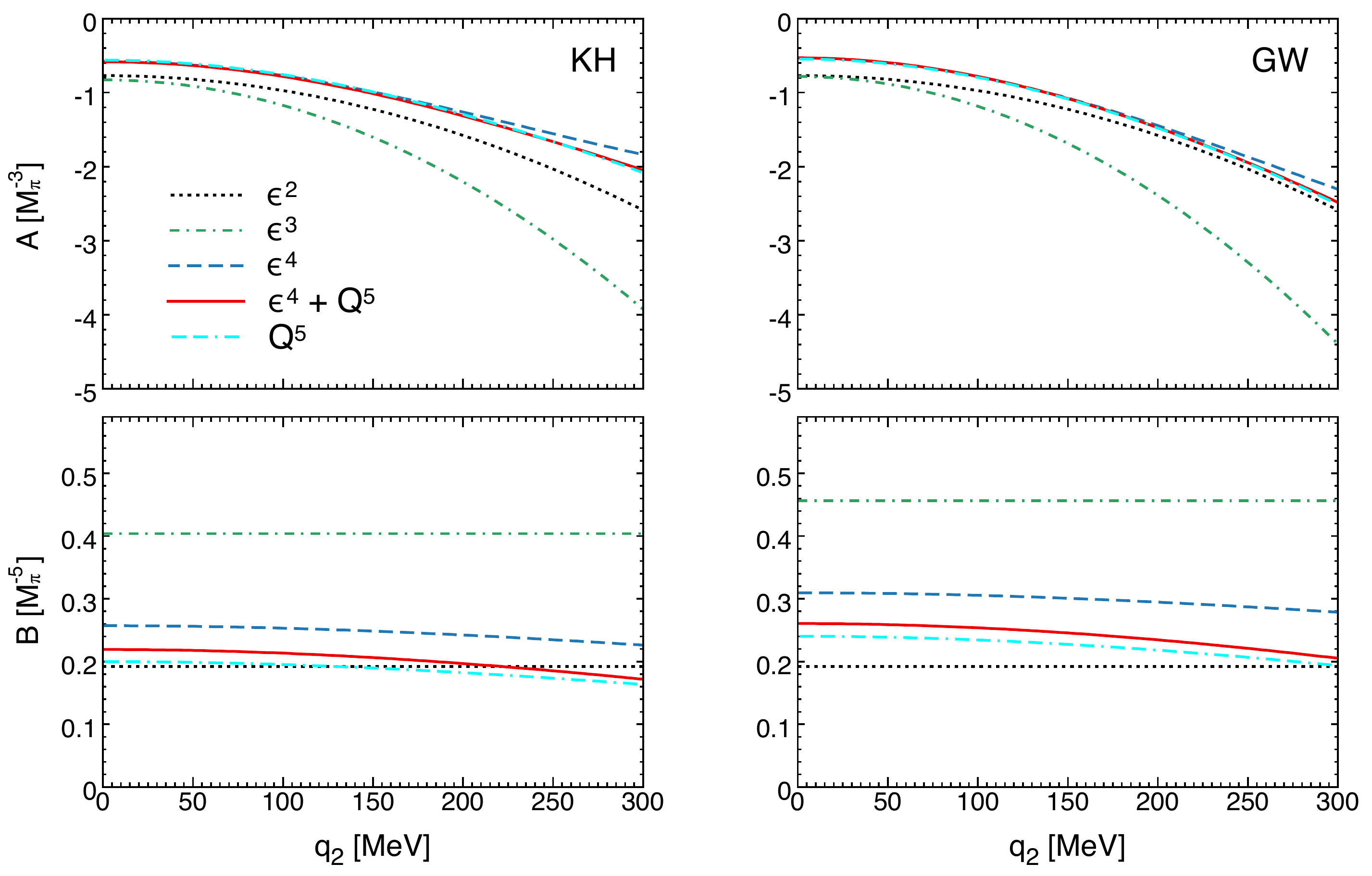}
    \caption{(Color online) Chiral expansion of the functions ${\cal A} (q_2)$ and ${\cal B} (q_2)$
entering the two-pion exchange 3NF in Eq.~(\ref{2pi_general}) in the $\Delta$-full and 
$\Delta$-less theories. Left (right)
panel shows the results obtained with the LECs determined from the fit
to the KH \cite{Koch:1985bn} (GW \cite{Arndt:2006bf}) partial wave
analysis of pion-nucleon scattering as explained in the text.  
\label{fig:AB} 
 }
\end{figure}

Having constructed explicitly the two-pion-exchange $\Delta$-full 
three-nucleon force and having determined all the relevant low-energy constants,
we are now in the position to analyze the convergence of chiral expansion 
for the long-range part of the 3NF.
In Fig.~\ref{fig:AB}, we show the results for the functions ${\cal A} (q_2)$ and ${\cal B} (q_2)$
for small values of the momentum transfer $q_2$, $q_2 < 300$ MeV 
at various orders in the small scale expansion. In addition to the $\epsilon^2$, $\epsilon^3$ and $\epsilon^4$
results we show also the results, where the purely nucleonic contributions of order $Q^5$ are added to the
$\epsilon^4$ result in order to compare it with the $\Delta$-less N$^4$LO calculation from Ref.~\cite{Krebs:2012yv}
(double-dashed dotted lines in Fig.~\ref{fig:AB}).
One should, however, keep in mind that this is  not a complete $\epsilon^5$ result. 
We use here  at all orders the low-energy constants  $c_i$, $\bar d_i$ and $\bar e_i$
determined from the order-$\epsilon^3+Q^4$ fit to the KH and GW partial wave analyses 
as described in section~\ref{sec:piN} and  listed in Table~\ref{table:parameters}.
We also adopt the same conventions regarding the LECs as in the case of pion-nucleon 
scattering, see  Eqs.~(\ref{ei_conv}) and (\ref{ga_conv}). Notice that
although some of the $\bar e_i$-constants ($\bar e_{15,16,18}$) are rather sensitive to a particular choice of the partial wave 
analysis in pion-nucleon scattering, see Table \ref{table:parameters},
the functions ${\cal A}$ and ${\cal B}$ depend only on the
LECs $\bar e_{14,17}$, which are quite stable.  

One observes a fairly slow convergence for the functions ${\cal A} (q_2)$ and ${\cal B} (q_2)$
when going from order $\epsilon^2$ to $\epsilon^4$. On the other hand, the difference
between the results at orders $\epsilon^4$ and $\epsilon^4+Q^5$  is
small for the function ${\cal B} (q_2)$ and almost negligible for the
function ${\cal A} (q_2)$, 
which may indicate that convergence is reached at this order.
Making a more definite statement about the convergence would, however,
require performing a complete $\epsilon^5$ derivation of the 3NF.

It is also comforting to see that the results at order $\epsilon^4+Q^5$ are very close 
to the $\Delta$-less calculation at order $Q^5$.  This indicates 
that the contributions of the $\Delta$-isobar to the two-pion exchange
3NF topology can be well represented in terms of  resonance saturation of the LECs 
$c_i$, $\bar d_i$ and $\bar e_i$ at N$^4$LO in the $\Delta$-less approach.
This also indicates that nucleonic terms at order $Q^6$ and higher saturated by the 
double and triple delta excitations are small.

\begin{figure}[tb]
\includegraphics[width=\textwidth,keepaspectratio,angle=0,clip]{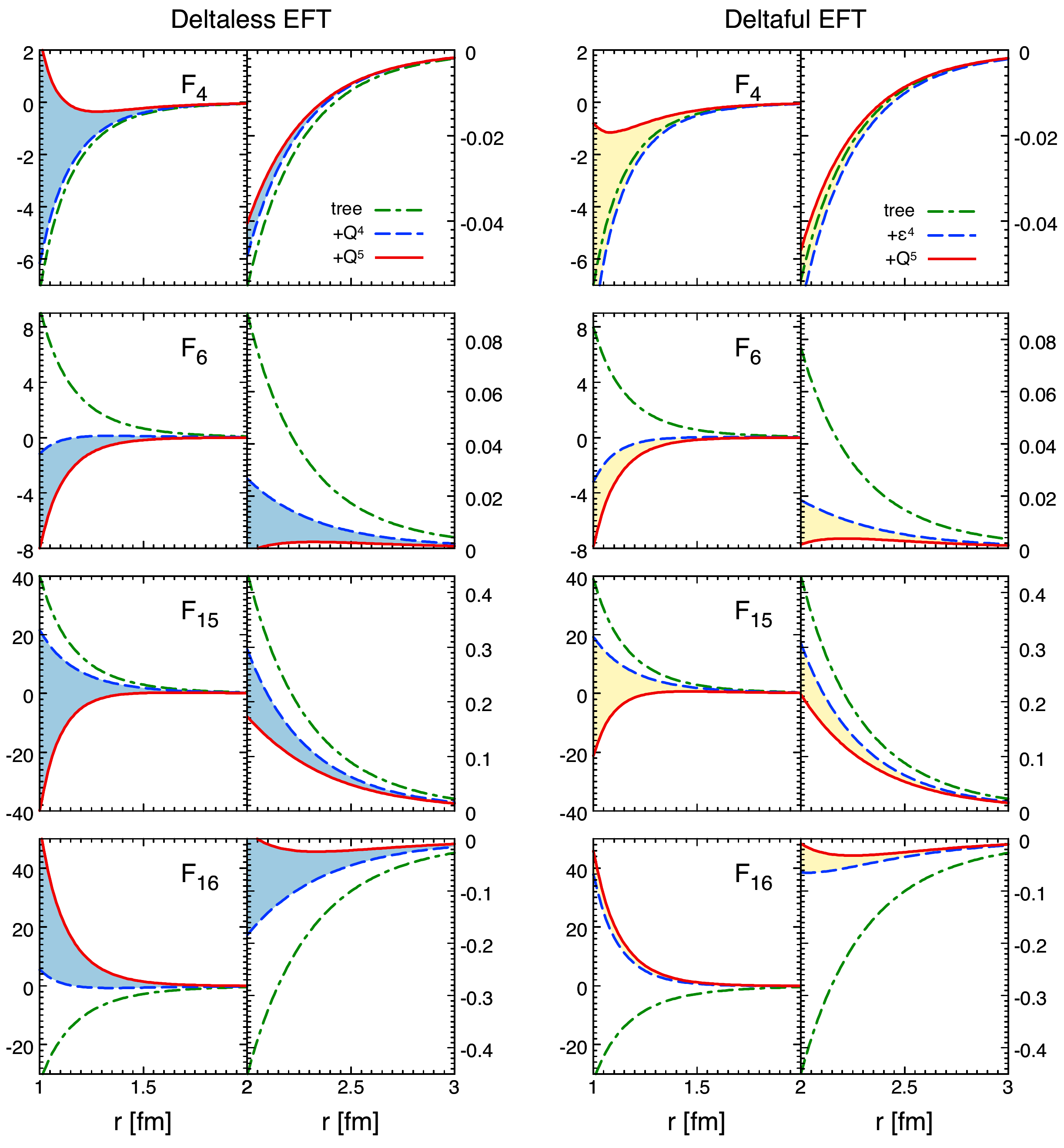}
\caption{(Color online) Profile functions $\mathcal{F}_{4}(r)$,
  $\mathcal{F}_{6}(r)$, $\mathcal{F}_{15}(r)$, $\mathcal{F}_{16}(r)$
  in units of MeV generated by the
two-pion exchange 3NF topology in the $\Delta$-less approach of Ref.~\cite{Krebs:2012yv} (left panel) 
and in the $\Delta$-full approach of the current work(right panel).
The dash-dotted, dashed and solid lines are the results of the calculation at order $Q^3$($\epsilon^3$),
$Q^4$($\epsilon^4$) and $Q^5$($\epsilon^4+Q^5$), respectively. The bands indicate the purely
nucleonic contribution at order $Q^5$.}
\label{fig:F4_6_15_16} 
\end{figure}

\begin{figure}[tb]
\includegraphics[width=\textwidth,keepaspectratio,angle=0,clip]{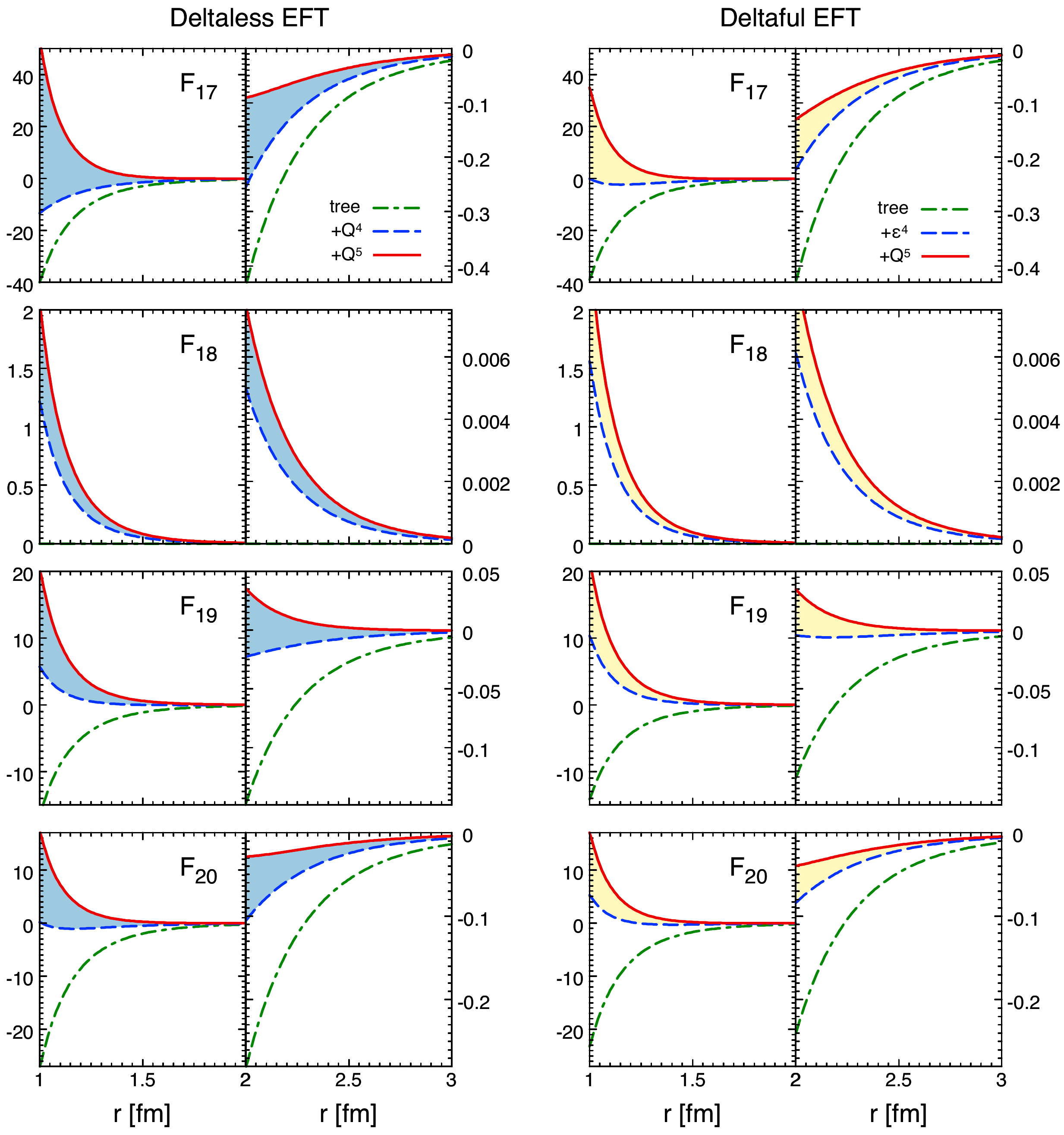}
\caption{(Color online) Profile functions $\mathcal{F}_{17}(r)$,
  $\mathcal{F}_{18}(r)$, $\mathcal{F}_{19}(r)$, $\mathcal{F}_{20}(r)$
  in units of MeV generated by the
two-pion exchange 3NF topology in the $\Delta$-less approach of Ref.~\cite{Krebs:2012yv} (left panel) 
and in the $\Delta$-full approach of the current work(right panel).
The dash-dotted, dashed and solid lines are the results of the calculation at order $Q^3$($\epsilon^3$),
$Q^4$($\epsilon^4$) and $Q^5$($\epsilon^4+Q^5$), respectively. The bands indicate the purely
nucleonic contribution at order $Q^5$.}
\label{fig:F17_18_19_20} 
\end{figure}

\begin{figure}[tb]
\includegraphics[width=\textwidth,keepaspectratio,angle=0,clip]{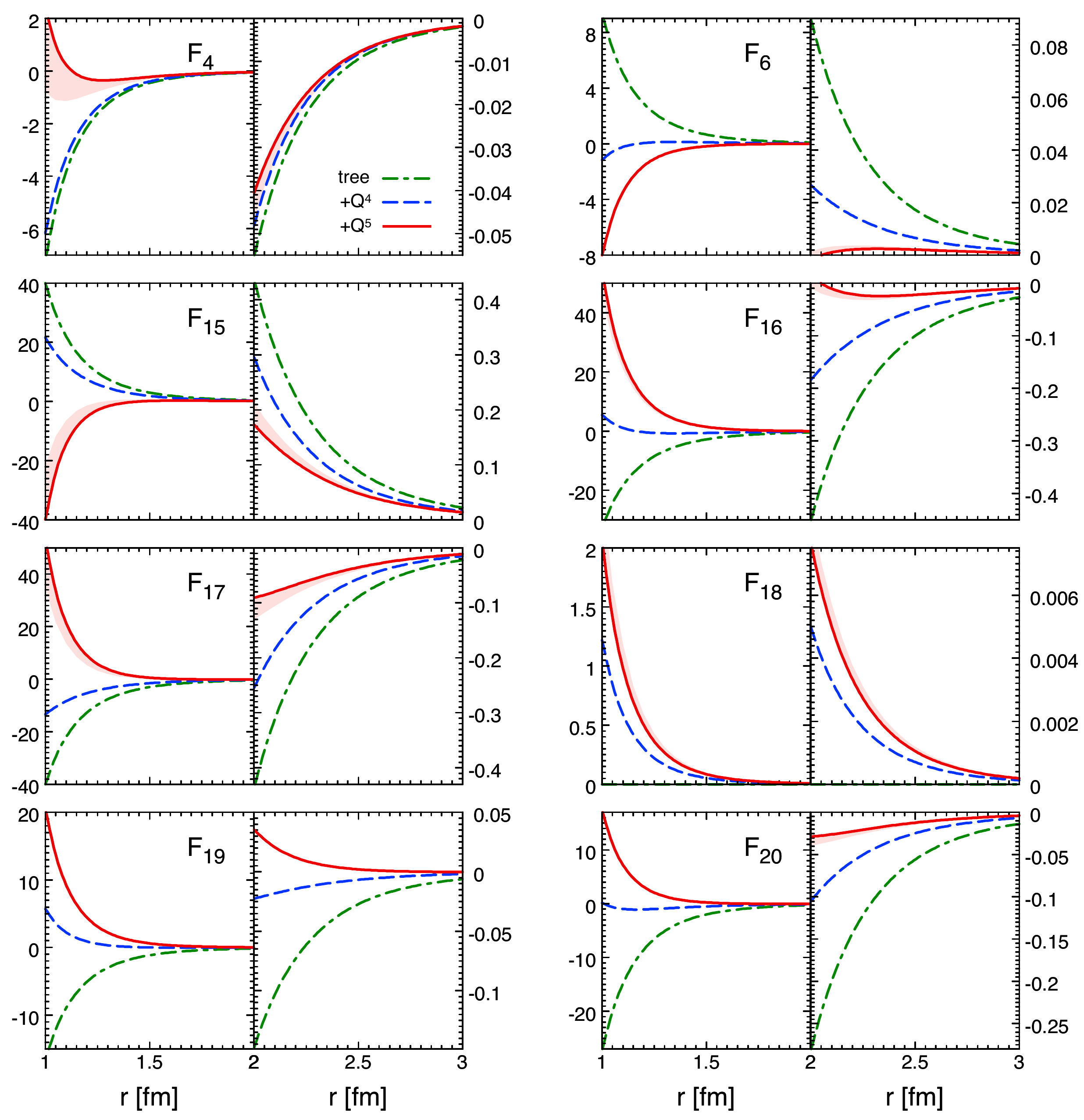}
\caption{(Color online) Profile functions $\mathcal{F}_i(r)$ in units
  of MeV generated by the
two-pion exchange 3NF topology in the $\Delta$-less approach of Ref.~\cite{Krebs:2012yv}.
The dash-dotted, dashed and solid lines are the results of the calculation at order $Q^3$,
$Q^4$ and $Q^5$, respectively.
The bands indicate the difference between the $\Delta$-less-$Q^5$ result and the 
$\Delta$-full result at order $\epsilon^4+Q^5$.}
\label{fig:DeltaBeyondSaturation} 
\end{figure}

Another instructive way to quantitatively analyze the obtained three-nucleon forces is to 
look at the structure functions ${\cal F}_i(r_{12},  r_{23}, r_{31})$ for the equilateral triangle configuration of the nucleons
given by the condition $r_{12}= r_{23}= r_{31} =r$ \cite{Krebs:2013kha}.
Structure functions ${\cal F}_i(r_{12},  r_{23}, r_{31})$ are the coefficients in the
expansion of a general local three-nucleon force in the basis of $20$ operators $\tilde {\cal G}_i$ (for their explicit form see Ref.~\cite{Epelbaum:2014sea}):
\beq
V_{\rm 3N}^{\rm full} = \sum_{i=1}^{20}\tilde {\cal G}_i{\cal F}_i (r_{12},  r_{23},
r_{31})+5\,{\rm permutations}.
\label{strf20}
\eeq
Only $8$ out of $20$ structure functions do not vanish for the
two-pion-exchange topology, namely  
${\cal F}_4$, ${\cal F}_6$, ${\cal F}_{15}$, ${\cal F}_{16}$, ${\cal
  F}_{17}$, ${\cal F}_{18}$, ${\cal F}_{19}$ and ${\cal F}_{20}$.
Our results for these structure functions are visualized in Figs.~\ref{fig:F4_6_15_16},~\ref{fig:F17_18_19_20},~\ref{fig:DeltaBeyondSaturation}.

We first comment on the convergence pattern of the chiral expansion for the
$\Delta$-less ($\Delta$-full) scheme, see Figs.~\ref{fig:F4_6_15_16},~\ref{fig:F17_18_19_20},
by looking at the tree-level results and at the results at orders $Q^4$ ($\epsilon^4$) and 
$Q^5$  ($\epsilon^4+Q^5$). 
We observe a better convergence of the $\Delta$-full approach, which is reflected
in significantly smaller bands on the right panels of
Figs.~\ref{fig:F4_6_15_16} and \ref{fig:F17_18_19_20}, which indicate
the size of  the purely nucleonic contributions at order $Q^5$. This
means that the large loop contributions at order $Q^5$ in the
$\Delta$-less theory reported in Refs.~\cite{Krebs:2012yv,Epelbaum:2014sea} are, to a considerable
extent, saturated  by the lower-order ($\epsilon^4$) contributions in the $\Delta$-full scheme. 
As the distance increases to $r\sim 2.5-3.0$ fm, the results at all orders 
get closer together for both the $\Delta$-less and $\Delta$-full
approaches fully in line with the general expectation that the chiral 
expansion converges most rapidly at large distances.

It is also instructive to compare with each other the results within
the $\Delta$-less and $\Delta$-full approaches at the highest
considered orders $Q^5$ and $\epsilon^4+Q^5$, respectively. As shown
in  Fig.~\ref{fig:DeltaBeyondSaturation}, the $\Delta$-isobar
contributions to the 
structure functions
${\cal F}_{6}$, ${\cal F}_{16}$, ${\cal F}_{18}$, ${\cal F}_{19}$ and
${\cal F}_{20}$ are almost
completely given by the  resonance saturation of the corresponding
LECs. Indeed, the bands indicating  the difference between the
order-$Q^5$ $\Delta$-less and order-$(\epsilon^4+Q^5)$ $\Delta$-full
results are almost invisible in those cases even at relatively short
distances of $r \sim 1.0 - 1.5$~fm. For the functions ${\cal F}_{15}$ and ${\cal F}_{17}$ 
the saturation at this chiral order explains only a part of the
$\Delta$-contributions.  
For the ${\cal F}_{4}$-function,  the $\Delta$-less and $\Delta$-full
results turn out to be of a different sign at short distances.  
Notice, however, that the corresponding structure function is rather
small in magnitude as compared to other ones.
For larger distances of $r \sim 2.5-3.0$~fm, the saturation pattern
improves and holds true for  
all structure functions. This means that 
$\Delta$-resonance saturation of the $\Delta$-less contributions
at orders beyond $Q^5$, emerging from the considered diagrams at order
$\epsilon^4 + Q^5$, leads to small effects for the two-pion exchange
3NF topology. This may be considered as
yet another indication of the convergence of the theory at orders
$Q^5$ and   $\epsilon^4 + Q^5$. 

\section{Summary and conclusions}
\def\theequation{\arabic{section}.\arabic{equation}}
\label{sec:summary}

In this paper, we have analyzed the longest-range contribution to the
three-nucleon force at N$^3$LO utilizing the heavy-baryon formulation of
chiral EFT with pions,  nucleons and deltas as the only explicit degrees of
freedom. The pertinent results of our study can be summarized as
follows.
\begin{itemize} 
\item 
We worked out in detail renormalization of the lowest-order effective
chiral Lagrangian at the one-loop level. 
\item
Employing renormalization conditions which maintain the explicit
decoupling of the $\Delta$ isobar, we derived the delta contributions
to those LECs $c_i$, $\bar d_i$ and $\bar e_i$ from the effective
Lagrangians $\mathcal{L}_{\pi N}^{(2)}$,    $\mathcal{L}_{\pi
  N}^{(3)}$ and  $\mathcal{L}_{\pi N}^{(4)}$ which contribute to
pion-nucleon scattering at the considered order.  
\item
In order to determine the LECs $c_i$, $\bar d_i$ and $\bar e_i$ 
contributing to the $2\pi$-exchange 3NF, we re-analyzed 
pion-nucleon scattering at order $\epsilon^3 + Q^4$ employing the same power
counting scheme as in the derivation of the nuclear forces and using
the same fitting protocol as in the $\Delta$-less analysis of Ref.~\cite{Krebs:2012yv}. We used
the available partial wave analyses of the pion-nucleon scattering
data to determine all relevant LECs. The resulting values turn out to
be rather stable and consistent with our $\Delta$-less
analysis reported in Ref.~\cite{Krebs:2012yv}.  
\item
We worked out the N$^3$LO delta contributions to the $2\pi$-exchange 3NF. The
unitary ambiguity of the Hamilton operator is parametrized by $50$ 
additional unitary transformations. After imposing the
renormalizability constraint, i.e. the requirement  that the resulting
3NF matrix elements are finite, the expressions for the 3NF appear to
be defined unambiguously. These
findings pave the way for the derivation of the remaining 3NF contributions
at the same order, which are not considered in this paper. 
\item
The obtained results for the $2\pi$-exchange 3NF at N$^3$LO
of the SSE are in a good agreement with the N$^4$LO calculations
of Ref.~\cite{Krebs:2012yv} within the $\Delta$-less approach. The
agreement becomes even better when adding the nucleonic contributions 
at order N$^4$LO to the expressions at N$^3$LO of the SSE. This
indicates that the effects of the $\Delta$ isobar for this particular
topology are well represented by resonance saturation of the LECs 
$c_i$, $\bar d_i$ and $\bar e_i$ at N$^4$LO in the $\Delta$-less
approach.      
\end{itemize}
The presented calculations should be extended to the
intermediate-range topologies, where we do expect
significant contributions of the $\Delta$ to be still missing in the N$^4$LO analysis of
Ref.~\cite{Krebs:2013kha} within the $\Delta$-less framework, as well as to short-range
contributions. Work along these lines is in progress.

\section*{Acknowledgments}

We are grateful to Ulf-G.~Mei{\ss}ner for sharing his insights into the
considered topics and  useful discussions. 
This work was supported by BMBF (contract No.~05P2015 - NUSTAR R\&D)
and by DFG through funds provided to the Sino-German CRC 110
``Symmetries and the Emergence of Structure in QCD'' (Grant No.~TRR110).

\appendix

\def\theequation{\Alph{section}.\arabic{equation}}
\setcounter{equation}{0}
\section{Unitary ambiguity of the 3NF and constraints imposed by the
  renormalizability requirement}
\label{app1}

To derive the effective potential we use the method of unitary transformation. A detailed
discussion of this approach including the explicit form of the 
unitary operator at low orders in the chiral expansion can be 
found in Ref.~\cite{Epelbaum:2007us}. As explained in section
\ref{sec:TPE}, this method can be straightforwardly extended to carry
out calculations within  the $\Delta$-full chiral EFT approach. In this
appendix we discuss the restrictions on the choice of the unitary
transformation imposed by the condition that the resulting nuclear
Hamiltonian is renormalizable. 

We first specify our notation.  The
indices in the interaction vertices $H_{a,b,c,d,e}^{(\kappa)}$
have the following meaning:
\beqa
a&=&{\rm Number\, of\,pion\,fields}\nn
b&=&{\rm Number\,of\,outgoing \,nucleons}\nn
c&=&{\rm Number\,of\,outgoing \,deltas}\nn
d&=&{\rm Number\,of\,incoming \,nucleons}\nn
e&=&{\rm Number\,of\,incoming \,deltas}\nn
\kappa&=&d + \frac{3}{2}(b+c+d+e)+a - 4,
\eeqa
where $d$ is the number of derivatives at a given  vertex. We also
introduce the projection operators $\eta$ and $\lambda$  onto the purely nucleonic and the
remaining parts of the Fock space, respectively.  These operators
satisfy the usual relations $\eta^2 = \eta$, $\lambda^2 = \lambda$, $ \eta \lambda 
= \lambda \eta = 0$ and $\lambda + \eta = {\bf 1}$. We also need to 
differentiate the states from the $\lambda$-subspace by introducing
the operators  $\lambda^{a,b}$, where $a$ and $b$ refer to  the number of pions and deltas  in the
corresponding intermediate state, respectively. The total energy of the 
pions and deltas in the corresponding state will be denoted by
$E_{\pi\Delta} = \mathcal{O}(\epsilon )$ .   

As pointed out in section \ref{sec:TPE}, renormalizability of the
nuclear Hamiltonian is achieved by performing all possible
$\eta$-space unitary transformations after decoupling of pions and
deltas by means of the (minimal) Okubo-type unitary
transformation. Such additional unitary operators have a general form
\beqa
U=e^{S},
\eeqa 
where $S$ is an antihermitian operator ($S^\dagger=-S$) acting on the $\eta$-space. We
parametrize the operator $S$ as
\beqa
S&=&\sum_{i}\alpha_i S_i + \sum_{i}\alpha_i^\Delta S_i^\Delta\,,
\eeqa
where $\alpha_i$ and $\alpha_i^\Delta$ are real numbers.  
The operators $S_i$ include only nucleon degrees of freedom and have
already been discussed earlier~\cite{Epelbaum:2007us,Krebs:2012yv}.
We now give $50$ operators $S_i^\Delta$ which include
delta contribution:
\beqa
S_{1}^\Delta&=& \eta  H_{1,1,0,0,1}^{(1)} 
\frac{\lambda^{{1,1}} }{{E_{\pi\Delta}}^2} H_{1,1,0,1,0}^{(1)}  
\frac{\lambda^{{0,1}} }{{E_{\pi\Delta}}} H_{1,1,0,1,0}^{(1)}  
\frac{\lambda^{{1,1}} }{{E_{\pi\Delta}}} H_{1,0,1,1,0}^{(1)} \eta -{hc},\nn 
S_{2}^\Delta&=& \eta  H_{1,1,0,0,1}^{(1)}  
\frac{\lambda^{{1,1}} }{{E_{\pi\Delta}}^2} H_{1,1,0,1,0}^{(1)} 
\frac{\lambda^{{2,1}} }{{E_{\pi\Delta}}} H_{1,1,0,1,0}^{(1)} 
\frac{\lambda^{{1,1}} }{{E_{\pi\Delta}}} H_{1,0,1,1,0}^{(1)} \eta -{hc},\nn 
S_{3}^\Delta&=& \eta  H_{1,1,0,1,0}^{(1)} 
\frac{\lambda^{{1,0}} }{{E_{\pi\Delta}}^2} H_{1,1,0,0,1}^{(1)}  
\frac{\lambda^{{0,1}} }{{E_{\pi\Delta}}} H_{1,0,1,1,0}^{(1)} 
\frac{\lambda^{{1,0}} }{{E_{\pi\Delta}}} H_{1,1,0,1,0}^{(1)} \eta -{hc},\nn 
S_{4}^\Delta&=& \eta  H_{1,1,0,1,0}^{(1)} 
\frac{\lambda^{{1,0}} }{{E_{\pi\Delta}}^2} H_{1,1,0,0,1}^{(1)} 
\frac{\lambda^{{2,1}} }{{E_{\pi\Delta}}} H_{1,0,1,1,0}^{(1)} 
\frac{\lambda^{{1,0}} }{{E_{\pi\Delta}}} H_{1,1,0,1,0}^{(1)} \eta -{hc},\nn 
S_{5}^\Delta&=& \eta  H_{1,1,0,0,1}^{(1)} 
\frac{\lambda^{{1,1}} }{{E_{\pi\Delta}}^2} H_{1,1,0,1,0}^{(1)} 
\frac{\lambda^{{0,1}} }{{E_{\pi\Delta}}} H_{1,0,1,1,0}^{(1)} 
\frac{\lambda^{{1,0}} }{{E_{\pi\Delta}}} H_{1,1,0,1,0}^{(1)} \eta -{hc},\nn 
S_{6}^\Delta&=& \eta  H_{1,1,0,0,1}^{(1)} 
\frac{\lambda^{{1,1}} }{{E_{\pi\Delta}}} H_{1,1,0,1,0}^{(1)} 
\frac{\lambda^{{0,1}} }{{E_{\pi\Delta}}^2} H_{1,0,1,1,0}^{(1)} 
\frac{\lambda^{{1,0}} }{{E_{\pi\Delta}}} H_{1,1,0,1,0}^{(1)} \eta -{hc},\nn 
S_{7}^\Delta&=& \eta  H_{1,1,0,0,1}^{(1)}
\frac{\lambda^{{1,1}} }{{E_{\pi\Delta}}} H_{1,1,0,1,0}^{(1)} 
\frac{\lambda^{{0,1}} }{{E_{\pi\Delta}}} H_{1,0,1,1,0}^{(1)} 
\frac{\lambda^{{1,0}} }{{E_{\pi\Delta}}^2} H_{1,1,0,1,0}^{(1)} \eta -{hc},\nn 
S_{8}^\Delta&=& \eta  H_{1,1,0,0,1}^{(1)} 
\frac{\lambda^{{1,1}} }{{E_{\pi\Delta}}^2} H_{1,1,0,1,0}^{(1)} 
\frac{\lambda^{{2,1}} }{{E_{\pi\Delta}}} H_{1,0,1,1,0}^{(1)} 
\frac{\lambda^{{1,0}} }{{E_{\pi\Delta}}} H_{1,1,0,1,0}^{(1)} \eta -{hc},\nn 
S_{9}^\Delta&=& \eta  H_{1,1,0,0,1}^{(1)} 
\frac{\lambda^{{1,1}} }{{E_{\pi\Delta}}} H_{1,1,0,1,0}^{(1)} 
\frac{\lambda^{{2,1}} }{{E_{\pi\Delta}}^2} H_{1,0,1,1,0}^{(1)} 
\frac{\lambda^{{1,0}} }{{E_{\pi\Delta}}} H_{1,1,0,1,0}^{(1)} \eta -{hc},\nn 
S_{10}^\Delta&=& \eta  H_{1,1,0,0,1}^{(1)} 
\frac{\lambda^{{1,1}} }{{E_{\pi\Delta}}} H_{1,1,0,1,0}^{(1)} 
\frac{\lambda^{{2,1}} }{{E_{\pi\Delta}}} H_{1,0,1,1,0}^{(1)} 
\frac{\lambda^{{1,0}} }{{E_{\pi\Delta}}^2} H_{1,1,0,1,0}^{(1)} \eta -{hc},\nn
S_{11}^\Delta&=& \eta  H_{1,1,0,0,1}^{(1)} 
\frac{\lambda^{{1,1}} }{{E_{\pi\Delta}}^2} H_{1,0,1,1,0}^{(1)} 
\frac{\lambda^{{2,0}} }{{E_{\pi\Delta}}} H_{1,1,0,1,0}^{(1)} 
\frac{\lambda^{{1,0}} }{{E_{\pi\Delta}}} H_{1,1,0,1,0}^{(1)} \eta -{hc},\nn 
S_{12}^\Delta&=& \eta  H_{1,1,0,0,1}^{(1)}  
\frac{\lambda^{{1,1}} }{{E_{\pi\Delta}}} H_{1,0,1,1,0}^{(1)} 
\frac{\lambda^{{2,0}} }{{E_{\pi\Delta}}^2} H_{1,1,0,1,0}^{(1)} 
\frac{\lambda^{{1,0}} }{{E_{\pi\Delta}}} H_{1,1,0,1,0}^{(1)} \eta -{hc},\nn 
S_{13}^\Delta&=& \eta  H_{1,1,0,0,1}^{(1)} 
\frac{\lambda^{{1,1}} }{{E_{\pi\Delta}}} H_{1,0,1,1,0}^{(1)} 
\frac{\lambda^{{2,0}} }{{E_{\pi\Delta}}} H_{1,1,0,1,0}^{(1)} 
\frac{\lambda^{{1,0}} }{{E_{\pi\Delta}}^2} H_{1,1,0,1,0}^{(1)} \eta -{hc},\nn 
S_{14}^\Delta&=& \eta  H_{1,1,0,0,1}^{(1)} 
\frac{\lambda^{{1,1}} }{{E_{\pi\Delta}}^2} H_{1,0,1,0,1}^{(1)} 
\frac{\lambda^{{0,1}} }{{E_{\pi\Delta}}} H_{1,1,0,1,0}^{(1)} 
\frac{\lambda^{{1,1}} }{{E_{\pi\Delta}}} H_{1,0,1,1,0}^{(1)} \eta -{hc},\nn 
S_{15}^\Delta&=& \eta  H_{1,1,0,0,1}^{(1)} 
\frac{\lambda^{{1,1}} }{{E_{\pi\Delta}}^2} H_{1,1,0,1,0}^{(1)} 
\frac{\lambda^{{0,1}} }{{E_{\pi\Delta}}} H_{1,0,1,0,1}^{(1)} 
\frac{\lambda^{{1,1}} }{{E_{\pi\Delta}}} H_{1,0,1,1,0}^{(1)} \eta -{hc},\nn 
S_{16}^\Delta&=& \eta  H_{1,1,0,0,1}^{(1)} 
\frac{\lambda^{{1,1}} }{{E_{\pi\Delta}}^2} H_{1,0,1,0,1}^{(1)} 
\frac{\lambda^{{0,1}} }{{E_{\pi\Delta}}} H_{1,0,1,0,1}^{(1)} 
\frac{\lambda^{{1,1}} }{{E_{\pi\Delta}}} H_{1,0,1,1,0}^{(1)} \eta -{hc},\nn 
S_{17}^\Delta&=& \eta  H_{1,1,0,0,1}^{(1)} 
\frac{\lambda^{{1,1}} }{{E_{\pi\Delta}}} H_{1,0,1,0,1}^{(1)} 
\frac{\lambda^{{0,1}} }{{E_{\pi\Delta}}^2} H_{1,1,0,1,0}^{(1)} 
\frac{\lambda^{{1,1}} }{{E_{\pi\Delta}}} H_{1,0,1,1,0}^{(1)} \eta -{hc},\nn 
S_{18}^\Delta&=& \eta  H_{1,1,0,0,1}^{(1)} 
\frac{\lambda^{{1,1}} }{{E_{\pi\Delta}}^2} H_{1,0,1,0,1}^{(1)} 
\frac{\lambda^{{2,1}} }{{E_{\pi\Delta}}} H_{1,1,0,1,0}^{(1)} 
\frac{\lambda^{{1,1}} }{{E_{\pi\Delta}}} H_{1,0,1,1,0}^{(1)} \eta -{hc},\nn 
S_{19}^\Delta&=& \eta  H_{1,1,0,0,1}^{(1)} 
\frac{\lambda^{{1,1}} }{{E_{\pi\Delta}}^2} H_{1,1,0,1,0}^{(1)} 
\frac{\lambda^{{2,1}} }{{E_{\pi\Delta}}} H_{1,0,1,0,1}^{(1)} 
\frac{\lambda^{{1,1}} }{{E_{\pi\Delta}}} H_{1,0,1,1,0}^{(1)} \eta -{hc},\nn 
S_{20}^\Delta&=& \eta  H_{1,1,0,0,1}^{(1)} 
\frac{\lambda^{{1,1}} }{{E_{\pi\Delta}}^2} H_{1,0,1,0,1}^{(1)} 
\frac{\lambda^{{2,1}} }{{E_{\pi\Delta}}} H_{1,0,1,0,1}^{(1)} 
\frac{\lambda^{{1,1}} }{{E_{\pi\Delta}}} H_{1,0,1,1,0}^{(1)} \eta -{hc},\nn
S_{21}^\Delta&=& \eta  H_{1,1,0,0,1}^{(1)} 
\frac{\lambda^{{1,1}} }{{E_{\pi\Delta}}} H_{1,0,1,0,1}^{(1)} 
\frac{\lambda^{{2,1}} }{{E_{\pi\Delta}}^2} H_{1,1,0,1,0}^{(1)} 
\frac{\lambda^{{1,1}} }{{E_{\pi\Delta}}} H_{1,0,1,1,0}^{(1)} \eta -{hc},\nn 
S_{22}^\Delta&=& \eta  H_{1,1,0,0,1}^{(1)} 
\frac{\lambda^{{1,1}} }{{E_{\pi\Delta}}^2} H_{1,0,1,0,1}^{(1)} 
\frac{\lambda^{{0,1}} }{{E_{\pi\Delta}}} H_{1,0,1,1,0}^{(1)} 
\frac{\lambda^{{1,0}} }{{E_{\pi\Delta}}} H_{1,1,0,1,0}^{(1)} \eta -{hc},\nn 
S_{23}^\Delta&=& \eta  H_{1,1,0,0,1}^{(1)}
\frac{\lambda^{{1,1}} }{{E_{\pi\Delta}}} H_{1,0,1,0,1}^{(1)} 
\frac{\lambda^{{0,1}} }{{E_{\pi\Delta}}^2} H_{1,0,1,1,0}^{(1)} 
\frac{\lambda^{{1,0}} }{{E_{\pi\Delta}}} H_{1,1,0,1,0}^{(1)} \eta -{hc},\nn 
S_{24}^\Delta&=& \eta  H_{1,1,0,0,1}^{(1)} 
\frac{\lambda^{{1,1}} }{{E_{\pi\Delta}}} H_{1,0,1,0,1}^{(1)} 
\frac{\lambda^{{0,1}} }{{E_{\pi\Delta}}} H_{1,0,1,1,0}^{(1)} 
\frac{\lambda^{{1,0}} }{{E_{\pi\Delta}}^2} H_{1,1,0,1,0}^{(1)} \eta -{hc},\nn 
S_{25}^\Delta&=& \eta  H_{1,1,0,0,1}^{(1)} 
\frac{\lambda^{{1,1}} }{{E_{\pi\Delta}}^2} H_{1,0,1,0,1}^{(1)} 
\frac{\lambda^{{2,1}} }{{E_{\pi\Delta}}} H_{1,0,1,1,0}^{(1)} 
\frac{\lambda^{{1,0}} }{{E_{\pi\Delta}}} H_{1,1,0,1,0}^{(1)} \eta -{hc},\nn 
S_{26}^\Delta&=& \eta  H_{1,1,0,0,1}^{(1)} 
\frac{\lambda^{{1,1}} }{{E_{\pi\Delta}}} H_{1,0,1,0,1}^{(1)} 
\frac{\lambda^{{2,1}} }{{E_{\pi\Delta}}^2} H_{1,0,1,1,0}^{(1)}  
\frac{\lambda^{{1,0}} }{{E_{\pi\Delta}}} H_{1,1,0,1,0}^{(1)} \eta -{hc},\nn 
S_{27}^\Delta&=& \eta  H_{1,1,0,0,1}^{(1)} 
\frac{\lambda^{{1,1}} }{{E_{\pi\Delta}}} H_{1,0,1,0,1}^{(1)} 
\frac{\lambda^{{2,1}} }{{E_{\pi\Delta}}} H_{1,0,1,1,0}^{(1)} 
\frac{\lambda^{{1,0}} }{{E_{\pi\Delta}}^2} H_{1,1,0,1,0}^{(1)} \eta -{hc},\nn 
S_{28}^\Delta&=& \eta  H_{1,1,0,0,1}^{(1)}
\frac{\lambda^{{1,1}} }{{E_{\pi\Delta}}^3} H_{1,0,1,1,0}^{(1)} \eta  
H_{1,1,0,1,0}^{(1)} \frac{\lambda^{{1,0}} }{{E_{\pi\Delta}}} 
H_{1,1,0,1,0}^{(1)} \eta -{hc},\nn 
S_{29}^\Delta&=& \eta  
H_{1,1,0,0,1}^{(1)}  \frac{\lambda^{{1,1}} }{{E_{\pi\Delta}}^2} 
H_{1,0,1,1,0}^{(1)} \eta  H_{1,1,0,1,0}^{(1)} 
\frac{\lambda^{{1,0}} }{{E_{\pi\Delta}}^2} H_{1,1,0,1,0}^{(1)} \eta -{hc},\nn 
S_{30}^\Delta&=& \eta  H_{1,1,0,0,1}^{(1)}  
\frac{\lambda^{{1,1}} }{{E_{\pi\Delta}}} H_{1,0,1,1,0}^{(1)} \eta  
H_{1,1,0,1,0}^{(1)}  \frac{\lambda^{{1,0}} }{{E_{\pi\Delta}}^3} 
H_{1,1,0,1,0}^{(1)} \eta -{hc},\nn
S_{31}^\Delta&=& \eta  
H_{1,1,0,0,1}^{(1)}  \frac{\lambda^{{1,1}} }{{E_{\pi\Delta}}^3} 
H_{1,0,1,1,0}^{(1)} \eta  H_{1,1,0,0,1}^{(1)}  
\frac{\lambda^{{1,1}} }{{E_{\pi\Delta}}} H_{1,0,1,1,0}^{(1)} \eta -{hc},\nn 
S_{32}^\Delta&=& \eta  H_{2,1,0,1,0}^{(2)}  
\frac{\lambda^{{2,0}} }{{E_{\pi\Delta}}^2} H_{1,1,0,0,1}^{(1)}  
\frac{\lambda^{{1,1}} }{{E_{\pi\Delta}}} H_{1,0,1,1,0}^{(1)} \eta -{hc},\nn 
S_{33}^\Delta&=& \eta  H_{2,1,0,1,0}^{(2)}  
\frac{\lambda^{{2,0}} }{{E_{\pi\Delta}}} H_{1,1,0,0,1}^{(1)}  
\frac{\lambda^{{1,1}} }{{E_{\pi\Delta}}^2} H_{1,0,1,1,0}^{(1)} \eta -{hc},\nn 
S_{34}^\Delta&=& \eta  H_{1,1,0,0,1}^{(1)}  
\frac{\lambda^{{1,1}} }{{E_{\pi\Delta}}^2} H_{2,1,0,1,0}^{(2)}
\frac{\lambda^{{1,1}} }{{E_{\pi\Delta}}} H_{1,0,1,1,0}^{(1)} \eta -{hc},\nn 
S_{35}^\Delta&=& \eta  H_{2,1,0,0,1}^{(2)}  
\frac{\lambda^{{2,1}} }{{E_{\pi\Delta}}^2} H_{1,0,1,1,0}^{(1)}  
\frac{\lambda^{{1,0}} }{{E_{\pi\Delta}}} H_{1,1,0,1,0}^{(1)} \eta -{hc},\nn 
S_{36}^\Delta&=& \eta  H_{2,1,0,0,1}^{(2)} 
\frac{\lambda^{{2,1}} }{{E_{\pi\Delta}}} H_{1,0,1,1,0}^{(1)}  
\frac{\lambda^{{1,0}} }{{E_{\pi\Delta}}^2} H_{1,1,0,1,0}^{(1)} \eta -{hc},\nn 
S_{37}^\Delta&=& \eta  H_{2,1,0,0,1}^{(2)}  
\frac{\lambda^{{2,1}} }{{E_{\pi\Delta}}^2} H_{1,1,0,1,0}^{(1)}  
\frac{\lambda^{{1,1}} }{{E_{\pi\Delta}}} H_{1,0,1,1,0}^{(1)} \eta -{hc},\nn 
S_{38}^\Delta&=& \eta  H_{2,1,0,0,1}^{(2)} 
\frac{\lambda^{{2,1}} }{{E_{\pi\Delta}}} H_{1,1,0,1,0}^{(1)} 
\frac{\lambda^{{1,1}} }{{E_{\pi\Delta}}^2} H_{1,0,1,1,0}^{(1)} \eta -{hc},\nn 
S_{39}^\Delta&=& \eta  H_{2,1,0,0,1}^{(2)}  
\frac{\lambda^{{2,1}} }{{E_{\pi\Delta}}^2} H_{1,0,1,0,1}^{(1)}  
\frac{\lambda^{{1,1}} }{{E_{\pi\Delta}}} H_{1,0,1,1,0}^{(1)} \eta -{hc},\nn 
S_{40}^\Delta&=& \eta  H_{2,1,0,0,1}^{(2)}  
\frac{\lambda^{{2,1}} }{{E_{\pi\Delta}}} H_{1,0,1,0,1}^{(1)} 
\frac{\lambda^{{1,1}} }{{E_{\pi\Delta}}^2} H_{1,0,1,1,0}^{(1)} \eta -{hc},\nn
S_{41}^\Delta&=& \eta  H_{1,1,0,1,0}^{(1)}  
\frac{\lambda^{{1,0}} }{{E_{\pi\Delta}}^2} H_{2,1,0,0,1}^{(2)}  
\frac{\lambda^{{1,1}} }{{E_{\pi\Delta}}} H_{1,0,1,1,0}^{(1)} \eta -{hc},\nn 
S_{42}^\Delta&=& \eta  H_{1,1,0,1,0}^{(1)}  
\frac{\lambda^{{1,0}} }{{E_{\pi\Delta}}} H_{2,1,0,0,1}^{(2)} 
\frac{\lambda^{{1,1}} }{{E_{\pi\Delta}}^2} H_{1,0,1,1,0}^{(1)} \eta -{hc},\nn 
S_{43}^\Delta&=& \eta  H_{1,1,0,0,1}^{(1)} 
\frac{\lambda^{{1,1}} }{{E_{\pi\Delta}}^2} H_{2,0,1,0,1}^{(2)} 
\frac{\lambda^{{1,1}} }{{E_{\pi\Delta}}} H_{1,0,1,1,0}^{(1)} \eta -{hc},\nn 
S_{44}^\Delta&=& \eta  H_{1,1,0,0,1}^{(1)} 
\frac{\lambda^{{1,1}} }{{E_{\pi\Delta}}^2} H_{1,1,0,0,1}^{(1)} 
\frac{\lambda^{{2,2}} }{{E_{\pi\Delta}}} H_{1,0,1,1,0}^{(1)}
\frac{\lambda^{{1,1}} }{{E_{\pi\Delta}}} H_{1,0,1,1,0}^{(1)} \eta -{hc},\nn 
S_{45}^\Delta&=& \eta  H_{1,1,0,0,1}^{(1)} 
\frac{\lambda^{{1,1}} }{{E_{\pi\Delta}}^2} H_{1,1,0,0,1}^{(1)} 
\frac{\lambda^{{0,2}} }{{E_{\pi\Delta}}} H_{1,0,1,1,0}^{(1)} 
\frac{\lambda^{{1,1}} }{{E_{\pi\Delta}}} H_{1,0,1,1,0}^{(1)} \eta -{hc},\nn 
S_{46}^\Delta&=& \eta  H_{1,1,0,0,1}^{(1)} 
\frac{\lambda^{{1,1}} }{{E_{\pi\Delta}}^2} H_{1,0,1,1,0}^{(1)} 
\frac{\lambda^{{2,0}} }{{E_{\pi\Delta}}} H_{1,1,0,0,1}^{(1)}  
\frac{\lambda^{{1,1}} }{{E_{\pi\Delta}}} H_{1,0,1,1,0}^{(1)} \eta -{hc},\nn 
S_{47}^\Delta&=& \eta  H_{0,2,0,0,2}^{(2)} 
\frac{\lambda^{{0,2}} }{{E_{\pi\Delta}}^2} H_{1,0,1,1,0}^{(1)} 
\frac{\lambda^{{1,1}} }{{E_{\pi\Delta}}} H_{1,0,1,1,0}^{(1)} \eta -{hc},\nn 
S_{48}^\Delta&=& \eta  H_{0,2,0,0,2}^{(2)}  
\frac{\lambda^{{0,2}} }{{E_{\pi\Delta}}} H_{1,0,1,1,0}^{(1)}  
\frac{\lambda^{{1,1}} }{{E_{\pi\Delta}}^2} H_{1,0,1,1,0}^{(1)} \eta -{hc},\nn 
S_{49}^\Delta&=& \eta  H_{0,2,0,2,0}^{(2)} \eta
H_{1,1,0,0,1}^{(1)} \lambda^{{1,1}} \frac{1}{{E_{\pi\Delta}}^3} 
H_{1,0,1,1,0}^{(1)} \eta -{hc},\nn 
S_{50}^\Delta&=& \eta  
H_{1,1,0,0,1}^{(1)} \frac{\lambda^{{1,1}} }{{E_{\pi\Delta}}^2} 
H_{0,1,1,1,1}^{(2)}  \frac{\lambda^{{1,1}} }{{E_{\pi\Delta}}} 
H_{1,0,1,1,0}^{(1)} \eta-hc.
\label{generators}
\eeqa
The requirement that the delta contributions to the 3NF are renormalizable
leads to the  following constraints on the
coefficients $\alpha_i^\Delta$:
\beqa
\alpha_{7}^\Delta-\alpha_{13}^\Delta&=&0,\nn 
-\alpha_{13}^\Delta+
\alpha_{8}^\Delta+\frac{1}{2}&=&0, \nn
\alpha_{9}^\Delta&=&0, \nn
\alpha_{10}^\Delta-\alpha_{13}^\Delta&=&0,\nn 
\alpha_{13}^\Delta+2 \alpha_{29}^\Delta-\frac{1}{4}&=&0, \nn
\alpha_{18}^\Delta-\alpha_{25}^\Delta&=&0, \nn
\alpha_{21}^\Delta-
\alpha_{26}^\Delta&=&0, \nn
\alpha_{19}^\Delta+\alpha_{26}^\Delta-\alpha_{27}^\Delta+
\frac{1}{2}&=&0,\nn 
\alpha_{15}^\Delta-\alpha_{25}^\Delta&=&0, \nn
\alpha_{17}^\Delta+\alpha_{25}^\Delta-\alpha_{27}^\Delta+
\frac{1}{2}&=&0, \nn
\alpha_{14}^\Delta+\alpha_{26}^\Delta-\alpha_{27}^\Delta+
\frac{1}{2}&=&0, \nn
\alpha_{24}^\Delta-\alpha_{27}^\Delta&=&0, \nn
\alpha_{1}^\Delta-\alpha_{2}^\Delta&=&0, \nn
-\alpha_{11}^\Delta+\alpha_{2}^\Delta+2 
\alpha_{29}^\Delta+\alpha_{5}^\Delta+\frac{1}{4}&=&0,\nn 
-2 \alpha_{29}^\Delta+\alpha_{3}^\Delta+\frac{1}{4}&=&0, \nn
-2 
\alpha_{29}^\Delta+\alpha_{4}^\Delta+\frac{1}{4}&=&0,\nn 
\alpha_{5}^\Delta-\alpha_{11}^\Delta&=&0, \nn
-\alpha_{11}^\Delta-2 
\alpha_{29}^\Delta+\alpha_{6}^\Delta-\frac{1}{4}&=&0, \nn
\alpha_{12}^\Delta+4 \alpha_{29}^\Delta&=&0, \nn
\alpha_{35}^\Delta-\alpha_{37}^\Delta&=&0, \nn
\alpha_{36}^\Delta-
\alpha_{37}^\Delta+\alpha_{38}^\Delta+\frac{1}{2}&=&0, \nn
\alpha_{37}^\Delta-\alpha_{38}^\Delta+\alpha_{41}^\Delta-
\frac{1}{2}&=&0, \nn 
\alpha_{30}^\Delta-\frac{1}{2}&=&0.
\label{constraints}
\eeqa

\def\theequation{\Alph{section}.\arabic{equation}}
\setcounter{equation}{0}
\section{Delta-isobar contributions to the invariant $\pi N$ amplitudes $g^\pm (\omega, t)$ and $h^\pm (\omega, t)$}
\label{appendix:piN_amplitude}
\allowdisplaybreaks[1]

In this Appendix we present the explicit expressions for the invariant 
amplitudes $g^\pm (\omega, t)$ and $h^\pm (\omega, t)$ which parametrize 
the pion-nucleon scattering matrix at first three orders in  $\epsilon$ expansion
(for an earlier calculation see Ref.~\cite{Fettes:2000bb}).
We give only contributions due to intermediate delta excitations,
which have to be added to the nucleonic terms calculated within the
$\Delta$-less theory and listed in Ref.~\cite{Krebs:2012yv}.

\underline{Contributions at order $\epsilon^1$:}
 \begin{eqnarray}
g^{+} & = & \frac{4\Delta
            h_{A}^{2}\left(-2M_{\pi}^{2}+t+2\omega^{2}\right)}{9F_{\pi}^{2}(\Delta-\omega)(\Delta+\omega)}\,,\nn
 g^{-}&=&-\frac{2h_{A}^{2}\omega\left(-2M_{\pi}^{2}+t+2\omega^{2}\right)}{9F_{\pi}^{2}(\Delta-\omega)(\Delta+\omega)}\,,\nonumber \\
h^{+} & = &
            \frac{4h_{A}^{2}\omega}{9F_{\pi}^{2}\left(\omega^{2}-\Delta^{2}\right)}\,,\nn 
h^{-}&=&\frac{2\Delta h_{A}^{2}}{9F_{\pi}^{2}\left(\Delta^{2}-\omega^{2}\right)}\,.\end{eqnarray}

\underline{Contributions at order $\epsilon^2$:} 
\begin{eqnarray}
g^{+} & = & 0\,,\ g^{-}=0\,,\ h^{+}=0\,,\ h^{-}=0\,.
\end{eqnarray}

\underline{Contributions at order $\epsilon^3$:}
\beqa
g^{\pm} & = & g^{\pm}_{{\rm SL}} + g^{\pm}_{1/m_N}, \quad h^{\pm}\,= \, h^{\pm}_{{\rm SL}} + h^{+}_{1/m_N},
\eeqa
with static limit contributions given by
\beqa
g^{+}_{{\rm SL}} &=&\frac{{h_A}^2}{486 F_\pi^4 \omega ^2} {\bar{J}_0}(-\Delta ) \left(\left(81 
{g_A}^2-50 {g_A} {g_1}+25 {g_1}^2\right) \left(M_\pi^2-\Delta ^2
\right) \left(-2 M_\pi^2+t+2 \omega ^2\right)-108 \omega ^2 
\left(M_\pi^2-2 t\right)\right)\nn
&+&\frac{{h_A}^2 
}{4374 F_\pi^4 \omega ^2 (\Delta -\omega 
)^2}{\bar{J}_0}(\omega -\Delta ) \left(M_\pi^2-(\Delta -\omega )^2\right) 
\left(2 M_\pi^2-t-2 \omega ^2\right) \left(9 \Delta ^2 \left(81 
{g_A}^2-50 {g_A} {g_1}+25 {g_1}^2\right)\right.\nn
&-&\left. 10 \omega  (\Delta -\omega ) 
(9 {g_A}-5 {g_1})^2\right) + \frac{  {h_A}^2}{36 \pi ^2 
F_\pi^4} {D}\left(\sqrt{-t}\right) \Delta\left(M_\pi^2-2 t
\right) \left(-2 \Delta ^2+2 M_\pi^2-t\right)\nn
&+&\frac{8 {h_A}^2 }{243 F_\pi^4 \left(\Delta ^2-\omega ^2
\right)^2}{\bar{J}_0}(\omega ) \left(\omega ^2-M_\pi^2
\right) \left(-2 M_\pi^2+t+2 \omega ^2\right) \left(9 {g_A}^2 
\left(\Delta ^2-\omega ^2\right)+{h_A}^2 \left(5 \Delta ^2+4 \omega 
^2\right)\right)\nn
&+&\frac{{h_A}^2 }{34992 \pi ^2 F_\pi^4 \left(\Delta 
^2-\omega ^2\right)^2}\left(\left(2 M_\pi^2-t-2 \omega ^2\right) 
\left(\left(81 {g_A}^2-50 {g_A} {g_1}+25 {g_1}^2\right) \left(9 
\Delta  \left(\Delta ^2-\omega ^2\right)^2\right.\right. \right.\nn
&+&\left.\left. \left. \pi  M_\pi^3 \left(17 
\Delta ^2+\omega ^2\right)+9 \Delta  M_\pi^2 \left(\omega ^2-\Delta 
^2\right)\right)+400 \pi  {g_A} {g_1} M_\pi^3 \left(\Delta ^2-\omega 
^2\right)\right)-486 \Delta  \left(\Delta ^2-\omega ^2\right)^2 
\left(M_\pi^2-2 t\right)\right)\nn
&+&\frac{\Delta  {h_A}^2 \log \left(\frac{2 
\Delta }{M_\pi}\right) \left(10 (9 {g_A}-5 {g_1})^2 \left(2 
M_\pi^2-t-2 \omega ^2\right)+729 \left(M_\pi^2-2 t\right)
\right)}{8748 \pi ^2 F_\pi^4}+\frac{2 \Delta  {h_A}^2 {\bar{I}_{20}}(t) 
\left(M_\pi^2-2 t\right)}{9 F_\pi^4}\nn
& + & (\omega\rightarrow -\, \omega),
\eeqa
\beqa
g^{-}_{{\rm SL}} &=&
\frac{ 
{h_A}^2}{69984 F_\pi^4 \pi ^2 \Delta  \omega  \left(\Delta ^2-\omega 
^2\right)^2}\left(\left(2 M_\pi^2-2 \omega ^2-t\right) \left(-81 \left(2 
\pi  \left(12 \Delta ^4-35 \omega ^2 \Delta ^2+20 \omega ^4\right) 
M_\pi^3\right.\right.\right. \nn
&-&\left.\left.\left.   3 \left(6 \Delta ^5-13 \omega ^2 \Delta ^3+7 \omega ^4 \Delta 
\right) M_\pi^2+9 \Delta  \left(\Delta ^2-\omega ^2\right)^2 \left(4 
\Delta ^2+3 \omega ^2\right)\right) {g_A}^2\right.\right.  \nn
&+&\left. \left. 150 {g_1} \Delta  \left(6 
\pi  \Delta  \omega ^2 M_\pi^3-3 \left(2 \Delta ^4-3 \omega ^2 \Delta 
^2+\omega ^4\right) M_\pi^2+\left(\Delta ^2-\omega ^2\right)^2 
\left(12 \Delta ^2+13 \omega ^2\right)\right) {g_A}\right.\right.  \nn
&+&\left. \left. 25 {g_1}^2 
\left(-23 \Delta  \omega ^6+\left(8 \pi  M_\pi^3-3 \Delta  M_\pi^2+34 
\Delta ^3\right) \omega ^4+\Delta ^2 \left(-26 \pi  M_\pi^3-3 \Delta  
M_\pi^2+\Delta ^3\right) \omega ^2+6 \Delta ^5 \left(M_\pi^2-2 \Delta 
^2\right)\right)\right)\right. \nn
&-&\left. 18 \Delta  \omega ^2 \left(\Delta ^2-\omega 
^2\right) \left(4 {h_A}^2 \left(2 M_\pi^2-2 \omega ^2-t\right) 
\left(3 M_\pi^2-4 \Delta ^2-2 \omega ^2\right)-3 \left(24 M_\pi^2-72 
\Delta ^2-19 t\right) \left(\Delta ^2-\omega
^2\right)\right)\right)\nn
&+&\frac{\Delta ^2 \left(-2 M_\pi^2+2 \Delta ^2+t\right) 
\omega  {D}\left(\sqrt{-t}\right) {h_A}^2}{18 F_\pi^4 \pi 
^2}+\frac{\left(8 M_\pi^2-12 \Delta ^2-5 t\right) \omega  
{\bar{I}_{20}}(t) {h_A}^2}{27 F_\pi^4}\nn
&+&\frac{ {h_A}^2}{4374 
F_\pi^4 \Delta  \omega  \left(\Delta ^2-\omega ^2\right)^2}\left(\left(2 M_\pi^2-2 
\omega ^2-t\right) \left(36 {h_A}^2 \left(4 \Delta ^4-10 \omega ^2 
\Delta ^2+M_\pi^2 \left(5 \Delta ^2+\omega ^2\right)\right) \omega 
^2\right.\right.  \nn
&+&\left.\left.  \left(\Delta ^2-\omega ^2\right) \left(81 \left(-19 \Delta ^4+15 
\omega ^2 \Delta ^2+M_\pi^2 \left(12 \omega ^2-8 \Delta ^2\right)
\right) {g_A}^2+1350 {g_1} \Delta ^2 \left(\Delta ^2-\omega ^2\right) 
{g_A}\right. \right.\right.  \nn
&+&\left. \left.\left.  25 {g_1}^2 \left(4 M_\pi^2-13 \Delta ^2\right) \left(\Delta ^2-
\omega ^2\right)\right)\right)-972 \Delta ^2 \omega ^2 \left(\Delta 
^2-\omega ^2\right)^2\right) {\bar{J}_0}(-\Delta )\nn
&-&\frac{2 
\left(2 M_\pi^2-2 \omega ^2-t\right) \left(\omega ^2-M_\pi^2\right) 
\left(2 {h_A}^2 \omega  \left(\Delta ^2-12 \omega  \Delta +2 \omega 
^2\right)-9 {g_A}^2 (3 \Delta -\omega ) \left(\Delta ^2-\omega ^2
\right)\right) {\bar{J}_0}(\omega ) {h_A}^2}{243 F_\pi^4 \omega  \left(
\Delta ^2-\omega ^2\right)^2}\nn
&+&\frac{
{h_A}^2}{8748 F_\pi^4 (\Delta -\omega )^2 \omega ^2}\left(M_\pi^2-(\Delta -\omega )^2
\right) \left(6 \left(243 {g_A}^2-150 {g_1} {g_A}+25 {g_1}^2\right) 
\Delta ^2-5 (9 {g_A}-5 {g_1})^2 (4 \Delta -\omega ) \omega \right) \nn
&\times&
\left(2 M_\pi^2-2 \omega ^2-t\right) {\bar{J}_0}(\omega -\Delta ) \nn
&+&\frac{\omega  
\left(\left(243 {g_A}^2+450 {g_1} {g_A}-288 {h_A}^2-125 {g_1}^2
\right) \left(2 M_\pi^2-2 \omega ^2-t\right)-810 t\right) \log 
\left(\frac{2 \Delta }{M_\pi}\right) {h_A}^2}{34992 F_\pi^4 \pi ^2}
\nn
& - & (\omega\rightarrow -\, \omega),
\eeqa
\beqa
h^{+}_{{\rm SL}}&=&
-\frac{{h_A}^2 }{4374 F_\pi^4 \omega ^2 (\Delta 
-\omega )^2}{\bar{J}_0}(\omega -\Delta ) \left(M_\pi^2-(\Delta 
-\omega )^2\right) \left(6 \Delta ^2 \left(243 {g_A}^2-150 {g_A} 
{g_1}+25 {g_1}^2\right)\right. \nn
&+&\left.(5 \omega^2-20\Delta \omega) (9 {g_A}-5 {g_1})^2\right)\nn
&+&\frac{{h_A}^2 }{2187 
\Delta  F_\pi^4 \omega  \left(\Delta ^2-\omega ^2\right)^2}{\bar{J}_0}(-\Delta ) \left(\left(\Delta 
^2-\omega ^2\right) \left(81 {g_A}^2 \left(19 \Delta ^4-3 \omega ^2 
\left(5 \Delta ^2+4 M_\pi^2\right)+8 \Delta ^2
M_\pi^2\right)\right. \right. \nn
&-&\left. \left. 1350 
\Delta ^2 {g_A} {g_1} \left(\Delta ^2-\omega ^2\right)-25 {g_1}^2 
\left(\Delta ^2-\omega ^2\right) \left(4 M_\pi^2-13 \Delta ^2\right)
\right)\right. \nn
&-&\left. 36 {h_A}^2 \omega ^2 \left(4 \Delta ^4-10 \Delta ^2 \omega 
^2+M_\pi^2 \left(5 \Delta ^2+\omega ^2\right)\right)\right)\nn
&-&\frac{4 
{h_A}^2 {\bar{J}_0}(\omega ) \left(M_\pi^2-\omega ^2\right) \left(2 
{h_A}^2 \omega  \left(\Delta ^2-12 \Delta  \omega +2 \omega ^2
\right)-9 {g_A}^2 (\Delta -\omega ) (3 \Delta -\omega ) (\Delta +\omega 
)\right)}{243 F_\pi^4 \omega  \left(\Delta ^2-\omega ^2
\right)^2}\nn
&+&\frac{{h_A}^2 }{34992 \pi ^2 
\Delta  F_\pi^4 \omega  \left(\Delta ^2-\omega ^2
\right)^2}\left(\left(\Delta ^2-\omega ^2\right) 
\left(\Delta  \left(\Delta ^2-\omega ^2\right) \left(729 {g_A}^2 
\left(4 \Delta ^2+3 \omega ^2\right)-150 {g_A} {g_1} \left(12 \Delta 
^2+13 \omega ^2\right)\right. \right. \right. \nn
&+&\left. \left. \left. 25 {g_1}^2 \left(12 \Delta ^2+23 \omega ^2
\right)\right)-72 \Delta  {h_A}^2 \omega ^2 \left(4 \Delta ^2-3 
M_\pi^2+2 \omega ^2\right)\right)+81 {g_A}^2 \left(2 \pi  M_\pi^3 
\left(12 \Delta ^4-35 \Delta ^2 \omega ^2+20 \omega
  ^4\right)\right. \right. \nn
&-&\left. \left. 3 
M_\pi^2 \left(6 \Delta ^5-13 \Delta ^3 \omega ^2+7 \Delta  \omega ^4
\right)\right)-150 \Delta  {g_A} {g_1} \left(6 \pi  \Delta  M_\pi^3 
\omega ^2-3 M_\pi^2 \left(2 \Delta ^4-3 \Delta ^2 \omega ^2+\omega ^4
\right)\right)\right. \nn
&+&\left. 25 {g_1}^2 \left(\pi  M_\pi^3 \left(26 \Delta ^2 
\omega ^2-8 \omega ^4\right)+3 \Delta  M_\pi^2 \left(-2 \Delta 
^4+\Delta ^2 \omega ^2+\omega ^4\right)\right)\right)\nn
&+&\frac{{h_A}^2 \omega  \left(-243 {g_A}^2-450 {g_A} {g_1}+288 
{h_A}^2+125 {g_1}^2\right) \log \left(\frac{2 \Delta }{M_\pi}
\right)}{17496 \pi ^2 F_\pi^4}
\nn
& - & (\omega\rightarrow -\, \omega),
\eeqa
\beqa
h^{{-}}_{{\rm SL}}&=&
\frac{\Delta  {h_A}^2 {D}\left(\sqrt{-t}\right) \left(4 \Delta ^2-4 
M_\pi^2+t\right)}{144 \pi ^2 F_\pi^4} -\frac{\Delta  
{h_A}^2 {\bar{I}_{20}}(t)}{9 F_\pi^4}\nn
&+&\frac{{h_A}^2 
}{4374 F_\pi^4 \omega ^2 \left(\Delta ^2-\omega ^2
\right)^2}{\bar{J}_0}(-\Delta ) \left(10 \left(\Delta ^2-\omega ^2\right)^2 
\left(81 {g_A}^2-90 {g_A} {g_1}+5 {g_1}^2\right) \left(M_\pi^2-\Delta 
^2\right)\right. \nn
&+&\left. 648 \Delta ^2 {g_A}^2 \left(\Delta ^2-\omega ^2\right) 
\left(M_\pi^2-\Delta ^2\right)+9 \omega ^2 \left(27 \left(\Delta 
^2-\omega ^2\right)^2+32 {h_A}^2 \left(2 \Delta ^2 \left(\Delta ^2-2 
\omega ^2\right)+M_\pi^2 \left(\Delta ^2+\omega ^2\right)\right)
\right)\right)\nn
&-&\frac{4 {h_A}^2 {\bar{J}_0}(\omega ) \left(M_\pi^2-\omega ^2
\right) \left({h_A}^2 \omega  \left(9 \Delta ^2-8 \Delta  \omega +8 
\omega ^2\right)-9 {g_A}^2 (\Delta -2 \omega ) \left(\Delta ^2-\omega 
^2\right)\right)}{243 F_\pi^4 \omega  \left(\Delta ^2-\omega ^2
\right)^2}\nn
&+&\frac{{h_A}^2 }{69984 \pi ^2 
F_\pi^4 \left(\Delta ^2-\omega ^2\right)^2}\left(\Delta  \left(\Delta ^2-\omega ^2\right) 
\left(M_\pi^2 \left(81 {g_A}^2-450 {g_A} {g_1}-576 {h_A}^2+225 
{g_1}^2\right)+1152 \Delta ^2 {h_A}^2\right)\right. \nn
&+&\left. \Delta  \left(\Delta ^2-
\omega ^2\right)^2 \left(-7047 {g_A}^2+3150 {g_A} {g_1}+25 {g_1}^2
\right)+\pi  M_\pi^3 (5 {g_1}-9 {g_A}) \left(\Delta ^2 (297 {g_A}-85 
{g_1})-\omega ^2 (279 {g_A}+5 {g_1})\right)\right)\nn
&-&\frac{\Delta  {h_A}^2 
\left(2349 {g_A}^2-2250 {g_A} {g_1}+1152 {h_A}^2+125 {g_1}^2+972
\right) \log \left(\frac{2 \Delta }{M_\pi}\right)}{34992 \pi ^2 
F_\pi^4}\nn
&-&\frac{{h_A}^2 (9 {g_A}-5 {g_1}) {\bar{J}_0}(\omega -\Delta ) 
\left(M_\pi^2-(\Delta -\omega )^2\right) \left(-360 \Delta  
\omega {g_A} +4 \Delta ^2 (81 {g_A}-5 {g_1})+5 \omega ^2 (9 {g_A}-5 {g_1})
\right)}{8748 F_\pi^4 \omega ^2 (\Delta -\omega )^2}
\nn
& + & (\omega\rightarrow -\, \omega),
\eeqa
and $1/m_N$ corrections given by
\beqa
g^{+}_{1/m_N} &=&  -  \frac{h_{A}^{2}}{9F_{\pi}^{2}m_N(\Delta^2-\omega^2)^{2}}\bigg\{16\Delta^{4}\omega^{2}-4\Delta^{3}\omega\left(-4M_{\pi}^{2}+t+4\omega^{2}\right)\nonumber\\
 & + & \Delta^{2}\left(4M_{\pi}^{4}-4M_{\pi}^{2}\left(t+6\omega^{2}\right)+t^{2}+8t\omega^{2}-12\omega^{4}\right)-2\Delta\omega\left(8M_{\pi}^{4}-2M_{\pi}^{2}\left(3t+4\omega^{2}\right)+t\left(t+4\omega^{2}\right)\right)\nonumber\\
 & + &
 \omega^{2}\left(4M_{\pi}^{4}-4M_{\pi}^{2}\left(t-2\omega^{2}\right)+t^{2}+4\omega^{4}\right)\bigg\}+\frac{16\,
 h_A^2 \,\omega^2}{9 F_\pi^2 m_N},\nonumber\\
g^{-}_{1/m_N} &=&  -  \frac{h_{A}^{2}}{18F_{\pi}^{2}m_N(\Delta^2-\omega^2)^{2}}\bigg\{-4\Delta^{3}\omega\left(-4M_{\pi}^{2}+t+4\omega^{2}\right)\nonumber \\
 & + & \Delta^{2}\left(8M_{\pi}^{4}-2M_{\pi}^{2}\left(3t+16\omega^{2}\right)+t^{2}+10t\omega^{2}+24\omega^{4}\right)\nonumber \\
 & - &
 2\Delta\omega\left(4M_{\pi}^{4}-4M_{\pi}^{2}t+t^{2}+2t\omega^{2}-4\omega^{4}\right)+\omega^{2}\left(8M_{\pi}^{4}-6M_{\pi}^{2}t+t^{2}+2t\omega^{2}-8\omega^{4}\right)\bigg\}\nonumber\\
&-&\frac{h_A^2}{9 F_\pi^2 m_N \Delta}\omega \big(t + 8 (\omega^2-M_\pi^2)\big),\nonumber\\
h^{+}_{1/m_N} &=&
\frac{h_{A}^{2}\Big[4\Delta^{3}\omega+\Delta^{2}\left(4M_{\pi}^{2}-t-8\omega^{2}\right)+2\Delta\omega\left(t-2M_{\pi}^{2}\right)+\omega^{2}\left(4M_{\pi}^{2}-t\right)\Big]}{9F_{\pi}^{2}m_N(\Delta^2-\omega^2)^{2}}
- \frac{4\,h_A^2\,\omega}{9 F_\pi^2 m_N \Delta},\nonumber
\\
h^{-}_{1/m_N}& = & \frac{h_{A}^{2}\left(4\Delta^{3}\omega+\Delta^{2}\left(2M_{\pi}^{2}-t-6\omega^{2}\right)+2\Delta\omega\left(-4M_{\pi}^{2}+t+2\omega^{2}\right)+\omega^{2}\left(2M_{\pi}^{2}-t+2\omega^{2}\right)\right)}{18F_{\pi}^{2}m_N(\Delta^2-\omega^2)^{2}}.
\eeqa
The loop functions are defined via:
\begin{eqnarray}
2 M_\pi^2\lambda_\pi &=& \mu^{4-d}\int \frac{d^d
  l}{(2\pi)^d}\frac{i}{l^2-M_\pi^2+ i \epsilon},\nn
\bar{J}_{0}(\omega) & = &\left\{\begin{array}{ll}
\frac{\omega}{8\pi^2} +\frac{\sqrt{\,\omega^{2}-M_{\pi}^{2}}}{4\pi^{2}}
\arccosh\left(-\frac{\,\omega}{M_{\pi}}\right) &{\rm for}\quad \omega < - M_\pi\\
\frac{\omega}{8\pi^2} -\frac{\sqrt{\,\omega^{2}-M_{\pi}^{2}}}{4\pi^{2}}
\left(\arccosh\left(\frac{\omega}{M_{\pi}}\right) - i\,\pi\right)
&{\rm for}\quad \omega >  M_\pi \\
\frac{\omega}{8\pi^2} -\frac{\sqrt{\,M_{\pi}^{2} - \omega^{2}}}{4\pi^{2}}
\arccos\left(-\frac{\,\omega}{M_{\pi}}\right) &{\rm for}\quad |\omega| < M_\pi
\end{array},\right.
\nonumber\\
D(x)&=&\frac{1}{\Delta}\int_{2M_{\pi}}^{\infty}\frac{d\mu}{\mu^2+x^2} \arctan \frac{\sqrt{\mu^2-4M_{\pi}^2}}{2\Delta}\,,\nonumber\\
\bar{I}_{20}(t) & = & \frac{1 }{16\pi^{2}} - \frac{\sqrt{1-4M_{\pi}^{2}/t}}{16\pi^{2}}
\log\frac{\sqrt{1-4M_{\pi}^{2}/t}+1}{\sqrt{1- 4M_{\pi}^{2}/t}-1}\,.
\end{eqnarray}

\end{document}